\documentclass[12pt,nofootinbib]{article}

\pdfoutput=1

\usepackage{slashed}
\usepackage{amsmath,amssymb,graphicx,multicol} 
\usepackage{epsf,color}
\usepackage{cite}
\usepackage{cancel}
\usepackage{framed}
\usepackage{multirow}

\newcommand{\nc}{\newcommand}
\nc{\postscript}[2]{\setlength{\epsfxsize}{#2\hsize}\centerline{\epsfbox{#1}}}

\newcommand\beq{\begin{eqnarray}}
\newcommand\eeq{\end{eqnarray}}
\def\bea{\begin{eqnarray}}
\def\eea{\end{eqnarray}}

\def\bit{\begin{itemize}}
\def\eit{\end{itemize}}

\def\l{\left}
\def\r{\right}

\def\baa{\begin{array}}
\def\eaa{\end{array}}

\def\lag{\mathcal L}

\def\simgt{\mathrel{\lower2.5pt\vbox{\lineskip=0pt\baselineskip=0pt
           \hbox{$>$}\hbox{$\sim$}}}}
\def\simlt{\mathrel{\lower2.5pt\vbox{\lineskip=0pt\baselineskip=0pt
           \hbox{$<$}\hbox{$\sim$}}}}

\begin{document}
\begin{titlepage}
\begin{flushright}
\end{flushright}

\renewcommand{\thefootnote}{\fnsymbol{footnote}}

\vspace{-0.5cm}
\begin{center} 
{\huge \bf  Electroweak Symmetry Breaking  \\[0.15cm] and the Higgs Boson: \\[0.5cm] Confronting Theories at Colliders}
\end{center}
\vskip0.8cm

\begin{center}
{\bf Aleksandr Azatov and Jamison Galloway\footnote{email:  aleksandr.azatov@roma1.infn.it, jamison.galloway@roma1.infn.it}
}
\end{center}

\begin{center}
{\it Dipartimento di Fisica, Universit\`a di Roma ``La Sapienza'' \\
{\rm and} INFN Sezione di Roma, I-00185 Rome, Italy} \\
\vspace*{0.2cm}
\end{center}

\vspace{0.5cm}

\vglue 0.2truecm

\begin{abstract}
\vskip 3pt \noindent
In this review, we discuss methods of parsing direct and indirect information from collider experiments regarding the Higgs boson and describe simple ways in which experimental likelihoods can be consistently reconstructed and interfaced with model predictions in pertinent parameter spaces.  Ultimately these methods are used to constrain a five-dimensional parameter space describing a model-independent framework for electroweak symmetry breaking.  We review prevalent scenarios for extending the electroweak symmetry breaking sector relative to the Standard Model and emphasize their predictions for nonstandard Higgs phenomenology that could be observed in LHC data if naturalness is realized in particular ways.    Specifically we identify how measurements of Higgs couplings can be used to imply the existence of new physics at particular scales within various contexts, highlighting some parameter spaces of interest in order to give examples of how the  data surrounding the new state can most effectively be used to constrain specific models of weak scale physics.\footnote{{\bf Data implemented}: March 2013 (following ``Rencontres de Moriond", $\int dt \, \mathcal L \leq 25 \, {\rm fb}^{-1}$ @ LHC)}
\end{abstract}

\vspace{1cm}

\begin{center}
{\it Review article prepared for International Journal of Modern Physics A}
\end{center}
 
\vspace{1cm} 

\end{titlepage}

\renewcommand{\thefootnote}{\arabic{footnote}}

\section{Introduction}
\label{sec:Intro} \setcounter{equation}{0} \setcounter{footnote}{0}
The field of elementary particle physics is entering an exhilarating new era, where the nature of the weak scale and possibly clues about its origins are being revealed experimentally.  We have mounting  evidence from the Large Hadron Collider (LHC) that electroweak symmetry breaking (EWSB) involves a scalar boson with mass near 125 GeV and a nonzero vacuum expectation value (VEV) \cite{ATLASdiscovery, CMSdiscovery}.  It could well be that this is the long-sought Higgs boson \cite{Higgs} of the Standard Model (SM) \cite{SM}, though decades of study regarding the theoretical implications of such a particle lead us naturally to hope that the most appropriate explanation for the new state will turn out to involve non-standard dynamics.  Most notably, obtaining a mass of 125 GeV for the Higgs boson in the SM requires its mass parameter to be tuned with a precision  that we have never before encountered in the study of elementary particles.

In this review, we recap the experimental results from the measurement of various Higgs properties, and describe in detail how these data can directly constrain constructions that go beyond the standard model (BSM).   We provide a detailed exposition of the methodology  used in assessing the data, collecting in a uniform way the tools that model builders will need in order to themselves confront new physics scenarios with input from the LHC.  Distilling the  data from experiments to their simplest, though complete, and  most user-friendly format in order to accomplish these tasks is the main goal of this work.  

We focus on two prevalent BSM scenarios to demonstrate the development and utility of the constraints from Higgs searches: first we look at cases where the Higgs is a composite pseudo Nambu-Goldstone boson (PNGB), and second where the Higgs is embedded into some implementation of supersymmetry (SUSY).  In order to thoroughly and accurately confront  theoretical hypotheses with experimental data, we review the basic Higgs phenomenology  within each context.  Both cases are devised to render the weak scale natural, with the Higgs mass shielded from the uncontrolled  quantum effects that plague the SM.  In this regard, we see careful measurement of the Higgs properties serve as indirect probes of the principle of naturalness.  

At the outset there are two simple statements about the cases we cover that can provide useful information about how naturalness may reveal itself, if indeed it is realized at all.  Both rely on the essential fact that deviations in tree-level Higgs couplings unequivocally signal an enlarged EWSB sector, while deviations in loop-level couplings signal the presence of additional non-SM matter fields:
\begin{itemize}
\item The composite Higgs scenarios rely largely  on the Higgs sector itself to restore naturalness.  In these cases then we would generically expect large deviations in interactions mediated at {\it tree-level}. Loop processes, as we'll see, are typically quite SM-like except for some non-generic cases.
\item In SUSY, it is the addition of new matter content that is responsible for restoring naturalness.  In these cases, we generically expect to find {\it loop-level} processes to deviate substantially from the SM.   Tree-level couplings can remain as in the SM without indicating a large departure from naturalness.
\end{itemize}
We thus discuss in each case the modifications of Higgs couplings at the level of zero and one loops.

We will employ a model-independent approach in the discussions to follow, with input from the underlying theories limited to only that which is truly essential.  The utility of  this formalism is that it highlights the pragmatic question  of whether or not the state observed at 125 GeV is single-handedly responsible for completing the physics below that scale, or rather if additional states associated with EWSB should be anticipated.  
We know for instance that tree-level couplings of a light Higgs in extended EWSB sectors adhere to simple sum rules applying for a particular species:
\beq
\label{eq:sumrule}
\sum_i g_{VV h_i}^2 = g_{VVh^{\rm (SM)}} ^2, \qquad
 \sum_i g_{f\bar f h_i} g_{VV h_i} = g_{f \bar f h^{\rm (SM)}} g_{VVh^{\rm (SM)}},
\eeq
i.e. the magnitude of couplings' deviations from the SM must be inversely related to the masses of additional fields.   We thus have very  well-defined questions---how SM-like do the couplings of the light state appear?---with practical implications, but that can also  begin to put naturalness to the test experimentally for the first time.

We emphasize that our intent is not to provide an exhaustive review of alternative scenarios for EWSB nor to  give a complete bibliography  for these alternatives, though we do provide accessible entry points to the literature.  Further we stress that even now, the results we present must be taken somewhat provisionally and improved upon as statistical significance of the LHC data increases.    

We organize the review as follows.  In Section~\ref{sec:Theory} we give an overview of an effective theory of EWSB that will serve as the formalism for all cases we consider.  In Section~\ref{sec:Experiment} we review the experimental input from the LHC, and discuss how it can directly probe the effective theory.  In Section~\ref{sec:Composite} we turn to the case study of the composite PNGB Higgs, and in Section~\ref{sec:SUSY} to the case of multi-Higgs models in SUSY.  We conclude in Section~\ref{sec:Conclusions}.  Various details of the precise experimental input and its application, along with a brief review of some necessary theoretical formalism, are relegated to four appendices.

\section{Theoretical Overview: An Effective Theory of EWSB}
\label{sec:Theory} \setcounter{equation}{0} 

We begin by reviewing the bottom-up construction of a theory of EWSB that will encompass the models discussed below.   We make no mention of the SM as an input to the theory: the Lagrangian we present below is meant as a description of low-energy physics that can eventually be understood as part of a more complete UV dynamics  \cite{lagrangian}.  In different models, the operators we focus on will have coefficients with forms that are characteristic of the particular model.   To  thoroughly explore the weak scale, however, it is convenient to proceed in a model-independent way.   The SM emerges only as a particular limit of the general framework, albeit the only limit where this framework by itself is really complete to arbitrarily high scales when gravity is neglected.

In the absence of a Higgs boson, our low-energy weak scale physics is described by a nonlinear sigma model (NLSM), and one of the most pressing questions for the LHC is whether or not the theory is fully linearized below the strong-coupling cutoff at $4 \pi v \sim  {\rm TeV}$.    Prior to any data taken at the LHC, we already had compelling hints from precision data collected at LEP \cite{LEPprecision} that this strong coupling limit would not be reached without the intervention of new physics: logarithmic corrections to precision  parameters \cite{PT} in particular were found to prefer a  cutoff well below the onset  of strong coupling within the NLSM.  LEP seems to have been advocating for the SM with a light ($\lesssim 200\, {\rm GeV}$) Higgs, and the LHC is now testing this directly.

Experimental results from LEP did not unequivocally identify the Higgs boson of the SM as the only possibility for completing weak scale physics, but it was made clear that a {\it Higgs-like} state would be a welcome new participant in the chiral lagrangian.  Our starting assumption is thus simply that the NLSM is appended by a custodial singlet scalar with  unspecified couplings.  In unitary gauge, we have \cite{plhc}
\beq
\lag = \lag^{(2)} + \lag^{(4)},
\eeq
where
\beq
\label{eq:general}
\mathcal{L}^{(2)} &=&  \frac{1}{2} (\partial_\mu h)^2 -\frac{1}{2} m_h^2 h^2  
 - \sum_{\psi = u,d,l} m_{\psi} \, \bar \psi \psi  \left (1+c_\psi \frac{h}{v} + \mathcal O(h^2) \right) \nonumber \\
&& - \left( m_W^2 W_\mu W^\mu  + \frac{1}{2} m_Z^2 Z_\mu Z^\mu \right) \left(1 + 2a \frac{h}{v} +  \mathcal O(h^2) \right);  \\
\mathcal{L}^{(4)} &=& \frac{\alpha_{em}}{4\pi}  \left(  \frac{c_{WW}}{s_W^2}\,   W_{\mu\nu}^+ W_{\mu\nu}^-  + \frac{c_{ZZ}}{s_W^2 c_W^2}\, Z_{\mu\nu}^2 + 
\frac{c_{Z\gamma}}{s_W c_W}\, Z_{\mu\nu} \gamma_{\mu\nu}   + c_{\gamma\gamma}\, \gamma_{\mu\nu}^2 \right) \frac{h}{v} \nonumber \\
&& +  \frac{\alpha_s}{4\pi} \, c_{gg}\, G_{\mu\nu}^2 \ \frac{h}{v}.
\eeq
The SM in this framework corresponds to the singular limit $a=c=1$, with $\mathcal L^{(4)}$ vanishing.  The weak gauge bosons and top quark remain in the theory, so loop-induced  couplings of $h$ involving these fields can be derived by appropriately rescaling the expressions derived in the SM, using factors that we summarize in appendix~\ref{sec:Loops}.

\begin{figure}[htb]
\begin{center}
\includegraphics[width=5cm]{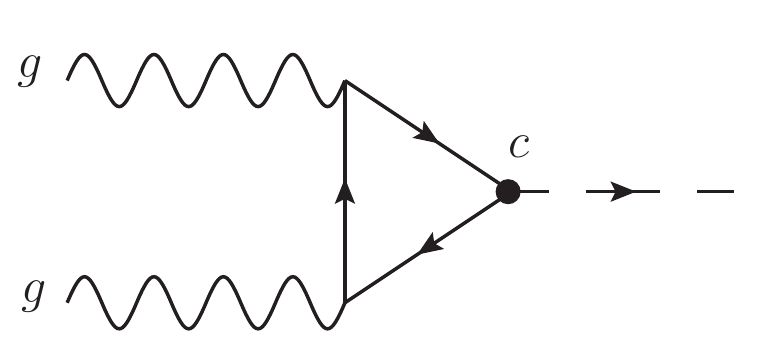}    \quad
\includegraphics[width=4.5cm]{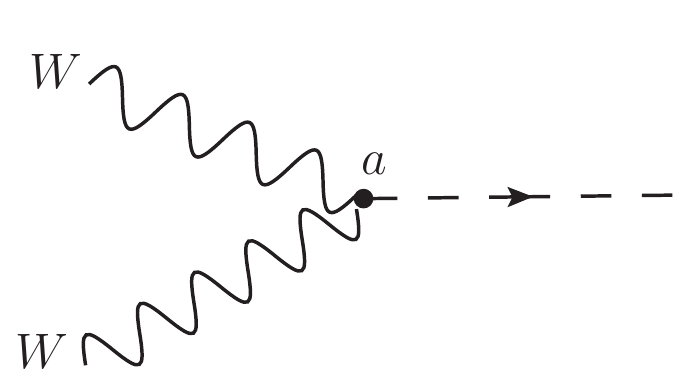}    \quad
\includegraphics[width=4.2cm]{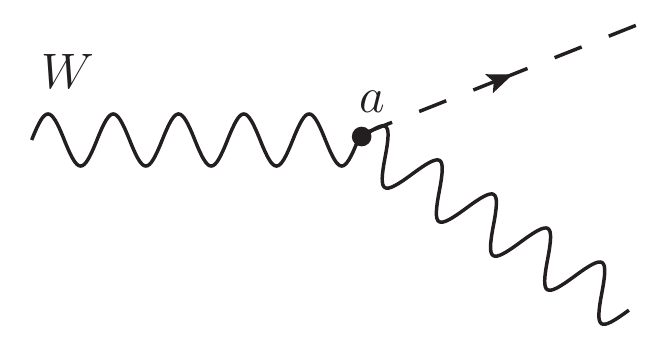}
\caption{\small Dominant single-Higgs production from   $\mathcal O (p^2)$ terms of Eq.~(\ref{eq:general}).  Gluon fusion at this level is rescaled by $c^2$ compared to the SM, VBF and VH by $a^2$.}
\label{fig:Production}
\end{center}
\end{figure}

From the Lagrangian, Eq.~(\ref{eq:general}), we see that the dominant single-Higgs production mechanisms shown in Fig.~\ref{fig:Production} are related in simple ways to their SM counterparts.
Given these couplings, we can make concrete predictions in the generic space for the event rates one would observe experimentally.  Comparing directly to results presented from the LHC collaborations then allows constraints to be constructed in any model of interest.  Many such applications of the data have appeared in the literature, cf. \cite{fits,sfitter,fits2}.

\section{Experimental Input}
\label{sec:Experiment} \setcounter{equation}{0}

\subsection{Hadron Colliders}
We focus primarily on Higgs searches and direct data regarding its existence coming from the LHC and Tevatron.  At present there are five primary final states used in the Higgs search.  In order of increasing sensitivity to a 125 GeV SM Higgs, these channels are $h \to b\bar b$, $h \to \tau \tau$, $h \to WW$, $h \to \gamma \gamma$, $h \to ZZ$.   In each case, we'd like to know how observed event yields compare to those predicted in a generic BSM scenario.  

The simplest way to proceed with this task is to rescale the well-known rates of the SM by appropriate factors.  We express the rescaling factor for each channel accounting for cuts and their associated efficiencies, $\zeta$, for a Higgs decaying to state $f$ as
\beq
R_f =
\frac{ \sum_p \sigma \left(p p \to h + X^{(p)}\right) \times \zeta_f^{(p)} \times {\rm BR}(h \to f)  }{ \sum_p  \sigma \left(p p \to h+X^{(p)}\right)_{\rm SM} \times \zeta_f^{(p)} \times {\rm BR}(h \to f)_{\rm SM}} .
\eeq
The sum is performed over all production modes for the final state involving $f$, and we see that knowledge of the $\zeta_f^{(p)}$ is critical.  Assuming for example a final state $f = W^+W^-$,  $x\%$ of which are produced via gluon fusion and $(1-x)\%$ via vector boson fusion, the rescaling (neglecting $c_{gg}$ of Eq.~(\ref{eq:general}) and assuming gluon fusion is dominated by the top quark) is
\beq
R &\simeq&  \\
&& \hspace{-1.3cm} \underbrace{ \big[x \times c_t^2+ (1-x) \times a^2 \big]}_{\rm production} 
 \times \underbrace{   \bigg[\frac{a^2}{a^2 \, {\rm BR}(h \to WW,ZZ) +c_b^2 \, {\rm BR}(h \to b \bar b) +c_\tau^2 \, {\rm BR}(h \to \tau \tau) }   \bigg]}_{\rm decay}. \nonumber 
\eeq
Thus cut efficiencies are what allow us to appropriately identify the weight of each coupling in a particular channel's production rescaling.

From the experimental data, a likelihood $L_f$ can now be constructed for each channel $f$ as a function of $R_f$ through which the model parameters enter.  The crucial task for which $L$ is used can be phrased in a Bayesian language as answering the question of how likely an underlying theory prediction of $n$ events is given an outcome of $n_{\rm obs}$ measured events.  
Denoting the predicted number of events in terms of a number $n_{\rm B}$ of background events,  and a number of signal events $n_{\rm S}$ given in units of the SM prediction, i.e. $n = n_{\rm BG} + R \, n_{\rm S}$, the likelihood---modeled as a Poisson distribution---is as follows:
\beq
\label{eq:Poisson}
L_f (R_f) &\propto & n^{n_{\rm obs}} \exp \left[- (n_{\rm B} + R_f \, n_{\rm S} - n_{\rm obs}) \right] \\
&\stackrel{{\rm large \ } n}{\longrightarrow} & \exp \left[ \frac{-(n_{\rm B} + R\, n_{\rm S} -n_{\rm obs})^2}{2 n_{\rm obs}^2} \right].
\eeq
The total likelihood is then the simple product, 
\beq
L_{\rm tot} = \prod_f L_f.
\eeq  
Combining channels in this way neglects correlations between channels, which is presently necessary since these correlations are not publicly available.  Until uncertainties are dominated by systematics, however, this does not amount to a significant practical limitation, and we find it preferable and more accurate to use individual channel data at the expense of ignoring correlations.  We demonstrate this explicitly in Appendix~\ref{sec:Reconstruction}. 

In the simplest presentation of likelihoods, one assumes that all SM rates are rescaled by a single factor (signal strength modifier) $\mu$.  From Eq.~(\ref{eq:general}) this  corresponds to modifying just the production, via
\beq
a, \, c \to \sqrt \mu.
\eeq
This allows constraints to be placed in the one-dimensional parameter space defined by $\mu$.  More generally we'd like to know how the functions $L_f$ vary in multi-dimensional spaces.   A simple way to do this is to use constraints on the signal strength from either exclusion plots or best fits, reconstructing the likelihood using one of the methods reviewed in appendix~\ref{sec:Reconstruction}.  In Tables~\ref{tab:ATLAS}-\ref{tab:Tevatron} of that appendix, we summarize the experimental results from the LHC and Tevatron Higgs searches, all of which are used in the total likelihood we use for fits presented below.

\subsection{Electroweak Precision Tests}
\label{sec:EWPT}
We gain further insight into the nature of the Higgs boson from indirect measurements, in particular from the precision data taken at LEP \cite{LEPprecision}.  We briefly review here the parametrization of these measurements and the ways in which they can constrain the BSM scenarios examined below.

The measurements of interest  are captured by the oblique parameters $S$ and $T$  \cite{PT}.
Below the scale of EWSB, these parameters enter the effective theory and describe respectively a kinetic mixing between the hypercharge and $W_L^3$ gauge bosons, and a custodial-violating splitting of the physical $W$ and $Z$ masses.  These are marginal operators in the effective theory and so receive  logarithmic corrections from low mass states.  In particular the Goldstone bosons contribute to the running of these effective parameters, giving
\beq
\Delta S(m_Z) &=& \frac{1}{12 \pi} \log \left( \frac{\Lambda^2}{m_Z^2} \right), \\
\Delta T(m_Z) &=& -\frac{3}{16 \pi \cos^2 \theta_W} \log  \left( \frac{\Lambda^2}{m_Z^2} \right),
\eeq
where $\Lambda$ is the scale at which we match onto a more fundamental theory.  Within the SM, the Higgs boson exactly cancels these divergences, leaving only a finite logarithmic correction.  Allowing however for variable coupling of the Higgs to SM weak gauge bosons, this cancellation is spoiled and some residual dependence on the cutoff is retained.  Summing the one-loop contributions of Higgs and Goldstone bosons, we have
\beq
\label{eq:SIR}
S_{\rm IR}(m_Z) &=&\frac{1}{12 \pi} \left[ \log \left( \frac{m_h^2}{m_Z^2} \right)  +\left(1-a^2\right) \log \left( \frac{\Lambda^2}{m_h^2} \right) \right] , \\
\label{eq:TIR}
T_{\rm IR}(m_Z) &=&  \frac{-3}{16 \pi \cos^2 \theta_W}  \left[ \log \left( \frac{m_h^2}{m_Z^2} \right)  +\left(1-a^2\right) \log \left( \frac{\Lambda^2}{m_h^2} \right) \right] .
\eeq
The first term in each is fixed by the Higgs mass and so is included in the central values used below.  The second terms of Eqs.~(\ref{eq:SIR}, \ref{eq:TIR}) however provide a new constraint on the vector coupling, which we can illustrate in coupling fits.  Approximating the cutoff  with respect to the vector coupling as
\beq
\Lambda = \frac{4 \pi v}{\sqrt{1-a^2}},
\eeq
allows construction of an additional contribution to the global likelihood which is a function only of $a$ if we assume that threshold corrections from additional heavy states are small.\footnote{The assumption of negligible threshold corrections to the oblique parameters is supported in the case of composite models by the observation that current fits would indicate a very high scale for strong dynamics and thus suggest typically  large masses for resonances whose contributions  scale like $S,T \sim m_{\rm res.}^{-2}$;  cf. Fig.~\ref{fig:MCHM} below.   A current analysis of finite threshold corrections in this context can be found in \cite{RychkovEWPT}.}   Using the results of Gfitter \cite{gfitter} summarized in appendix \ref{sec:Data}, we find a likelihood that places significant constraints on $a$.  Taken on its own and assuming that no additional states contribute appreciably to the precision fit, we find a preferred value $a =1.01\pm 0.06$ at 95\% CL from this information alone under the conservative assumptiong that $U$ is simply marginalized in the likelihood.  In global fits below, we will   show the results with precision data included in the total combination separately in order to distinguish the effects of lepton and hadron colliders.

\section{Case Study I: Warped and Composite Models}
\label{sec:Composite} \setcounter{equation}{0}
We turn now to concrete cases that may realize naturalness, and that can be accommodated within the general framework of Eq.~(\ref{eq:general}).  In each case we will see that the demands of naturalness can be expressed with the functional forms for the Higgs couplings.  Tree-level couplings probe the EWSB sector directly, while loop-level couplings probe the presence of additional matter content outside the Higgs sector.  In the case of composite Higgs, which we discuss in this section, the latter are required in order to give SM fields their masses;  in minimal SUSY (discussed in the following section) they are desired for naturalness.

The first class of models that we will focus on is that of warped and composite setups.  
A single fundamental scalar that linearizes the low-energy theory would be a truly unique discovery, making the Higgs model of the SM a tremendous  possibility.  It is also however a singular limit of a much broader framework that can begin to complete the low energy's NLSM, and a composite origin for the Higgs takes advantage of this fuller theory space.  It thus serves as a concrete example of BSM physics at the weak scale, and provides a useful illustrative example of how a bottom-up approach can lead to powerful model-independent constraints on EWSB.

We follow the guiding principle (reviewed for instance in \cite{ContinoTASI}) that the Higgs is a Goldstone state of some new strong dynamics, providing an extension of earlier composite models  \cite{cohen georgi}.  The Goldstone nature imposes a shift symmetry on this state, allowing only derivative couplings for the Higgs (forbidding a mass term in the absence of symmetry-breaking spurions).  Such a  symmetry is assumed here since the Higgs is observed to be light relative to the implied confinement scale and the mass of other resonances.  In four dimensions this situation can be easily understood in analogy with QCD and its dual sigma model in the IR.  The generalization we need to make amounts simply to enlarging  the inventory of Goldstone bosons (pions) to include a Higgs-like state.

The picture of the effective dynamics here afforded by five-dimensional constructions, stemming from the initial work in \cite{RS}, has the EWSB sector residing on or near a 4D brane (`TeV brane') whose presence breaks the model's would-be conformality and gives rise to states with weak-scale masses.  The extra dimension is bounded by a second (`Planck') brane  where the elementary states are localized.  Here  we will identify a 4D setup of sufficient clarity and completeness to illustrate crucial symmetry properties of the IR physics in 4D {\it and} 5D, so will not expound on particular details of extra-dimensional models.

\begin{figure}[htb]
\begin{center}
\includegraphics[width=6.5cm]{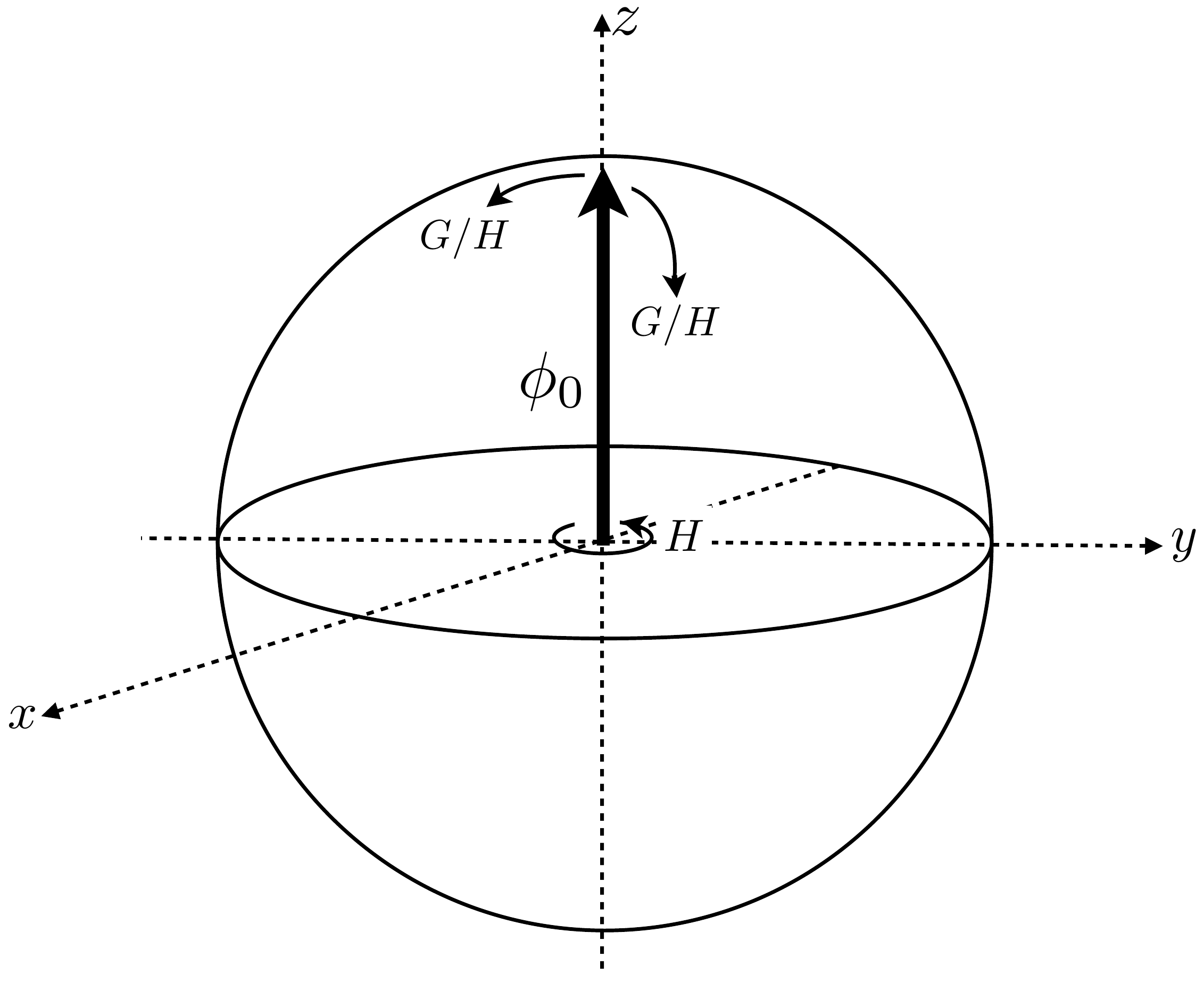} \hspace{1cm}
\includegraphics[width=6.5cm]{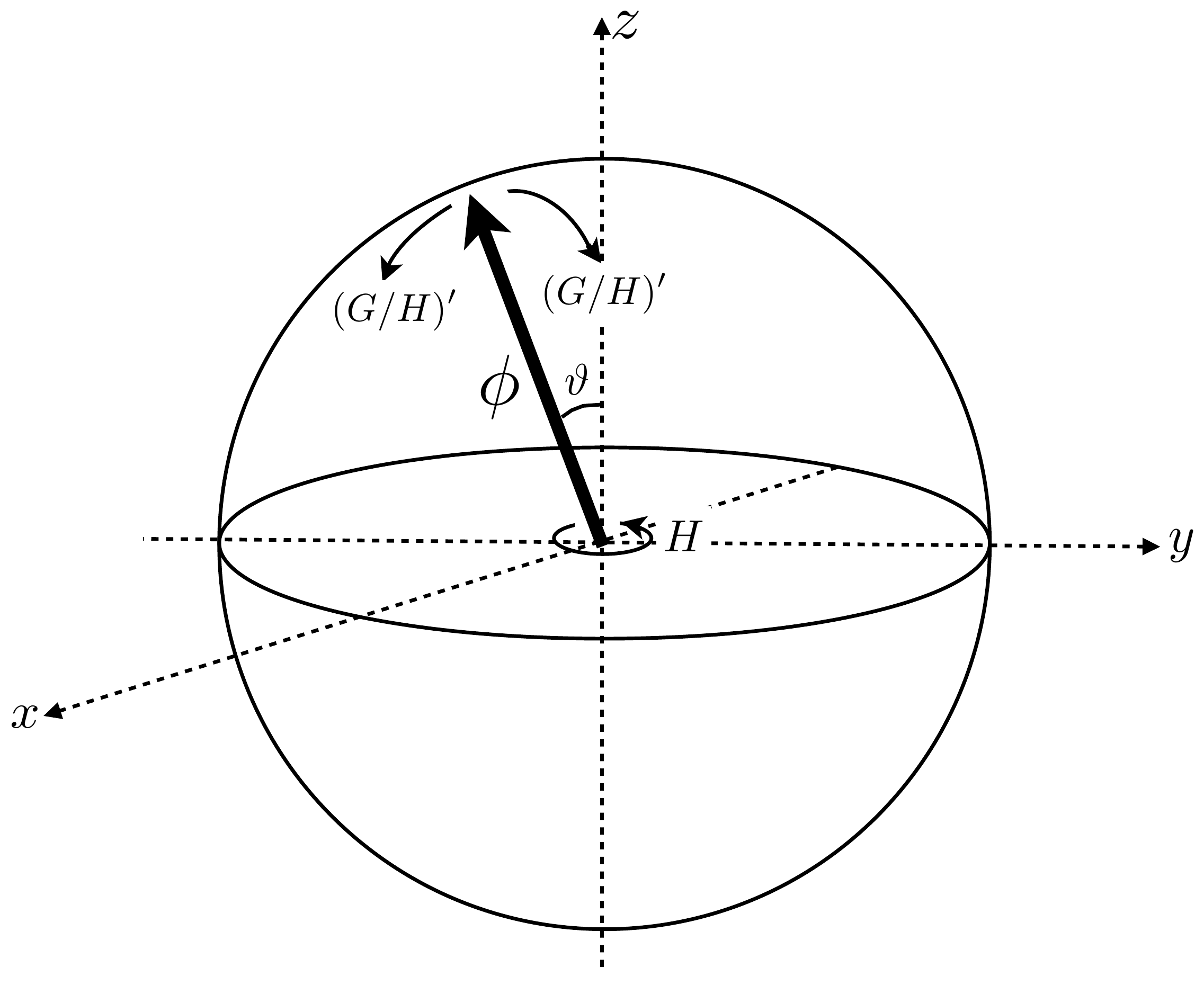}
\caption{\small Illustration of vacuum misalignment and spontaneous breaking of gauged directions of $H$ induced from a rotation of $\phi_0$, by angle $\vartheta$ along a VEVed  direction of the Goldstone space $G/H$.}
\label{fig:sphere}
\end{center}
\end{figure}

We will make use of the formalism of Callan, Coleman, Wess, and Zumino (CCWZ) \cite{CCWZ} throughout this section, and review the formalism in Appendix~\ref{sec:CCWZ}. In general we consider some new confining dynamics that respects a global symmetry $G$, with a vacuum state that breaks this symmetry to a subgroup $H$ at a scale $\Lambda \sim 4 \pi f$.  The Goldstones span the coset space $G/H$, and the low-energy states are classified in multiplets of $H$.
We can imagine then gauging the entire subgroup $H$, and inducing a misalignment between it and the coset $G/H$ by giving a VEV to one of the Goldstones; we illustrate this in Fig.~\ref{fig:sphere} for the simple case of breaking spherical  to rotational symmetry.  
Any gauge directions with nonzero projection onto the misaligned coset will become massive.  We can use this fact to correctly  produce EWSB by partially aligning the directions corresponding to the $W$ and $Z$ bosons with a coset space that has been rotated by a composite Goldstone Higgs with nonzero VEV.  The mass of the Higgs itself arises at loop level from terms of some generic coupling $\lambda_{\cancel{\rm G}}$ that break the shift symmetry,  schematically  giving
\beq
m_h^2 \sim \frac{\lambda_{\cancel{\rm G}}^2}{16 \pi^2} \times \Lambda^2.
\eeq
The lightness of the Higgs is thus understood as a loop suppression between it and the  scale of the new dynamics.

Specializing to the case of  $SO(N)/SO(N-1)$  for the sake of demonstration, we can begin with a `standard vacuum' as in Fig.~\ref{fig:sphere}
\beq
\langle \phi_0 \rangle = \begin{pmatrix} 0, 0, \dots, f \end{pmatrix}^T
\eeq
and identify the Goldstones $\pi(x)$ as fluctuations of the vacuum $\phi(x)$ about the angular directions, denoted by $\hat a$:\footnote{Normalization factors appearing in  deriving  these expressions depend on the choice of normalization for the group generators.  In our example we've taken ${\rm tr} \left(T^A T^B\right) = \delta^{AB}$.  For explicit forms of generators in the minimal coset $SO(5)/SO(4)$, we refer to the appendix of \cite{ContinoTASI}.}
\beq
\phi(x)  =  \exp \left(i\sqrt 2 \pi^{\hat a}(x) T^{\hat a}/f\right) \phi_0.
\eeq
After the Higgs acquires a VEV and breaks electroweak symmetry, we find  the rotated vacuum state:
\beq\label{eq:vacuum}
\langle \phi \rangle = \begin{pmatrix} 0, \dots, f \sin \vartheta, \dots, f \cos \vartheta \end{pmatrix}^T,
\eeq
where the newly nonzero component corresponds to direction $\hat a_h$ to which we associate the Higgs.  We see then explicitly how the Higgs VEV acts as a measure of the vacuum misalignment,\footnote{The misalignment is coming from some dynamics that we will leave unspecified; realistically we expect this to arise from some interplay of top loops, gauge loops, and other sources of  breaking of $G$.  A UV-complete example of this  can be found in \cite{MCTC}.}  i.e. Eq.~(\ref{eq:vacuum}) can be understood as a rotation of the standard vacuum:
\beq
\langle \phi_0\rangle \mapsto \langle \phi \rangle =  \Xi(x) \, \phi_0,
\eeq
where 
\beq
\Xi(x) = \exp \left[ \sqrt 2\, i\, T^{\hat a_h} \pi^{\hat a_h}(x) /f \right],
\eeq
and the Higgs direction is set to its VEV, i.e. $\langle \pi^{\hat a_h}(x) \rangle \to \vartheta f$.  This vacuum, Eq.~(\ref{eq:vacuum}), is the one we expand around in the discussions to follow.  Expanding in this way can be carried out by rotating all (and only) the broken generators  via
\beq
T^{\hat a} \mapsto \Xi(x)\, T^{\hat a}\, \Xi^T(x).
\eeq
Following this for a specific model gives all the ingredients needed to describe the Higgs sector in isolation, and allows us to identify the composite Higgs $h(x) \equiv \pi^{\hat a_h} (x) + \vartheta f$ and its conventional VEV as
\beq
\label{eq:fsintheta}
v  = f \sin \vartheta.
\eeq

\subsection{Couplings of a Composite Higgs: Tree Level}
The tree-level couplings of a composite Higgs are somewhat model dependent.  The couplings of the fermions depend on the assumptions that are made regarding global symmetries of the composite sector, and the representations that are chosen for the SM fields.  However, the couplings at tree-level already serve to probe the level of naturalness in any of these theories: couplings are determined by the confinement scale of the new strong dynamics, which has a simple relationship to the level of tuning required to obtain a weak-scale VEV for the composite Higgs.

Couplings of a composite Higgs to weak gauge bosons are nearly model-independent (with the notable exception of cases involving extended Higgs representations, cf. \cite{agtr1}).   From  Eq.~(\ref{eq:fsintheta}), expressing the direct coupling of two vectors to the strong sector as 
\beq
\frac{\partial m_V^2}{\partial f } = \frac{1}{2} \, g^2 f \sin^2 \vartheta  = g_{VVh^{\rm (SM)}} \times \frac{f}{v} \times \sin^2 \vartheta  \nonumber
\eeq
 we can find the Higgs coupling directly from Eq.~(\ref{eq:sumrule}):
\beq
\label{eq:aComp}
 g_{VVh}^2 +
\left( \frac{\partial m_V^2}{\partial f} \right)^2
=g_{VVh^{\rm (SM)}}^2 
\quad \Longrightarrow \quad a = \frac{g_{VVh}}{g_{VVh^{\rm (SM)}}} = \sqrt{1 - \frac{v^2}{f^2}}.
\eeq
More explicitly we note that the  couplings can be easily computed in the CCWZ construction from the two-derivative term containing the masses of the broken gauge fields:
\beq
\label{eq:ddComp}
\Delta \lag_{\rm eff}  = \frac{f^2}{4}  {\rm tr} \left(d_\mu d^\mu\right) .
\eeq
As can be seen from the expansion of $d$ in terms of pion fields (cf. Appendix~\ref{sec:CCWZ}) the coefficient  of Eq.~(\ref{eq:ddComp}) is determined simply by fixing canonically normalized kinetic terms for the pions.  From the expansion, and defining
\beq
G_\mu^{\hat  a}(x)&=& A_\mu^A(x)\, \pi^{\hat b}(x)\, {\rm tr} \left(T^A \,T^{\hat b}\, T^{\hat a}\right), \\
\hookrightarrow \quad d_\mu^{\hat a} d^{\hat a \, \mu}   &=& \left[ A_\mu^{\hat a}(x) + i \frac{\sqrt 2}{f} G_\mu^{\hat a}(x) \right]^2,
\eeq
one straightforwardly recovers Eq.~(\ref{eq:aComp}).
This form highlights the fact that the composite Higgs setup interpolates between a technicolor limit where $v=f$ and the Higgs state decouples entirely, and the SM where $f \to \infty$ and the hierarchy problem fully returns.

The form Eq.~(\ref{eq:aComp}) also gives a first estimate of the scale of new physics we might anticipate given a measurement of the coupling itself.  Following \cite{SILH}, we expect vector resonances $\rho$ that couple generically with strength $g_\rho$ to other composite fields to enter with masses
\beq 
m_\rho^2 =  g_\rho^2 f^2 \sim \frac{16 \pi^2}{N} f^2,
\eeq
where $N$ is the number of colors associated with the confining  gauge group.  Thus
\beq
m_\rho \sim \frac{4 \pi v}{\sqrt{(1-a^2) \times N}}.
\eeq
Large (small) deviations from $a=1$ correspond to the presence of typically light (heavy) vectors, and we see explicitly  the connection between the decoupling of additional states from the EWSB sector and the return to SM-like values of Higgs couplings.

\paragraph{Minimal Composite Higgs 4 and 5 (MCHM4, MCHM5):} In order to see how fermion couplings are expressed as functions of the scale of the confining dynamics, we discuss two models based on a minimal coset that endows the theory with a custodial symmetry.  These are the `Minimal Composite Higgs' models \cite{MCHM}, with $G/H = SO(5)/SO(4)$ giving four Goldstone bosons, one of which is identified as a composite (pseudo-Goldstone) Higgs.  We will not cover specific model details here, but rather quote the important results and refer to the original literature for more complete discussions.  

\begin{figure}[htb]
\begin{center}
\includegraphics[width=8cm]{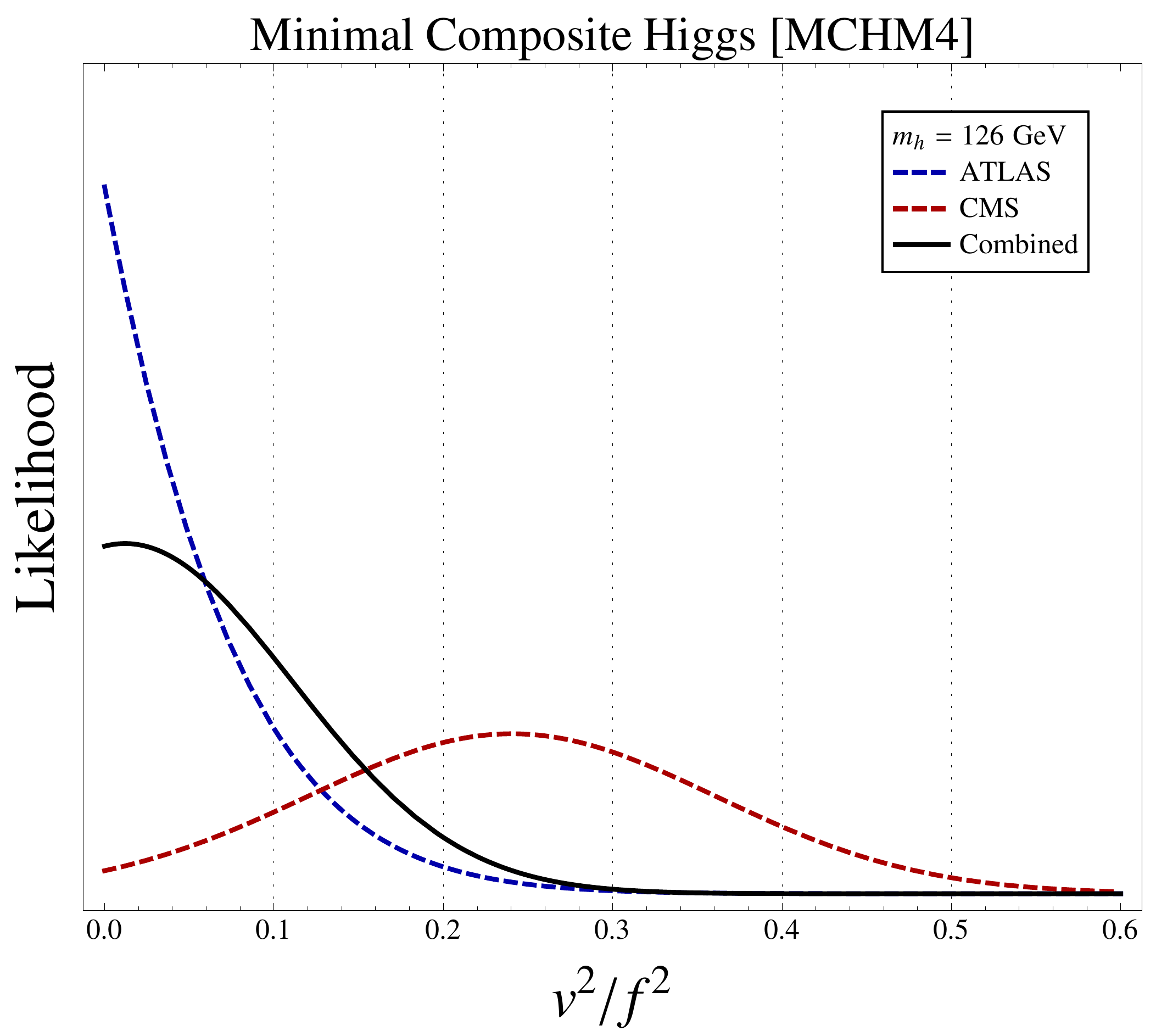} \hspace{0cm}
\includegraphics[width=8cm]{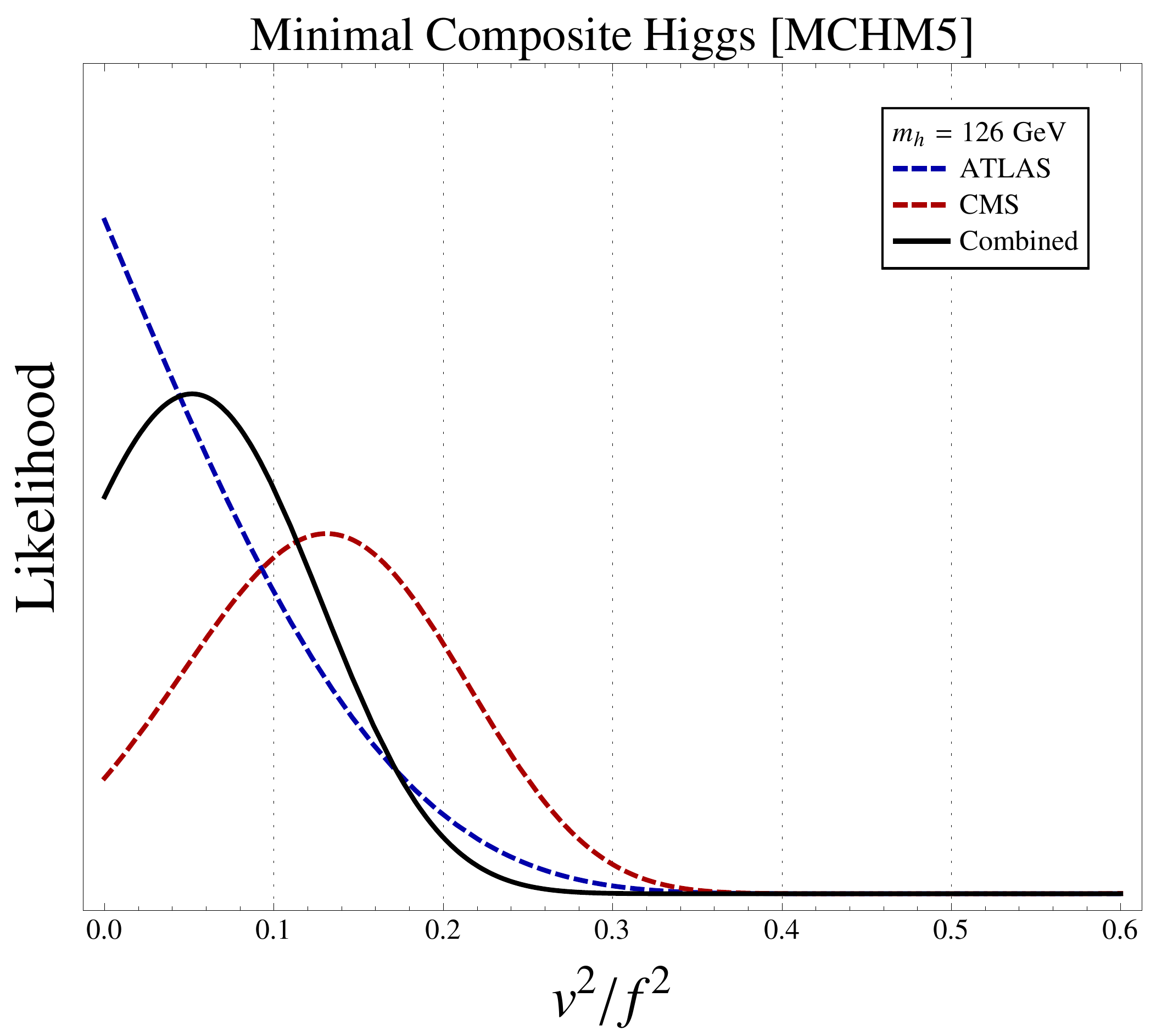}
\caption{\small LHC likelihoods for Minimal Composite Higgs models and total combination including Tevatron data.  Any preference for  deviations from the SM would come with the implication of a very high scale for the strong dynamics.}
\label{fig:MCHM}
\end{center}
\end{figure}

In MCHM4, SM fermions are embedded in the spinor {\bf 4} representation of the global $SO(5)$; in MCHM5 they are embedded in the fundamental {\bf 5}.  These different setups give different relations for the fermion couplings relative to the confinement scale:
\beq\label{eq:c4}
c^{(4)} &=& \sqrt{1-v^2/f^2}, \hspace{1cm} \mbox{(flavor universal)}\\
\label{eq:c5}
c^{(5)} &=& \frac{1-2\, v^2/f^2}{\sqrt{1-v^2/f^2}}.  \hspace{1cm} \mbox{(flavor universal)}
\eeq
In these models, then, {\it all} tree-level Higgs couplings are suppressed by the strong dynamics.   We can thus construct likelihoods as a function of $v/f$, which we show in Fig.~\ref{fig:MCHM}.

\begin{figure}[htb]
\begin{center}
\includegraphics[width=10cm]{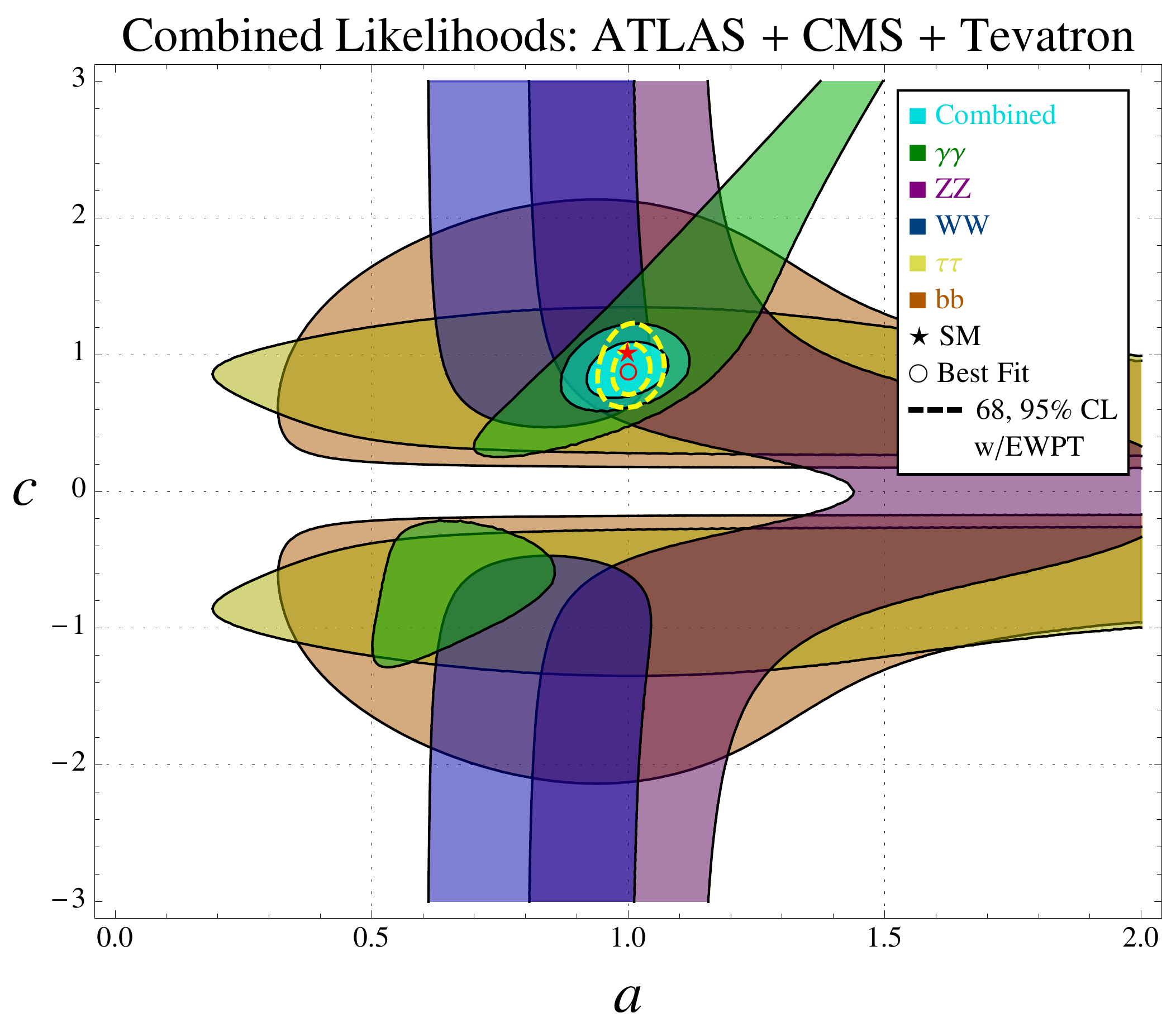}
\caption{\small Breakdown and combination of all final states participating in the global fit.  Individual contours are shown at 95\% CL and the combination is shown at 68\% and 95\%.  Shown separately as a yellow dashed contour we illustrate the combination including the effects of precision electroweak measurements after marginalizing over $U$ as discussed in appendix~\ref{sec:Data}.}
\label{fig:ac}
\end{center}
\end{figure}

\paragraph{The `Generic Composite Higgs'}  
When all SM fermions are embedded in equivalent representations we will find a flavor-universal rescaling of their couplings, but the precise form of this rescaling depends on the symmetry group itself.  We thus consider a `generic composite Higgs', in which the fermion rescaling is assumed universal but is otherwise unspecified (see \cite{GenCH} for more details along these lines).  Such a picture suggests examination of the general space spanned by the vector coupling, $a$, and the fermion coupling, $c$.  Likelihoods from LHC, Tevatron, and LEP data combined are shown in such a space along with a breakdown of channels in Fig.~\ref{fig:ac}.

\subsection{Couplings of a Composite Higgs: Loop Level}
\paragraph{Partial Compositeness}
Before turning to results for loop-level modifications of Higgs couplings, we must discuss  a mechanism by which flavor can be generated in the composite Higgs scenario.  At issue is the essential point that SM fermions must be linked somehow to the strong dynamics that breaks electroweak symmetry.   We consider three options: {\it i}) the fermions couple bilinearly to the strong sector through heavy mediators, as in extended and bosonic technicolor and their variants \cite{ETC,BTC}; {\it ii}) the fermions couple  to the strong sector linearly by mixing with fermionic composites as in \cite{twosite}; or {\it iii}) the fermions themselves reside in the strong sector \cite{RS,compfermions}.  In the first case the fermions are elementary into the far UV, while in the latter two there will be some degree of compositeness for these fields revealed at low scales.  Fully composite SM fermions have been disfavored by LEP \cite{LEPfermions}, while models with fully elementary fermions face  tension in obtaining the top mass without some additional strong dynamics entering at relatively low scales \cite{topcolor}.

We choose a simplified setup that describes both the 4D and 5D pictures and their associated low-energy phenomenology, and corresponds to the second option above for incorporating flavor.  The idea is that of partial compositeness, where the elementary (Planck brane) states and composite (TeV brane) states are considered to exist in two separate sectors coupled by a mass mixing \cite{twosite}.  Since the Higgs is considered to be fully composite, couplings between it and the SM fields require the latter to live partially in the composite sector as well.  Thus each elementary state is mixed with a composite  partner state of the same quantum numbers,  allowing the light mass eigenstates corresponding to the SM fields to become massive.

For some simple intuition of  partial compositeness and its necessary ingredients, we look to a very basic toy model.  We imagine an elementary sector to consist of only elementary top quarks $q_3$ and $t^c$, and suppose that there is a vector-like fermion $\mathcal Q$ in some representation of the strong sector's global symmetry $G$ with which the top can mix.  Provided  $\mathcal Q$ contains an electroweak doublet of hypercharge $1/6$ (denoted as a projection $Q \equiv P_D \mathcal Q$) and a singlet of hypercharge $2/3$ (denoted $T\equiv P_S \mathcal Q$), we can construct the following mass terms:
\beq
\label{eq:twositemixing}
\Delta \lag = M(\bar Q Q + \bar T T) + (\lambda_L q_L^\dagger Q_R + \lambda_R t_R^\dagger T_L +{\rm h.c.}).
\eeq
The mass eigenstates are thus defined with respect to a mixing angle, $\theta$, as
\beq
\begin{pmatrix} 
\tilde Q_L  \cr
q_L^{\rm SM}
\end{pmatrix} = 
\begin{pmatrix}  \cos \theta  & \sin \theta \cr
-\sin \theta & \cos \theta
\end{pmatrix}
\begin{pmatrix}
Q_L \cr
q_L
\end{pmatrix} ; \quad \tan \theta = \frac{\lambda_L}{M}.
\eeq
This gives a massless eigenstate $q_L^{\rm SM}$, and a state $\tilde Q_L$ with mass
\beq
M_{\tilde Q} = \frac{M}{\cos \theta} =  \frac{\lambda_L}{\sin \theta } .
\eeq
The SM fields are massless before EWSB and have the form
\beq
q_L^{\rm SM} = q_L \cos \theta -Q_L \sin \theta \, .
\eeq
As these fields become increasingly composite (i.e. as $\sin \theta \to 1$ by taking $\lambda_L$  large or $M$  small), they interact more strongly with the composite Higgs.  When the latter acquires a VEV, then, the spectrum of the SM fermions is dictated by their degree of compositeness.   Moreover, SM fields with a large degree of compositeness (e.g. the top) are expected to be accompanied by fermionic partners from the strong sector itself that may remain relatively light.  These new states have interesting phenomenological implications themselves, though here we consider their relevance only through implications they will have for the Higgs and its (loop-level) couplings.

\paragraph{Loop-Induced Couplings from Composites}
We might expect large corrections to loop-induced couplings of the light Higgs in the presence of partners with which SM matter and gauge fields mix in order to obtain mass.  In general both vector and fermion resonances can play a role in these effects, though for brevity here we consider only the effects of the composite fermions and refer to \cite{CompVectors} for discussions of composite vectors.   We first consider how the Higgs couplings to unbroken gauge directions can be modified, and then discuss how couplings involving at least one {\it broken} direction are affected. 

A useful set of tools for analyzing corrections to loop-induced Higgs couplings is that of the Higgs low-energy theorems \cite{Higgslowenergy}, the idea being to treat the Higgs as a non-propagating background field.  In this sense, the only appearance of the Higgs in the low-energy theory is through mass parameters of other fields to which it couples.  Explicitly separating out the EWSB part of a given  particle's mass, $m_{\cancel {\rm EW}} /m= \partial \log m /\partial \log v$, we perform the following redefinition of masses in the Lagrangian capturing both mass and Higgs interaction terms:
\beq
\label{eq:LETmass}
m  \to m \times   \left( 1 + \frac{h}{v} \frac{\partial \log m}{\partial \log v} \right).
\eeq
The Higgs couplings can thus be derived from lower order correlation functions by expanding (to a prescribed  order in $h$) the operators related to these redefined mass parameters.\footnote{See  \cite{GrojeanLET} for important corrections to couplings involving more than one power of the Higgs.}

An example of the utility of this scheme is the derivation of the fact that the coupling $h \to \gamma \gamma$ is proportional to the QED beta function in the limit $p_h \to 0$ \cite{Higgslowenergy}. 
At loop level the photon two-point function is modified, due to corrections of the kinetic term coming from heavy particles of mass $m_i$:
\beq
\Delta \lag_{\rm eff} = -\frac{1}{4} F_{\mu \nu} F^{\mu \nu}  \times \sum_i b_i\, \frac{e^2}{16 \pi^2} \log \left( \frac{\Lambda^2}{m_i^2} \right),
\eeq
where $b_i$ is  the appropriate factor given the spin of particle $i$.
Treating  the masses as in Eq.~(\ref{eq:LETmass}) and expanding  to linear order in $h$ gives the interaction term
\beq
 \Delta \lag \stackrel{p_h \to 0}{=}  F_{\mu \nu} F^{\mu \nu}  \times \frac{e^2}{64 \pi^2} \frac{h}{v} \frac{\partial}{\partial \log v} \sum_i b_i\,  \log m_i^2.
\eeq
For a multiplet of some particular spin with mass matrix $M$ we can imagine working in the mass basis so that the sum becomes a simple trace, gaining ultimately a basis-independent expression: 
\beq\label{eq:LogDet}
c_{\gamma \gamma} \propto  \frac{\partial }{\partial \log v}  {\rm tr} \left(\log M^\dagger M  \right)
= \frac{\partial}{\partial \log v} \log \det M^\dagger M.
\eeq

\begin{figure}[htb]
\begin{center}
\includegraphics[height=2.5cm]{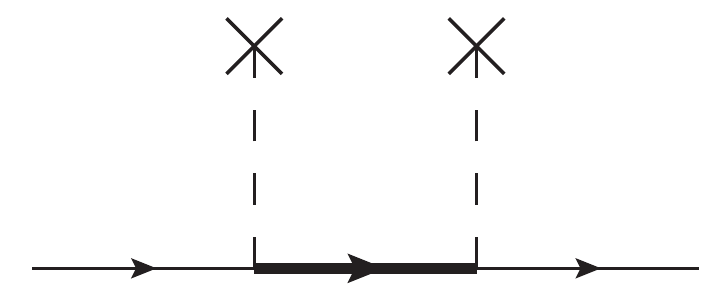}   \hspace{2cm}
\includegraphics[height=2.5cm]{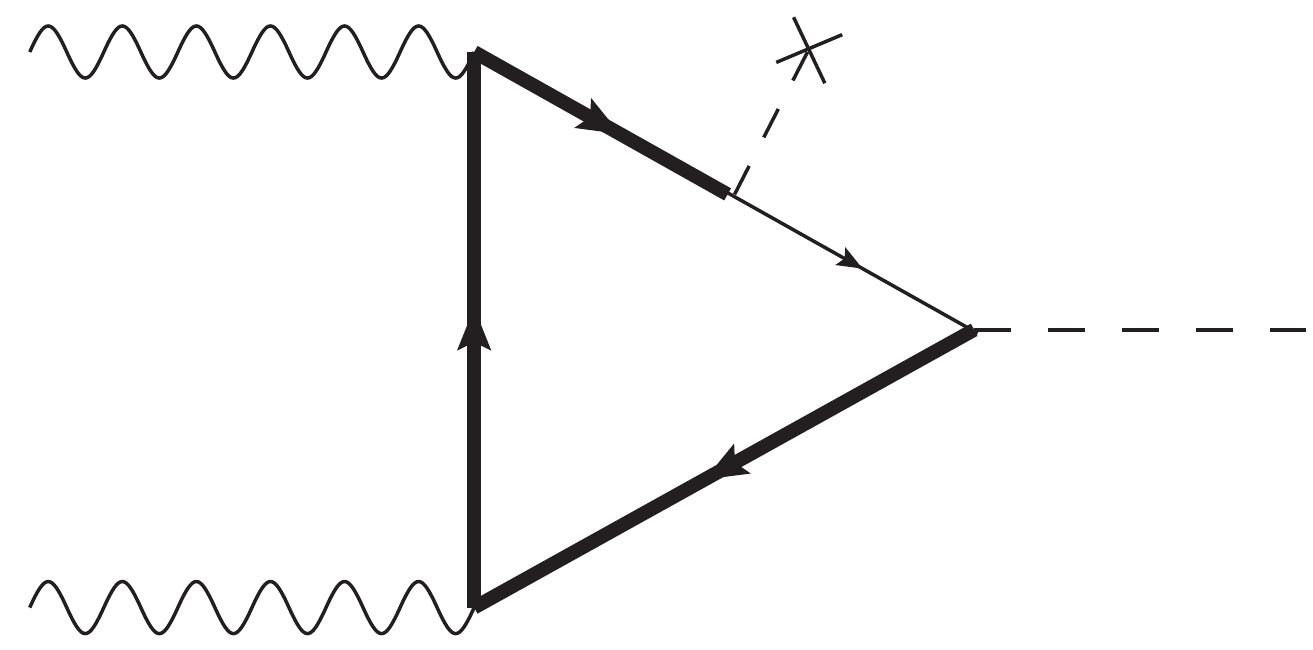}
\caption{\small Two conspiring effects in Higgs couplings to unbroken gauge directions coming from Higgs mixing of elementary and composite fermions.  {\bf Left:} Wavefunction renormalization from integrating out heavy fields (bold lines), leading to modification of the light field's Yukawa coupling.  {\bf Right:} Direct loop contribution from the heavy fields.}
\label{fig:AG}
\end{center}
\end{figure}

From Eq.~(\ref{eq:LogDet}), we learn important facts about the loop-induced couplings of simple partial compositeness setups.   
Since each SM fermion's mass comes from mixing with a heavier fermion of the same quantum numbers, a naive estimate would be to realize $\mathcal O(1)$ corrections $c_{gg,\gamma \gamma}$.  Due to the assumption of a Higgs invariant under a shift symmetry, however, this naive estimate is invalid: the global symmetry $G$ within the strong sector itself is broken only spontaneously, so the composite matter content cannot generate the Goldstone-violating object $H^\dagger H$ to which one would couple $F_{\mu \nu} F^{\mu \nu}$ or $G_{\mu \nu} G^{\mu \nu}$ (see \cite{LowVichi,AGHGG} for discussions).  The Goldstone symmetry is however violated by mixing terms between the composite and elementary sectors, so mixing terms can be treated as spurions and loop effects analyzed accordingly.  Using the CCWZ language, the mass mixing terms of Eq.~(\ref{eq:twositemixing}) can be incorporated as follows:
\beq
\label{eq:twositeCCWZ}
\Delta \lag = \mathcal Q^\dagger (i \slashed \nabla - M)  \mathcal Q  + (\lambda_L q_L^\dagger P_D \mathcal Q_R + \lambda_R t_R^\dagger P_S \mathcal Q +{\rm h.c.}),
\eeq
and the picture clarified by making a rotation $\mathcal Q \mapsto \xi^\dagger(x) \mathcal Q$.\footnote{The utility of this basis change is in simplifications observed in $\mathcal Q$'s mass matrix that make symmetry properties more transparent.  
Full details can be found in \cite{AGHGG}.}  With this done, one generates interactions between a Higgs and composite and elementary fermions like those participating in Fig.~\ref{fig:AG}.

In the new basis of Eq.~\ref{eq:twositeCCWZ}, we can go a bit further in understanding deviations in $c_{\gamma \gamma}$ and $c_{gg}$ and their connection to the Goldstone symmetry within the framework of partial compositeness.  In particular, it is worth noting an  important conspiracy between the two effects illustrated in Fig.~\ref{fig:AG}.
The first diagram modifies the light quark's kinetic term and thus the Yukawa interaction:
\beq
\label{eq:twositeYukawa}
 i q^\dagger \slashed \partial \psi  
 \to i q^\dagger \slashed \partial \psi \times \left(1+\frac{v^2}{f^2}\frac{\lambda^2}{M^2} \right) \quad
 \Longrightarrow \quad
yh\bar q q 
\to yh\bar q q  \times  \left(1-\frac{1}{2}\frac{v^2}{f^2}\frac{\lambda^2}{M^2}\right).
\eeq

The second diagram of Fig.~\ref{fig:AG} also gets a mass ($M$)-dependent contribution, though in simple partial compositeness setups it exactly  cancels the effect of Eq.~(\ref{eq:twositeYukawa}) for the case of the top quark.  This is easily seen with use of Eq.~(\ref{eq:LogDet}), which accounts for both effects.  The overall correction  to $c_{gg}$ and $c_{\gamma \gamma}$ from the top and its partners thus comes entirely from rescaling of the top Yukawa coupling due to nonlinearities of the Higgs as in Eqs.~(\ref{eq:c4}, \ref{eq:c5}), and is independent of the spectrum of heavy fermions.  In these cases, loop effects are accounted for by fitting couplings in the space $(a,c)$ of Fig.~\ref{fig:ac}.  The bottom quark, on the other hand, plays a less significant role in loop couplings, but  remains in a multiplet with a top quark and so couples strongly to the composite sector.  As such, its partners can in principle mediate a large contribution especially to $c_{gg}$ \cite{AGHGG}.

The preceding result can be stated as an effect of the Goldstone symmetry:  the heavy matter fields respect $G$ and thus Goldstone-violating  interactions like $c_{gg}$ and $c_{\gamma \gamma}$ are largely insensitive to their properties.  A different outcome can arise for Higgs couplings to gauge bosons when one of the directions is partially aligned with the broken subgroup as in $h \to Z \gamma$ \cite{ZGamma}.  
In particular we observe  two terms constructed from the $SU(2)_{L,R}$ subgroups' field strengths:\footnote{The corresponding operators in the basis of the Strongly Interacting Light Higgs \cite{SILH}, where a doublet notation for the Higgs is manifest, are
$
O_{HW} = i \left(D^\mu H\right)^\dagger \sigma^i (D^\nu H) W^i_{\mu \nu} 
$
and 
$
O_{HB} = i \left(D^\mu H\right)^\dagger  (D^\nu H) B_{\mu \nu}.
$}
\beq
\Delta \lag =  c_L \, {\rm tr} \left(d_\mu d_\nu E^{\mu \nu}_{L}\right) 
+  c_R \, {\rm tr} \left(d_\mu d_\nu E^{\mu \nu}_{R}\right),
\eeq
which contain the desired general form
\beq
\label{eq:Genlag}
\Delta \lag = \frac{1}{\sqrt 2 f}   A_{[\, \mu}^{\hat a_3}(x) \, \partial_{\nu\, ]} h(x)  \times \left\{  c_L \, {\rm tr} \left[ [T^{\hat a_3},T^{\hat a_h}]\, T_L^3\right] W_{L}^{3\, \mu \nu}
+  c_R \, {\rm tr} \left[ [T^{\hat a_3},T^{\hat a_h}]\, T_R^3\right] B^{\mu \nu} \right\}
\eeq
where we use $\hat a_3$ to denote the broken direction along which the physical $Z$ has nonzero overlap, and $\hat a_h$ as above to denote the direction associated with the composite Higgs.  In the case of the MCHM coset discussed above, we can expand  the traces with explicit forms for the generators $T_{L, R}$ to find that this contains the coupling term
\beq
\Delta \lag 
=
\label{eq:hZgam}
i \frac{a\, v}{2\sqrt 2 f^2} 
 \left(c_L - c_R \right) 
 \times Z_{[\, \mu}(x) \, \partial_{\nu\, ]} h(x)  F_\gamma^{\mu \nu} ,
\eeq
where $a$ is the vector coupling, Eq.~(\ref{eq:aComp}), in these models.  

What we find then is the possibility to realize a correction to $h\to Z\gamma$ which doesn't  violate the Goldstone symmetry (the Higgs enters with a derivative), but does  require some breaking of the left-right ($P_{LR}$) symmetry of the composite sector such that $c_L$  in Eq.~(\ref{eq:hZgam}) can differ from $c_R$.  Due to the preservation of the Goldstone symmetry, however, effects obtained from integrating out heavy fermions can be substantial, and $\mathcal O(1)$ effects become possible \cite{ZGamma}.  Explicit calculations must be carried out without the simplifying use of Eq.~(\ref{eq:LogDet}), since mass matrices can mix states with differing eigenvalues of the broken directions' generators $T_{L,R}^3$.   In the minimal coset, we identify the $\bf 10 =\bf (2,2) + \bf (1,3) + \bf (3,1)$ as the simplest composite representation contributing to $c_{Z\gamma}$ in the absence of spurions from mixing with elementary states; this cannot be obtained from a $\bf 5 = \bf 4 + \bf 1$ which does not by itself break $P_{LR}$.  Distinct masses can be assigned to the $\bf (1,3)$ and $\bf (3,1)$ of the $\bf 10$, however, giving  
\beq
c_{Z \gamma} \propto N_{\rm gen} \times \frac{v^2}{f^2} \times \frac{m_{(1,3)} - m_{(3,1)}}{m_{(1,3)} + m_{(3,1)}},
\eeq
where $N_{\rm gen}$ is the number of composite generations.  Considering the fact that the same $P_{LR}$ was used to suppress large corrections to the $Z\bar b b$ coupling \cite{contino-pomarol}, simple models with large $P_{LR}$  breaking and large modifications of $h \to Z \gamma$ face some tension.  If $Z\bar b b$ is protected by an accidental symmetry, however, large corrections can persist and probe the flavor structure of the composite sector in a unique way.  These signals are beginning to become relevant at the LHC \cite{ZgammaExp}, but will nonetheless have to await further data for experimental verification due to their inherently meager rates \cite{ZGammaSig}.

 Overall what we see in the case of the composite Higgs is that tree-level couplings serve as direct probes of theory naturalness and the existence of additional low-scale EWSB structure, while the fact that the Higgs is light forces upon us a Goldstone symmetry that tends to suppress loop-induced couplings mediated by additional resonances of the strong sector.   If we expect the Higgs properties to convey hints regarding whether or not naturalness is a valid  principle for weak-scale physics, in the composite Higgs setup we would most urgently hope for deviations in the dominant tree-level production and decay channels. These obvious  hopes are facing some disappointing realities as we've seen above, though some subtle  effects with elegant explanations from compositeness may still be observed.
 
  We turn next to a second class of models---two-Higgs doublet models with SUSY---which turn out in effect to obey a reversal of the situation we've seen with composite Higgs.  In minimal SUSY, it is commonly the case that tree-level couplings  are relatively insensitive to the presence of additional states in the EWSB sector in minimal setups.  Loop-induced couplings, on the other hand, seem to have more insistent  things to say regarding the anticipation of new low-scale physics in  natural cases.

\section{Case Study II: Multi-Higgs Models and Supersymmetry}
\label{sec:SUSY} \setcounter{equation}{0}

We deal here with the two Higgs doublet construction that is central in supersymmetry and the minimal supersymmetric SM (MSSM).   We focus only on the Higgs sector of these theories, where for the most part it suffices to consider a simple type-II two-Higgs doublet model (2HDM).  We will make use of the fact that scalar top partners are involved in the MSSM in the interest of naturalness  and that quartic couplings are fixed by the symmetry, but this is the only place SUSY will enter our initial discussion.  The parameter space fits, for instance, are valid for both supersymmetric and non-supersymmetric type-II 2HDM.

\subsection{SUSY and Type-II 2HDM}
The presence of a second doublet in supersymmetry is mandated by the theory itself, as the superpotential  is holomorphic and the theory anomaly-free only when both fields are present.  One thus introduces a hypercharge $1/2$ doublet $H_u$ that couples to up-type fermions, and a hypercharge $-1/2$ doublet $H_d$ that couples to down-type; this is the defining setup of a type-II 2HDM.  Explicitly,
\beq
H_u = \begin{pmatrix} H_u^+ \cr H_u^0 \end{pmatrix},
\quad
H_d= \begin{pmatrix} H_d^0 \cr H_d^- \end{pmatrix} 
\eeq
where the real part of each's neutral component acquires a VEV, $v_u \equiv v \sin \beta $ and $v_d \equiv v \cos \beta$ respectively.  The light Higgs boson is thus minimally accompanied by an additional CP-even neutral field ($H$) as well as two charged states ($H^\pm$) and one CP-odd neutral state ($A^0$).   We focus on the CP-even neutral states here, whose mass states are conventionally expressed as
\beq
\begin{pmatrix} h \cr H \end{pmatrix}  =
\sqrt 2 \begin{pmatrix} -\sin \alpha & \cos \alpha \cr \cos \alpha & \sin \alpha \end{pmatrix} 
\begin{pmatrix} {\rm Re} H_d^0 \cr {\rm Re} H_u^0 \end{pmatrix} .
\eeq
The mixing angle $\alpha$ lies in the range $[-\pi/2,\pi/2]$, though typically in SUSY theories it has to be restricted to the negative range in order to obtain  a neutral CP-odd state $A^0$ with mass above that of the $Z$ boson.

The quartic couplings in a SUSY 2HDM  are fixed by the demands of holomorphy.  Superpotential terms involving $H_u H_d$ would need to couple to a gauge singlet---of which there is no suitable example in the MSSM---so quartics arise only from the K\"alher  potential through gauge-invariant kinetic terms after integrating out auxiliary  $D$ fields, e.g.
\beq
\label{eq:MSSMDterms}
\Delta \lag &=& \int d^4 \theta H_u^\dagger e^{V} H_u \supset \int d^4 \theta \, \frac{g}{2} H_u^\dagger \left(\theta^2 \bar \theta^2 T^a D^a\right) H_u, \\
 \hookrightarrow \quad \Delta \lag &=& \frac{g^2+g'^{\,2}}{8} \left|H_u\right|^4, \nonumber
\eeq
with analogous expressions for the other quartics.
Due to these requirements, the mass of the lightest Higgs boson is subject to an upper bound at tree level:
\beq
m_h^2 \leq m_Z^2 \cos^2 2 \beta  ,
\eeq
which is maximized at either very large or very small $\tan \beta$.  Thus the observation of a Higgs-like state with a mass of around 125 GeV necessitates  some additional source of mass for the light Higgs.   This mass can be accommodated within the MSSM itself, though at the cost of some theoretical discomfort: obtaining $m_h = 125 \, {\rm GeV}$ requires that a substantial fraction of the Higgs mass come from SUSY breaking through non-supersymmetric contributions to quartic couplings  \cite{MSSMHiggsMass}, and thus  reintroduces  fine-tuning into the theory.  The more natural option would be to induce some non-supersymmetric contribution to quartics from some other, non-minimal, source of breaking.  We will discuss this further below.

\subsection{Couplings of a SUSY Higgs: Tree Level}

\begin{figure}[htb]
\begin{center}
\includegraphics[width=8cm]{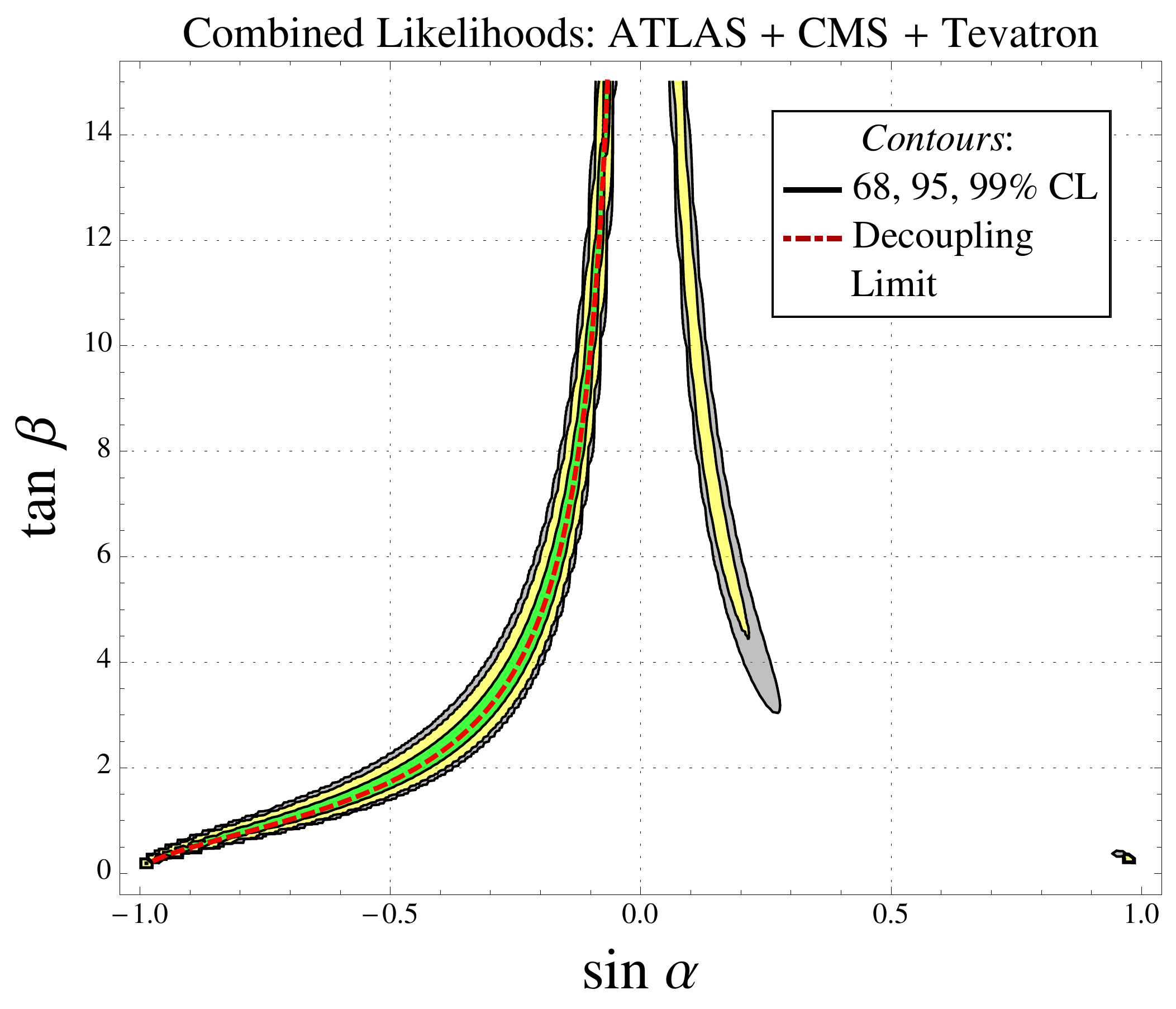} \hspace{0cm}
\includegraphics[width=8cm]{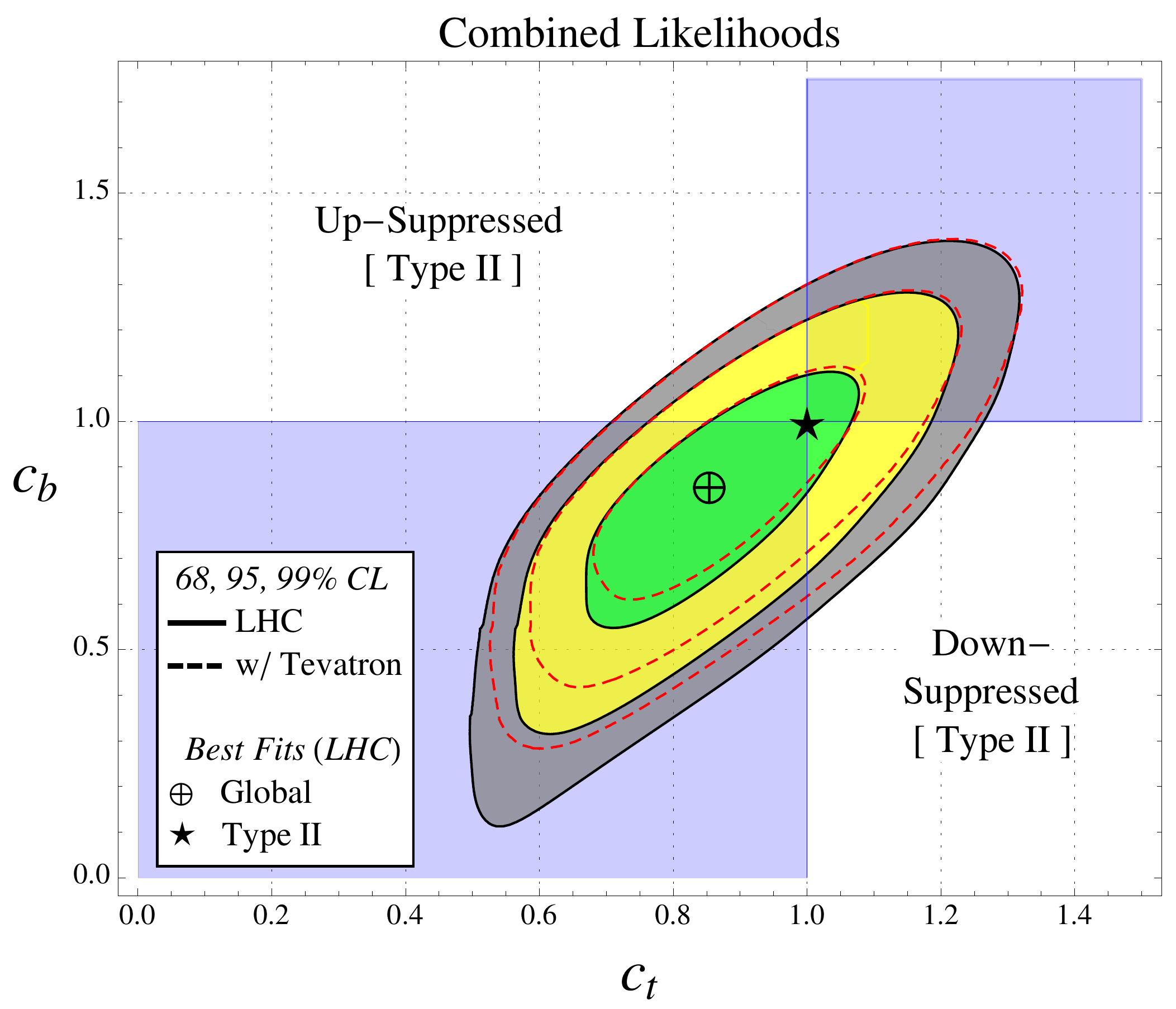}
\caption{\small Fits from LHC data to parameter space of type-II 2HDM.  {\bf Left:} The trigonometric space determining the Higgs properties.  The fit in this space for $\tan \beta >1$ is dominated almost entirely by fitting the bottom coupling alone, as it is the only state whose coupling varies significantly with deviations from the decoupling limit contour (cf. Eq.~(\ref{eq:cwrtmH})).  {\bf Right:} Fit in the space of bottom and top Yukawa couplings, marginalizing over the gauge coupling in the range $a = [0,1]$.}
\label{fig:SUSYfit}
\end{center}
\end{figure}

The tree-level coupling structure of the light Higgs in SUSY can be easily derived.  The top quark, for instance, couples to $H_u$ with a coefficient $y_t$ which in general differs from the SM Yukawa coupling $y_t^{\rm (SM)}$:
\beq
\Delta \lag = i \, y_t \,  H_u^T \epsilon \, Q\, t^c +{\rm h.c.}
\supset -\frac{y_t}{\sqrt 2} (v_u + H_u^0 + i A_u^0) t t^c + {\rm h.c.}
\eeq
Comparing to the mass term of the SM---$\Delta \lag = -y_t^{\rm (SM)} v\, t t^c/\sqrt 2$---we see that  $y_t = y_t^{\rm (SM)}/\sin \beta$ to obtain the correct top mass.  In the Higgs mass basis with $H_u^0 = h \cos \alpha + H \sin \alpha$, we see also that the coupling $htt^c$ is related to that of the SM in a simple trigonometric way.  Following equivalent reasoning for down-type fermions and gauge bosons, one obtains
\beq
a=\sin (\beta - \alpha), \quad
c_u = \frac{\cos \alpha}{\sin \beta}, \quad c_d = \frac{-\sin \alpha}{\cos \beta};
\eeq
the form of $c_u$ and $c_b$ holds at tree-level for all up- and down-type fermions.  A sensible space to analyze for the 2HDM of minimal SUSY is therefore that of $\sin \alpha$-$\tan \beta$:  we show this in the left panel of Fig.~\ref{fig:SUSYfit}.  From this illustration it seems that a decoupling limit---the limit where all additional scalars of the EWSB sector are integrated out well above the weak scale and $\alpha \to \beta - \pi/2$ returning the SM couplings---is preferred. 

The preference for decoupled heavy Higgs states is not  a case of unnaturalness, though perhaps we'd have wished for a low SUSY scale {\it generically} for new scalars, in which case one would look for deviations  from the decoupling contour in coupling fits.  A possible space of interest for this is shown in the right panel of Fig.~\ref{fig:SUSYfit}, where the plane of top and bottom Yukawa couplings is shown after marginalizing over the vector coupling.  Here we get a clearer sense of values for physical couplings that are still within acceptable ranges.

To address any remaining hope of additional low-scale Higgs states, we turn to a brief discussion of `slow decouplers', i.e. places where even as the additional EWSB states are made heavy we could still find sizable deviations from  tree-level  SM couplings; for a general treatment we refer to \cite{GHdecoupling}.  Here we treat $m_H^2$ as the decoupling parameter and fix $B\mu$ with the requirement of obtaining the correct total VEV.  Then defining $x =\pi/2-(\beta - \alpha)$ such that $x \to 0$  describes decoupling, we find for the tree-level MSSM
\beq
\left. m_H^2 \right|_{x \to 0}=  m_Z^2 \times \frac{\sin 4 \beta}{ 2 x}   .
\eeq
This gives simple expansions in $1/m_H$ for the  functions $\sin (\beta -\alpha)$ and  $\cos (\beta -\alpha)$:
\beq
\sin (\pi /2 - x) &\simeq& 1-\frac{1}{8} \frac{m_Z^2}{m_H^4} \sin^2 4 \beta; \\
\cos (\pi /2 - x) &\simeq& \frac{1}{2} \frac{m_Z^2}{m_H^2} \sin 4 \beta.
\eeq
Leading deviations in physical couplings thus have simple scaling behaviors:
\beq
\label{eq:cwrtmH}
\begin{split}
\delta a \ &\simeq \  \mathcal \sin^2 4 \beta \times \mathcal O(m_H^{-4}) + \dots, \\ 
\delta c_u \ &\simeq \  \cot \beta \sin 4 \beta \times \mathcal O(m_H^{-2}) +   \dots, \\
\delta c_d \ &\simeq  \ \tan \beta \sin 4 \beta \times \mathcal O(m_H^{-2}) + \dots,
\end{split}
\eeq
where we leave out higher order terms in $1/m_H$.  At large (small) values of $\tan \beta$, $c_d$ ($c_u$) will be the only coupling that varies as $m_H^{-2}$ rather than $m_H^{-4}$ (thus the `slow decoupler'), while the vector coupling remains relatively insensitive to $H$ throughout the parameter space.

We can see slow decoupling behavior  in another transparent way within the MSSM.  If we assume $\tan \beta>1$ with the light Higgs living primarily in $H_u$, then we can get a  direct sense of the physical decoupling scale in the 2HDM space.  We consider a simplified MSSM-inspired potential for the neutral fields of the form
\beq
\label{eq:RadMSSM}
\Delta V &=& m_{H_u}^2 \left| H_u^0 \right|^2 +  m_{H_d}^2 \left| H_d^0 \right|^2 - B\mu \left(H_u^0H_d^0 +{\rm c.c.}\right) \nonumber \\
&& \hspace{0.5cm} + \, \frac{1}{8} \left(g^2+g'^{\, 2} \right) \left[ (1+\delta \lambda_1) \left|H_u^0\right|^4 +   \left|H_d^0\right|^4 - 2 \left|H_u^0\right|^2 \left|H_d^0\right|^2\right];
\eeq
where $\delta \lambda_1$ is assumed to be the dominant source of  additional  Higgs mass.  Again  trading $B\mu$ for $v$, we can fix $\delta \lambda_1$ so as to obtain a light Higgs at 125 GeV:
\beq
\delta m_h^2 &=&  \delta \lambda_1 \times m_Z^2  \times  \frac{\cos \alpha \sin^3 \beta}{\sin \, (\beta -\alpha) }  \\
&\stackrel{\rm decoup.}{\longrightarrow} & \delta \lambda_1 \times m_Z^2 \sin^4 \beta. \nonumber 
\eeq 
This leaves free a two-dimensional parameter space, allowing us to examine contours of $m_H$ in simple ways.  Some aspects of these heavy states have been taken up recently in \cite{HeavyHiggs}.
\begin{figure}[htb]
\begin{center}
\includegraphics[width=10cm]{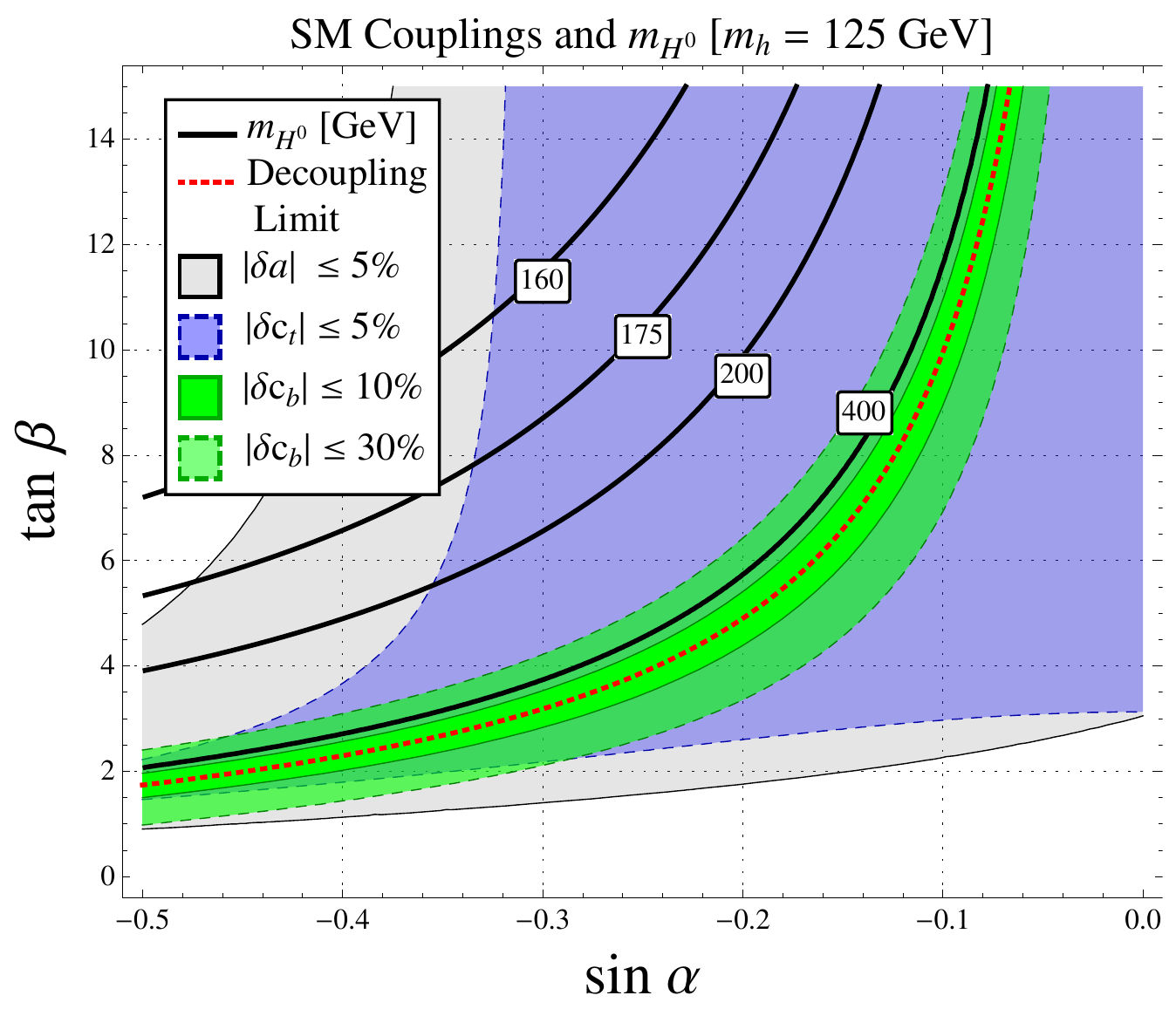}  
\caption{\small Deviations in couplings from their SM values with varying second Higgs mass, $m_H$, fixing the light Higgs to 125 GeV.  For $\tan \beta \gtrsim 2$, the bottom is the only state  state with slowly decoupling deviations that may indirectly probe the presence of additional Higgs scalars.  We assume here that corrections to $|H_u|^4$ dominate in the potential, such that a suppressed bottom coupling (the region below the decoupling line) is inaccessible in the simplified model \cite{ACCG}.}
\label{fig:decoupling}
\end{center}
\end{figure}

In Fig.~\ref{fig:decoupling}, we show how couplings of the light Higgs to SM fields change effectively as a function of the heavy Higgs mass $m_H$ in the space of $(\sin \alpha,\tan \beta)$.  The evident conclusion from this is that top and gauge couplings are almost entirely insensitive to the presence of the second Higgs: it is only in limits where $m_H \lesssim 200\, {\rm GeV}$ that we would expect to find deviations in these couplings exceeding $5\%$ of their SM values.  The bottom coupling (couplings of down-type fermions in general), on the other hand, can experience large deviations even in fairly decoupled situations  and thus serves as a sensitive probe of additional Higgs scalars in the absence of their direct detection. We refer to \cite{DawsonTASI,BlumPredicts} for further details and discussion.

The question of whether there are additional light scalar Higgs bosons in the spectrum remains open.  While acccurate measurement of the bottom coupling may provide the best indirect hints of their presence, this does not serve as indication of naturalness in the theory.  For a SUSY theory to remain natural in the simplest way, one would anticipate a relatively light top partner to soften the theory's UV sensitivity.  If there is indeed a light stop, however, one still has to understand how the Higgs acquires its $\gtrsim 35 \, {\rm GeV}$ of SUSY-breaking mass.  This leaves open two simple 
possibilities to consider:
\begin{itemize}
 \item Soft $A$-terms associated with the stop are chosen to maximize the SUSY breaking that enters low-energy quartic interactions;
 \item New non-minimal dynamics is added to the theory.  
 \end{itemize}
 We turn now to a  discussion of the important implications in each case.

\begin{figure}[htb]
\begin{center}
\includegraphics[width=5cm]{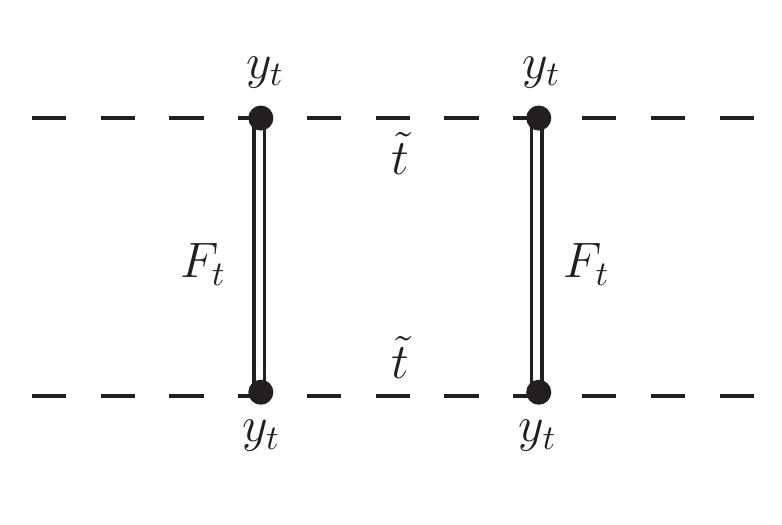}  \
\includegraphics[width=5cm]{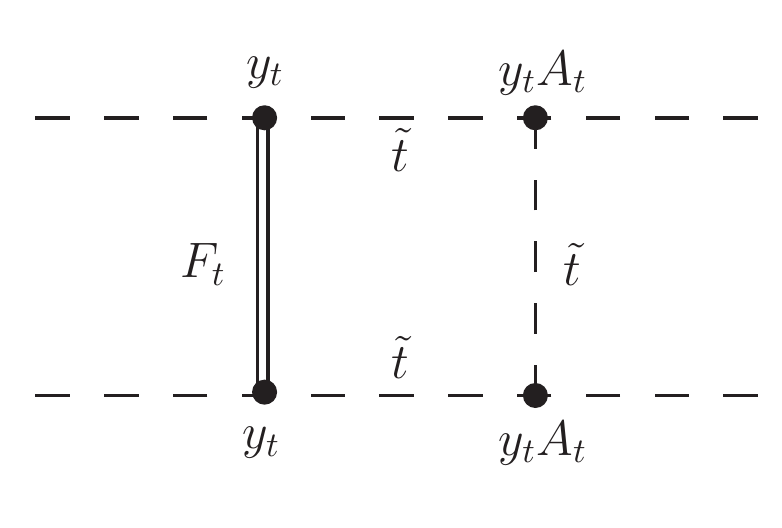}  \
\includegraphics[width=5cm]{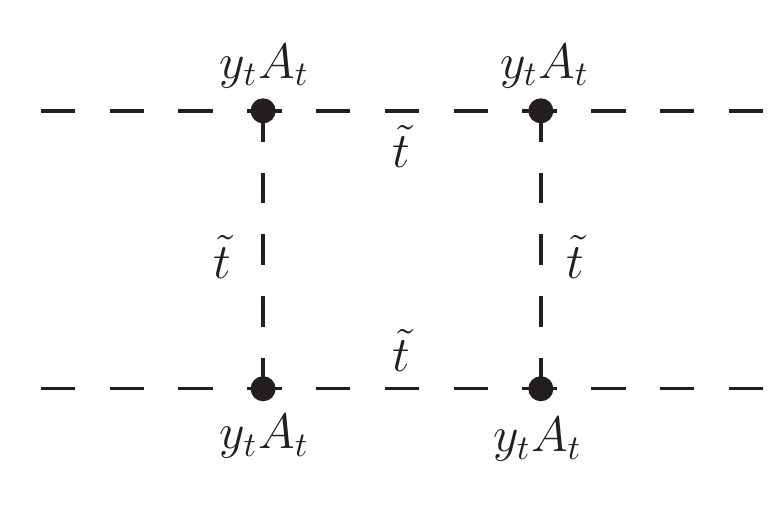}  
\caption{\small Representative diagrams contributing to SUSY-breaking quartics.  All external lines are understood in this case to denote $H_u^0$, and the auxiliary $F$ fields have been kept explicit for uniformity.  Coupling conventions are as in \cite{MSSMHiggsMass}.}
\label{fig:quartics}
\end{center}
\end{figure}

\paragraph{Light stops and Maximal Mixing:}   Hard SUSY breaking can be generated through RG running alone of SUSY-breaking mass splittings, or through threshold corrections directly proportional to soft terms.  In the MSSM these possibilities are generated from diagrams like those in Fig.~\ref{fig:quartics};  cf. \cite{MSSMHiggsMass,MSSMQuartics} for details.   In each case, the dimensionless SUSY-breaking parameter is identified as either $\bar A_x \equiv A_x/m_{\rm SUSY}$, $t \equiv \log (m_{\rm SUSY}^2)$, or  a combination thereof.

In the limit where the light Higgs lives primarily in $H_u^0$, the dominant radiative corrections to its mass enter through $\delta \lambda_1$ of Eq.~(\ref{eq:RadMSSM}).  With common soft terms and assuming large $\tan \beta$,  the correction generated by diagrams of Fig.~\ref{fig:quartics}  and their running is given by
\beq
\delta \lambda_1= \frac{3 y_t^4}{8 \pi^2}\bigg[\underbrace{ \left( \bar A_t^2 - \frac{\bar A_t^4}{12} \right)  \bigg(1}_{\rm threshold}+ \underbrace{\frac{3 y_t^2}{16 \pi^2}\times t \bigg) +t}_{\rm RG \; running}  \bigg],
\eeq
from which the optimal `maximal mixing' arrangement $\bar A = \sqrt 6$ giving maximal $\delta \lambda_1$ is identified.  The Higgs soft mass is quadratically sensitive to $A$, however, so in terms of naturalness this choice is tantamount to unmixed stops with masses of order TeV and percent level tuning.

Working at two-loop order one finds an upper bound $m_h \lesssim 135 \, {\rm GeV}$ in the minimal setup, for which one needs a large SUSY-breaking splitting of the top and stop mass and/or large $A$ terms.   Large mixing has interesting implications for couplings at one-loop level, which we'll discuss below.   We've seen though that a residual tuning (which SUSY had been hoped to render unnecessary) is required in these cases where hard breaking terms (non-SUSY quartics) are induced from soft terms $A$ and $m_0$.

\paragraph{Non-minimal SUSY sectors:}  A simple solution to the SUSY mass problem is to couple the Higgs to additional sources.   The remaining SUSY partners of SM fermions can then be relatively light and naturalness restored.  Some simple prospects are the following:
\begin{itemize}
\item Coupling $H$ to a gauge singlet $S$ with non-decoupling  $F$ terms
\beq
\label{eq:NMSSM}
\Delta W =  \lambda S H_u H_d.
\eeq
Including  various additional interactions for the singlet, one generates from this basic superpotential term a plethora of models extending the MSSM, the simplest example being the next-to-minimal model (NMSSM, as reviewed in \cite{NMSSM})   where an $S^3$ interaction is included, breaking a would-be Peccei-Quinn symmetry.  If the scalar component of the $S$  is given a large SUSY-breaking mass (it is not integrated out supersymmetrically), its $F$ term can generate a sizable contribution to $|H_u|^2 |H_d|^2$ in the potential with coefficient $\propto |\lambda|^2$,
and raises the tree-level mass bound on the Higgs \cite{NMSSMmass}:
\beq
m_h^2 \leq m_Z^2 \cos^2 2 \beta + \frac{\lambda^2 v^2}{2} \sin^2 2 \beta.
\eeq
See \cite{NMSSMrecent} for recent analyses with this construction.

\item Charging $H$ under an additional (broken) gauge symmetry with coupling $g_S$, giving new quartic interactions from non-decoupling $D$ terms \cite{NDDT} 
\beq
\label{eq:NDDT}
\Delta K = H^\dagger \exp \left(   -g_S \theta \sigma^\mu  \bar \theta \mathcal A_\mu + \dots + \frac{g_S}{2} \theta^2 \bar \theta^2 D_S    \right) H
\eeq
Integrating out the new auxiliary fields thus generates additional quartics $\propto g_S^2$  (compare Eq.~(\ref{eq:MSSMDterms})) provided the scalar mode of the massive vector multiplet is made massive.
The gauge symmetry must be broken above the weak scale, giving rise also to massive vector states associated with $\mathcal A$.    The important point in this discussion is however as in the case of the non-decoupling $F$ terms: the MSSM dynamics is not solely responsible for generating SUSY-breaking quartics, thereby allowing the SM's partner fields to have generically smaller masses than in the minimal model.

\item Coupling $H$ linearly to a separate EWSB sector containing  operators $\mathcal O_{u,d}$ of the same quantum numbers as the up- and down-type Higgses, inducing EWSB with tadpoles \cite{SCTC,GP,DSSM}
\beq
\label{eq:SCTC}
\Delta W = \lambda_u H_u \mathcal O_d + \lambda_d H_d \mathcal O_u.
\eeq
Coupling the Higgs fields linearly to composite operators as in Eq.~(\ref{eq:SCTC}) requires a strong sector with properties similar to conventional technicolor, though with a confinement scale $\Lambda \sim 4 \pi f$ satisfying $f \ll v$ to allow the bulk of EWSB to come from induced VEVs for the elementary fields:
\beq
v_{u,d} \sim \frac{\lambda_{u,d}}{4 \pi} \frac{\Lambda^2}{m_{H_{u,d}}^2} \times f.
\eeq
The mass squared for each field can now remain positive and directly control the physical Higgs mass.  The coupling terms $\lambda_{u,d}$ thus set $\tan \beta$, while $m_{H_{u,d}}^2$ can be freely used to tune the mixing angle $\alpha$.  This can lead to dramatic phenomenology since the trigonometric functions controlling SM fermion couplings are completely independent, very unlike the situation of the MSSM.  
\end{itemize}

 In  cases utilizing one of these options,\footnote{There are other interesting cases where new $F$ and $D$ terms can {\it both} contribute to the Higgs mass \cite{FatMagnetic}, or where light SUSY otherwise accompanies (partial) compositeness of the Higgs and/or gauge bosons \cite{SUSYpc}.} the Higgs sector of the MSSM can remain non-decoupled, allowing for the possibility of measuring large deviations in tree-level couplings particularly of down-type fermions for $\tan \beta >1$.

\subsection{Couplings of a SUSY Higgs: Loop Level}
We comment here briefly on the possibility for large corrections to loop-level Higgs couplings that may be generated by SUSY's additional matter content.  The derivation of these couplings follows the discussion of low-energy theorems in Section~\ref{sec:Composite}.   As in the case of a composite PNGB Higgs, a breaking of the symmetry protecting the Higgs mass can be beneficial for inducing large loop-level couplings.   Larger effects are typically possible here since naturalness, even in the minimal model, relies on the presence of new matter fields which we know break SUSY significantly.\footnote{Strictly speaking, we cannot yet be sure of this in the relevant cases: squarks that are nearly degenerate with their partners can evade detection via standard missing energy signatures \cite{stealth}.  Stops with small SUSY-breaking masses, for instance, may still be feasible.}

\begin{figure}[htb]
\begin{center}
\includegraphics[width=10cm]{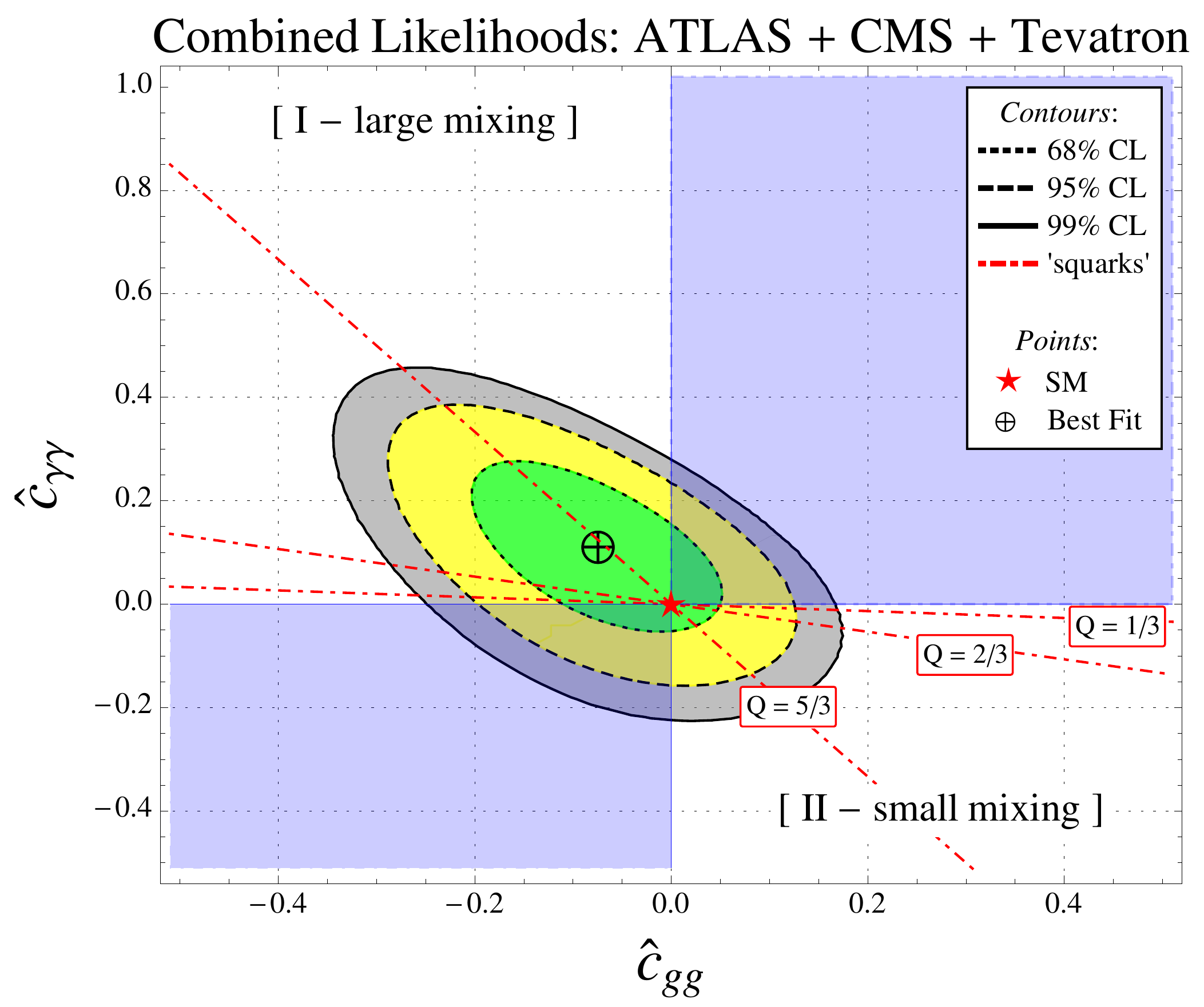}  
\caption{\small A fit to BSM contact operators $h \to \gamma \gamma$ and $h \to g g$, setting other Higgs couplings to their SM values.  We restrict priors to the range where effects of new matter fields do not `overshoot' the SM contribution \cite{ReeceSign}.  Hatted couplings $\hat c$ are normalized by corresponding SM quantities,
with positive $\hat c_{\gamma \gamma}$ corresponding to constructive interference with $W$ loops and positive $\hat c_{gg}$ to constructive interference with top loops.  We indicate lines along which new contributions from color fundamentals (denoted generically as `squarks') of varying electric charge would lie (cf. Eq.~(\ref{eq:cgslope})).  The shaded regions are inaccessible when only one species contributes significantly.}
\label{fig:cgg}
\end{center}
\end{figure}

To  recap the situation of a 125 GeV Higgs in SUSY, we have identified three possible outcomes:
\begin{enumerate}
\item All scalar partners are heavy and the Higgs mass is generated  by a large stop mass, yielding a very fine-tuned weak scale.
\item Stops are maximally mixed: they may be  light, but large off-diagonal terms in their mass matrices are necessary, corresponding to moderate tuning.
\item SUSY remains fully natural: the dimensionless SUSY-breaking parameters $m_{\tilde t}/m_t$ and $A_t/m_{\tilde t}$ are both small and the light Higgs acquires its mass elsewhere.
\end{enumerate}
In the first case, the light Higgs  may well appear indistinguishable from that of the SM if the additional Higgs states are also heavy.  In the third case we could reasonably expect to observe deviations in the tree-level couplings, primarily again those involving the bottom quark as discussed above.  We focus now on the second possibility, where loop effects would be most relevant (as has attracted much revived attention recently; cf. \cite{staus, Newgaga}).  In particular, corrections to the effective couplings $c_{gg}$ and $c_{\gamma \gamma}$ may be substantial.  The current best fit in this space with other couplings fixed to their SM values is shown in Fig~\ref{fig:cgg}, where some room remains for potentially important new contributions.  Note importantly that even highly-mixed stops do not improve the fit significantly for any arrangement of masses.\footnote{This assumes contributions remain in the SM neighborhood.  The space we show actually has four solutions, where the three not shown can be generated from the first by recognizing a reflection symmetry about the lines $\hat c_{gg} = -1$ and $\hat c_{\gamma \gamma} = -1$.}

In the presence of large mixing terms for the SM fields' partners, we can realize a situation that may seem worthwhile  with respect to Fig.~\ref{fig:cgg}.\footnote{For recent approaches to realizing modifications of loops, particularly in gluon fusion, cf. \cite{GluonFusion}.}  An improvement in fit can rely for instance on the form of the mass matrices for superpartners of generation $i$
\beq
M_{\tilde u_i}^2 &=&  \begin{pmatrix}
\tilde m_{Q_i}^2 +m_{u_i}^2 + \mathcal O(g^2 v^2)  & m_{u_i} X_{u_i}   \cr
m_{u_i} X_{u_i} &  \tilde m_{u_i}^2 +m_{u_i}^2 + \mathcal O(g^2 v^2) 
\end{pmatrix}, \\
M_{\tilde d_i}^2 &=&  \begin{pmatrix}
\tilde m_{Q_i}^2 +m_{d_i}^2 + \mathcal O(g^2 v^2)  & m_{d_i} X_{d_i}   \cr
m_{d_i} X_{d_i} &  \tilde m_{d_i}^2 +m_{d_i}^2 + \mathcal O(g^2 v^2) 
\end{pmatrix},
\eeq
and the following simple observation starting from Eq.~(\ref{eq:LogDet}):
\beq
\label{eq:cgMatter}
c_{gg}, c_{\gamma \gamma} 
&\propto&  \frac{v}{(m_{11}^2 m_{22}^2 - m_{12}^4)}\left(m_{22}^2  \frac{\partial m_{11}^2}{\partial v}  + m_{11}^2 \frac{\partial m_{22}^2}{\partial v}- \frac{\partial m_{12}^4}{\partial v}\right),
\eeq
i.e. large values of the mixing parameters, $X_{u_i} = A_{u_i} -\mu \cot \beta$ or $X_{d_i} = A_{d_i} -\mu \tan \beta$, can lead to contributions of the desired sign for both $c_{gg}$ and $c_{\gamma \gamma}$.  When off-diagonal terms dominate the diagonal terms in Eq.~(\ref{eq:cgMatter}) we access region I of Fig.~\ref{fig:cgg}, otherwise we access region II.   The shaded regions can be accessed only with the participation of more than one charge species.

The slope for contributions of any individual species is fixed, but more generally an optimized ratio of electric and color charges can be easily  devised for additional fields.    For a fundamental of $SU(3)_C$, a heavy state of electric charge $Q$ traces out a line in the space of $(\hat c_{gg},\hat c_{\gamma \gamma})$ shown in Fig.~\ref{fig:cgg} with slope
\beq
\label{eq:cgslope}
\frac{\hat c_{\gamma \gamma}}{\hat c_{gg}} \simeq -0.6  \times Q^2 \, ,
\eeq
the numerical prefactor arising from normalizing $c_{\gamma \gamma}$ by its SM value (compare Eq.~(\ref{eq:staucgam}) below; see also \cite{Moreau} for a similar analysis).\footnote{The sign of the slope can be understood from the low-energy theorems: matter fields enter the QED and QCD beta functions with the same sign, but interfere destructively with the dominant $W$ loops in the coupling with which we normalize $\hat c_{\gamma \gamma}$ (e.g. the top quark enhances gluon fusion but reduces $h \to \gamma \gamma$).}   New contributions along these contours can then be determined in terms of masses.  Taking squarks, as an example:
\beq
\label{eq:squarkcg}
\hat c_{gg}=  \frac{1}{2} \frac{b_0}{b_{1/2}}  \left( \frac{m_q^2}{m_{\tilde q_1}^2} + \frac{m_q^2}{m_{\tilde q_2}^2} - \frac{m_q^2 X_q^2}{m_{\tilde q_1}^2 m_{\tilde q_2}^2} \right),
\eeq 
where $b_i$ is the coefficient of a spin-$i$ particle's contribution to the beta function.  For uncolored doublets with physical masses $m_{\tilde i_{1,2}}$ and mass $m_i$ for the SM partner, we look directly at contributions to $h \to \gamma \gamma$.  Neglecting contributions from the $b$ quark, we find
\beq
\label{eq:staucgam}
\hat c_{\gamma \gamma} = \frac{1}{2} \frac{b_i}{b_{1/2}} \frac{Q^2_i}{3 Q^2_t}
\left( \frac{m_i^2 X_i^2}{m_{\tilde i_1}^2 m_{\tilde i_2}^2}
- \frac{m_i^2}{m_{\tilde i_1}^2} 
-\frac{m_i^2}{m_{\tilde i_2}^2} \right) 
\times 
\underbrace{ \left|
\frac{c_{\gamma \gamma}^{\rm (top)}}{c_{\gamma \gamma}^{(W)} +c_{\gamma \gamma}^{\rm (top)}} \right|
}_{\simeq \, 7/25}.
\eeq

\begin{figure}[htb]
\begin{center}
\includegraphics[width=8cm]{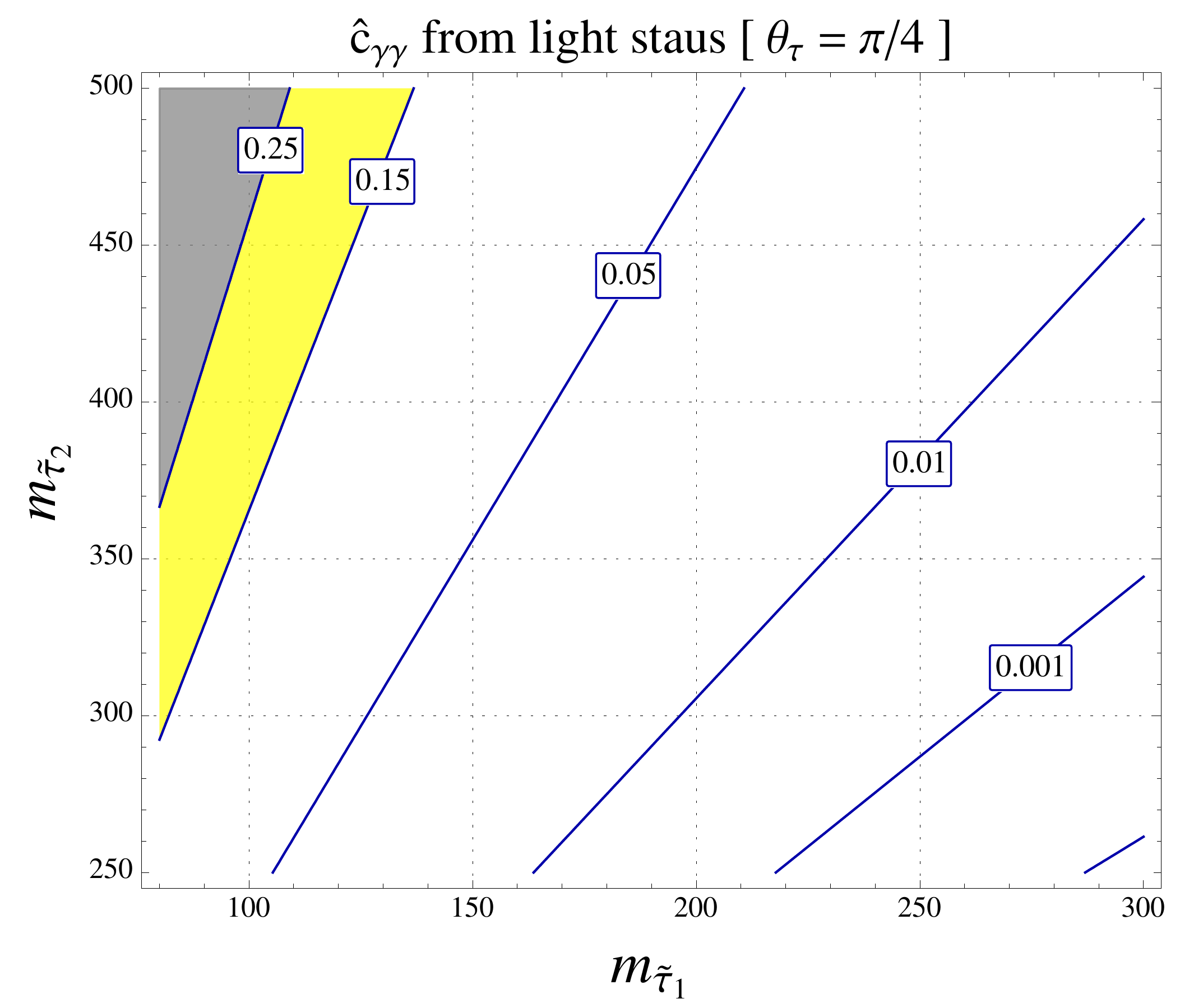}  \hspace{0cm}
\includegraphics[width=8cm]{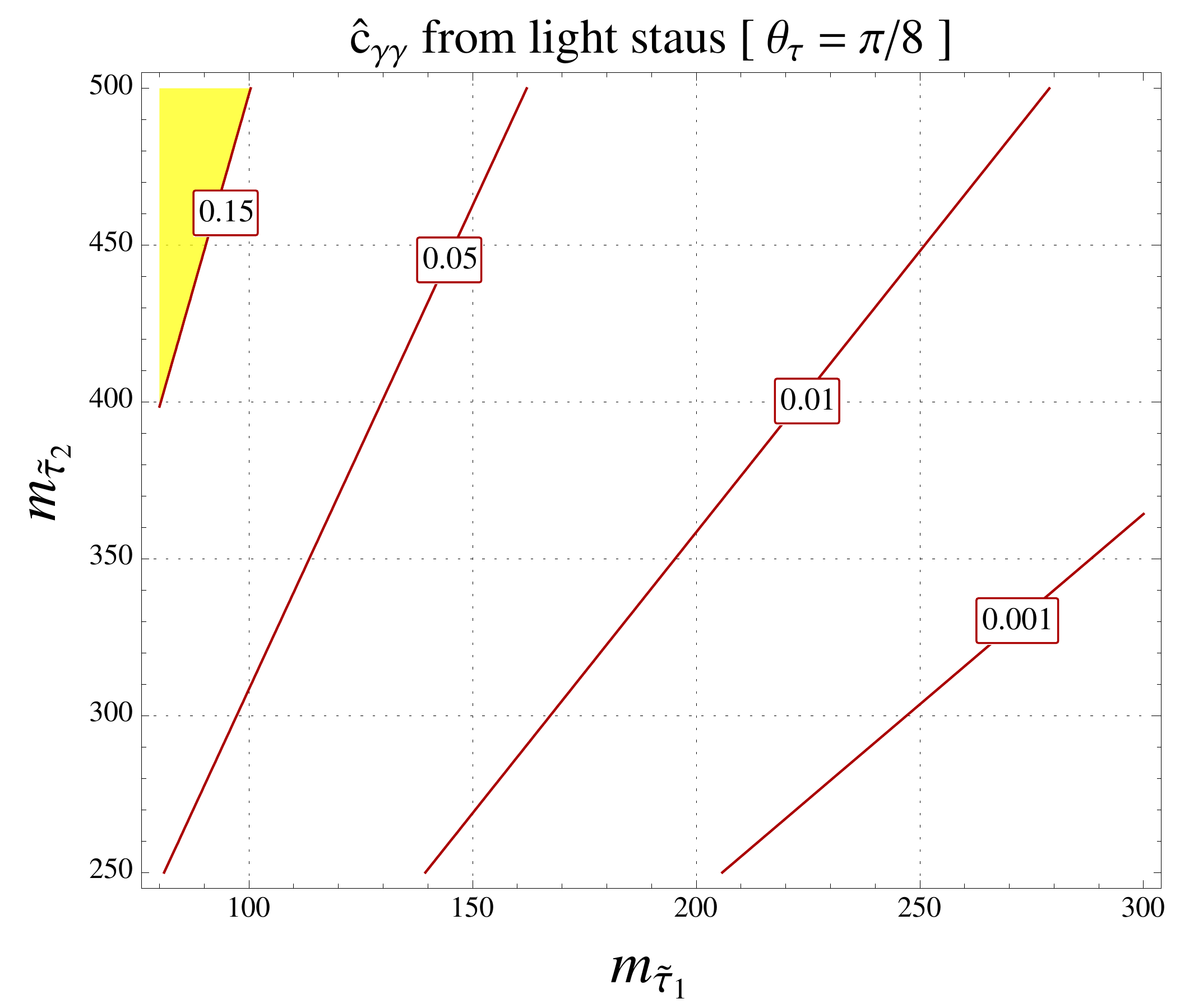}  
\caption{\small Contributions to $\hat c_{\gamma \gamma}$ from light staus.  {\bf Left}: Highly mixed ($\theta_\tau = \pi/4$) scenario.  {\bf Right}: Moderate mixing ($\theta_\tau = \pi/8$).  In both cases the unshaded region lies within the 68\% CL range, and the 95 (99)\% CL regions are shaded in yellow (gray).}
\label{fig:tau}
\end{center}
\end{figure}

We note the particular  possibility of invoking new contributions to $c_{\gamma \gamma}$ alone, which can provide an attractive option if excesses in the diphoton channel are observed while other couplings remain SM-like.  This was an especially enticing prospect with early data showing exactly this sort of behavior, though here with increased statistics we see that such a scenario is no longer required since the SM itself lies within th 68\% CL contour.  In minimal SUSY a contribution to $c_{\gamma \gamma}$ could nonetheless be relevant,  generated for instance  in the presence of highly-mixed light sleptons like the stau \cite{staus}.  We show this in Fig.~\ref{fig:tau}, starting with Eq.~(\ref{eq:staucgam}), and trading $X_\tau$ for a mixing angle $\theta_\tau$ via the identity 
\beq
\sin 2 \theta_\tau = \frac{2 m_\tau X_\tau}{m_{\tilde \tau_2}^2-m_{\tilde \tau_1}^2}.
\eeq
 Many interesting considerations---both experimental \cite{staupheno} and theoretical \cite{stauTheory}---attend this scenario, and we defer to those references for further details.  It is in any case important to keep in mind how these conclusions depends  on the treatment of all other couplings in the theory (cf. Table~\ref{tab:chi} of the following section).

\section{Conclusions and Outlook}
\label{sec:Conclusions} \setcounter{equation}{0} 
In the absence of any new physics beyond the Higgs discovered in the near future, significant  theoretical effort can be guided by the information provided by the Higgs itself.  As such, it would be useful to identify the directions in a global space where gradients of Higgs likelihoods are large: this corresponds to identifying the linear combinations of effective operators that most effectively minimize the total chi-squared of the fit.  The most promising directions for BSM physics would be those in which the deviation of the best fit point's likelihood compared to that of the SM point is largest.  There are few obvious prospects at present however, as we find with our best fit points of the global spaces summarized in Table~\ref{tab:chi}.
\begin{table}[htb]
\centering
\renewcommand{\arraystretch}{0.9}
\begin{tabular}{| l | l | c | c |} 
\hline
Space & \hspace{1.7cm} Best Fit &  $-2 (\Delta \log P)$ &  $-2 \log P/{\rm dof}$ \\
\hline \hline
SM &  \hspace{2.2cm} --- &  1.42  & 2.64\\ 
\hline
1D  & $\mu= 0.99 $ & 1.41&  2.73 \\ 
 & $c_{\gamma \gamma} = 0.06$ & 1.00 &  2.72 \\ 
\hline
2D & $(a,c) = (1.00,0.87)$ &  0.49 & 2.80 \\
 & $(c_{\gamma \gamma},c_{gg}) = (0.11,-0.07)$ & 0.21 & 2.79 \\ 
 \hline
3D &  $(a,c_b,c_t) = (1.05,0.99,0.89)$& 0.29& 2.89 \\
 \hline
5D & $(a,c_b,c_t,c_{\gamma \gamma} ,c_{gg})$  &  ---  &  3.11 \\
& $=(1.02, 0.96,1.2,0.15,-0.32) $  &  & \\
\hline
\end{tabular}
\caption{\small Best fit points and delta log likelihood  with respect to the global best fit in various parameter spaces. }
\label{tab:chi}
\end{table}
It appears that the experimental outcome is leaving little hint in terms of glaring discrepancies and directions for BSM physics in these terms.   Even the notable exception of an excess in the  $\gamma \gamma$ channels is at this point hard pressed to seriously compete with the SM: counting separately only grossly distinct topologies gives approximately 30 separate measurements from the LHC and Tevatron, from which the likelihood per degree of freedom is maximal for the case of a SM Higgs.

Many other scenarios can be, and have been, considered in similar ways to those we have covered here.  We note some other scenarios that have been studied in light of Higgs data for the sake of comparison:
\begin{itemize}
\item The signals observed at the LHC could be coming from a `Higgs impostor', as in the case of a light dilaton \cite{impostor}.
\item The Higgs could decay invisibly, as studied in Refs.~\cite{invisible}.  
\item The Higgs signals could overlap with additional states that mix directly with the Higgs, as in the `social Higgs' scenario studied in \cite{friends}.
\end{itemize}
Indeed many other possibilities could be explored.  Here we have focused on  giving simple demonstrations of how questions of the nature of the Higgs can be answered  in general.

The experimental situation regarding the Higgs boson will remain in flux as additional data is taken.  As such, it is too early to draw definitive conclusions.   It appears at this point that the SM, or some theory with fairly decoupled additional states, is emerging as a leading candidate.   We have reviewed here the ways in which this information bears on two landmark scenarios for naturalness at the weak scale.   For the case of a composite Higgs, the apparent lack of large deviations in tree-level couplings is already an indication that there would remain some tuning at the percent level in such theories, while loop-level couplings that would be most interesting in these cases  (e.g. $h \to Z \gamma$) must await future data.  For SUSY, on the other hand, interesting anomalies in loop-level couplings allow for the possibility of new physics that is on par with the goodness of fit of the SM.   In both cases, it is these loop-level couplings which can start to probe the details of the theory to which tree-level couplings are less sensitive.

Under the reasonable assumption that the state discovered  near 125 GeV is indeed a Higgs boson, or {\it the} Higgs boson, we have already learned a remarkable amount about the nature of weak scale physics and how electroweak symmetry is broken.   The presence of additional symmetries that might be operative  at this scale may take much longer to ascertain, but these symmetries will have to play a part in determining the more detailed properties of the Higgs if they are at all relevant for stabilizing its mass.   This makes for a sensational period in the field of particle physics, one in which we are well-advised to study every detail of experimental information that is made available to us in order to catch any hint of  new physics that the Higgs might be carrying with it.

\begin{center}
\line(1,0){200} 
\end{center}

\section*{Acknowledgments}
We are very happy to thank Spencer Chang, Roberto Contino, Nathaniel Craig, Daniele Del~Re, Marco Grassi, and Shahram Rahatlou for collaboration in studies on which this review is largely based.  We have benefited from discussions with Evan Friis, Eric Kuflik, Markus Luty, Adam Martin, Matthew McCullough, Daniel Stolarski, and Felix Yu  during the course of those studies and throughout the completion of this work.  We are particularly grateful to R.~Contino for continued collaboration and inspiration, and for useful comments on this manuscript.  We thank the Galileo Galilei Institute for Theoretical Physics for hospitality during the final stages of preparing this review.   This work was supported by the ERC Advanced Grant No.~267985
{\em Electroweak Symmetry Breaking, Flavour and ÊDark Matter: One Solution for Three Mysteries (DaMeSyFla)}.

\vspace{1cm}

\begin{appendix}

\section{(Re)Constructing Higgs Likelihoods}
\label{sec:Reconstruction} \setcounter{equation}{0} 
We review here three methods for constructing Higgs likelihoods given various inputs from experimental collaborations.  These likelihoods are minimally functions of three variables: the number of signal, background, and observed events.  We present methods of determining the approximate form of these functions in three different cases, which have thus far proven sufficient in assembling  all data into parameter space fits.  The distinct criteria for each method are as follows:
\begin{enumerate}
\item Event yields are quoted directly along with estimated signal and background rates together with their respective uncertainties.
\item A clear peak has yet emerged and only exclusion data are available.
\item A peak has emerged and best fit values for the signal strength modifier are given.  
\end{enumerate}
The total likelihood can be assembled simply by constructing and combining the likelihood for each individual (sub)channel in one of the ways outlined here.

\subsection{Constructing Likelihoods from Event Yields}
Likelihood profiles as functions of the signal strength modifier $\mu$---with its generalization to $R$ in multidimensional cases---can be constructed directly from event yields with known signal and background central values and uncertainties.  The likelihood is treated as a Poisson distribution, with background and signal values themselves distributed according to some appropriate function (typically a truncated Gaussian or a log-normal distribution).  In principle  one should marginalize not just over uncertainties in background and signal rates, but over many additional `nuisance parameters'.  We find in practice however that limiting the problem to just these two---signal and background---a suitably accurate approximation can be obtained.  

The underlying statistical reasoning is that of Bayes' theorem: we observe $n_{\rm obs}$ events in a given channel, and compute the posterior probability that such an observation is compatible with a predicted number $n= n_{\rm B} + \mu \, n_{\rm S}$, with $n_{\rm S}$ understood to denote the predicted number of signal events within the SM.  Bayes' theorem tells us that this can be computed as follows:
\beq
L(n | n_{\rm obs}) = \pi (\mu) \times L(n_{\rm obs} | n).
\eeq
Here $\pi(\mu)$ is the Bayesian prior, which is assumed to be flat in the cases we consider.  The final form of the likelihood can thus be written
\beq
L(\mu) =  \pi(\mu) \times (n_{\rm B} + \mu \, n_{\rm S})^{n_{\rm obs}} e^{-(n_{\rm B} + \mu \, n_{\rm S})},
\eeq
where now $\pi(\mu)$ absorbs  factors that do not introduce additional $\mu$-dependence, i.e. the prior amounts simply to an overall normalization.  For notational simplicity we will discard this factor in what follows with the understanding that all likelihoods will ultimately have to be properly normalized in the space under consideration.

Taking account of  uncertainties in signal and background, we promote $n_{\rm S,B}$ to variables $\theta_{\rm S,B}$ and integrate (marginalize) over their respective distributions $f(\theta_{\rm S,B},\sigma_{\rm S,B})$, i.e.
\beq
L(\mu )  =  \int d \theta_{\rm S} \, d \theta_{\rm B} \,
 (\theta_{\rm B} + \mu \, \theta_{\rm S})^{n_{\rm obs}} 
 \, e^{-(\theta_{\rm B} + \mu \, \theta_{\rm S})} 
 \, f (\theta_{\rm S},\sigma_{\rm S}) \, f(\theta_{\rm B},\sigma_{\rm B}).
\eeq
The choice of distribution functions, $f(\theta, \sigma)$, depends on the details of the given channel: if either of $n_{\rm S,B}$ is small and its uncertainty large, a log-normal distribution is used to ensure that the event counts remain positive.  In cases where counts are large and uncertainties relatively small (as is typically the case), a Gaussian distribution truncated at $\theta_{\rm S,B} = 0$ can be reliably used.  Note that the marginalization over uncertainties is bound to introduce {\it some} degree of non-Gaussianity in the likelihoods even when the event yields are large.

\subsection{Reconstructing Likelihoods from Exclusion Contours}
In cases where a clear peak in the data has not yet emerged, it may be the case that only exclusion data are available.  Such data provides only two constraints---the expected and observed exclusions---on our three-dimensional likelihood, giving a problem that we are naively not able to solve.  In practice, however, it is usually possible to obtain an accurate reconstruction of the full likelihood from this limited information.  Note a crucial point here that the likelihoods we reconstruct have already been fully marginalized, so we  deal only with the Poisson distribution in its Gaussian limit; cf.  \cite{fits2} for details.

The expected exclusion limit at a confidence level $\alpha$ corresponds to integrating the likelihood $L(n_{\rm B} + \mu \, n_{\rm S}|n_{\rm B})$ to a critical value $\mu = \mu_{\rm exp}^{(\alpha)}$ such that an integrated fraction $\alpha$ is captured, i.e. one assumes all observed events arise from background only and determines the value $\mu_{\rm exp}^{(\alpha)}$ where this hypothesis fails at the level $\alpha$.  In the Gaussian limit, we have
\beq\label{eq:alphaExp}
\alpha = \int_0^{\mu_{\rm exp}^{(\alpha)}} d \mu \ L(n_{\rm B} + \mu \, n_{\rm S} | n_{\rm B}) 
\simeq  \int_0^{\mu_{\rm exp}^{(\alpha)}}  d\mu \ \exp\left(\frac{-\mu^2 n_{\rm S}^2}{2 \, n_{\rm B}}\right).
\eeq
Properly normalizing the likelihood, one obtains
\beq
\label{eq:Erf}
\alpha = {\rm erf} \left( \frac{ \mu_{\rm exp}^{(\alpha)} \, n_{\rm S}}{\sqrt {2 \, n_{\rm B}}} \right),
\eeq
e.g. at 95\% CL we have
\beq
\label{eq:muExp95}
\mu_{\rm exp}^{(95\%)} \simeq 1.96 \times \frac{\sqrt {n_{\rm B}}}{n_{\rm S}}.
\eeq
Taking the expected exclusion limit from collaboration data thus tells us a specific ratio of signal to background events.

Incorporating the {\it observed} exclusion limit---again in the Gaussian limit---we find it convenient to recast the Gaussian and express the integral as follows:
\beq\label{eq:muObs}
\alpha  \simeq  \int_0^{ \mu_{\rm obs}^{(\alpha)}} d\mu \ \exp\left[-\frac{1}{2}\left(\mu \frac{n_{\rm S}}{\sqrt{n_{\rm B}}}   \frac{\sqrt{n_{\rm B}}}{\sqrt{n_{\rm obs}}} +\delta \right)^2\right],
\eeq
where we've defined 
\beq\label{eq:delta}
\delta = \frac{n_{\rm B}-n_{\rm obs}}{\sqrt{n_{\rm obs}}}.
\eeq
Now if we assume that the number of signal events is small compared to background, i.e.
\beq\label{eq:IntLimit}
\frac{n_{\rm obs} - n_{\rm B}}{n_{\rm B}} \ll1 \quad 
\Longrightarrow \quad L(\mu) \to \exp\left[-\frac{1}{2}\left(\mu \frac{n_{\rm S}}{\sqrt{n_{\rm B}}} +\delta \right)^2\right],
\eeq
we find from Eqs.~(\ref{eq:Erf}, \ref{eq:muExp95}, \ref{eq:muObs}) 
\beq
\label{eq:ObsFinal}
\alpha =  \int_0^{\mu_{\rm obs}^{(\alpha)}} d\mu \exp \left[ -\frac{1}{2} \left( \frac{ \sqrt 2\, \mu}{\mu_{\rm exp}}  \, {\rm erf}^{-1}( \alpha ) + \delta \right)^2 \right].
\eeq
Now setting $\mu_{\rm obs}^{(\alpha)}$ to its experimentally determined value, Eq.~(\ref{eq:ObsFinal}) can be solved numerically for $\delta$ and the likelihood in Eq.~(\ref{eq:IntLimit}) is fully determined as a function of $\mu$.

This method is useful, as advertised, in cases where best fits are not quoted.  In cases where we can explicitly check its accuracy (cases where event yields or best fits are also quoted) we find deviations from the `true' likelihood of order $15\%$ or less over the entire range of $\mu$; global fits performed using this approximation however turn out to agree with official results at the level of $\sim 10 \%$ in the Higgs searches for $m_h \lesssim 400 \, {\rm GeV}$.   Such approximations may thus be useful in assessing the properties of other significant fluctuations that may appear in the Higgs searches at higher scales.

\subsection{Constructing Likelihoods from Best Fits}
Reconstructing likelihoods becomes a particularly simple task when best fit values and error bands are given for the signal strength modifier of each channel.  At early stages of data taking, non-Gaussianities can be sizable, so we adopt an approach to accommodate asymmetric uncertainties.  
We  construct a likelihood by joining two separate (half) Gaussians, $L^\pm$, fit to either side of the central value of the signal strength modifier, $\hat \mu$.  Each distribution is defined by its respective variance, $\sigma_\pm$, corresponding to the appropriate uncertainty.  Explicitly,
\beq
\label{eq:TSG}
L_i^\pm (\mu) \propto \exp \frac{-(\mu - \hat \mu_i)^2}{2 (\sigma_i^\pm)^2}.
\eeq
This procedure provides an accurate reconstruction of likelihoods when other presentation of the data is not available.   Its reliability is demonstrated in Fig.~\ref{fig:CMSValidation}.  Albeit seemingly ad hoc, marginalization of nuisance parameters will inevitably give non-Gaussianities, so some sort of accommodation of asymmetric uncertainties is necessary.   The use of two-sided Gaussians is the simplest implementation of this requirement.

\section{Data Used in Fits}
\label{sec:Data} \setcounter{equation}{0} 

\subsection{Hadron Colliders}
We collect the LHC data used in all fits in Tables~\ref{tab:ATLAS}-\ref{tab:Tevatron}.  In all cases the likelihoods have been constructed or reconstructed using one of the three methods outlined above, and we note this accordingly in the tables.
\begin{table}[hhh]
\footnotesize
\centering
\renewcommand{\arraystretch}{1.1}
\begin{tabular}{| l | c | c | c | c | c |}
\hline
Channel & $\hat \mu$ (7 TeV) & $\zeta_i^{\rm (G, V, T)}$ (\%)&  $\hat \mu$ (8 TeV) & $\zeta_i^{\rm (G,V,T)}$ (\%) & Refs. \\
\hline\hline 
$b \bar b$ & comb. w/8 & --- & ${-0.42\pm 1.05}$ & $(0,100,0)$ & \cite{ATLASbb, ATLASbbT} \\
$b \bar b$ ($ttH$)& $3.81 \pm 5.78^{**}$ & $(0,30,70)$ & ---  & --- &\\
\cline{1-6}
$\tau \tau$ & comb. w/8 & --- & ${0.7\pm 0.7}^*$ & $(20,80,0)$& \cite{ATLAStautau} \\
\cline{1-6}
$WW\, (0j)$ & $0.06 \pm 0.60^*$ & inclusive & $0.92^{+0.63*}_{-0.49}$ & inclusive &    \\
$WW\, (1j)$ & $2.04^{+1.88*}_{-1.30}$ & inclusive & $1.11^{+1.20*}_{-0.82}$ & inclusive &   \cite{ATLASww} \\
$WW\, (2 j)$ & --- & --- & $1.79^{+0.94*}_{-0.75}$ & $(20,80,0)$ & \\
\cline{1-6}
$ZZ$ & comb. w/8  & --- & $1.7^{+0.5}_{-0.4}$ & inclusive& \cite{ATLASzz} \\
\cline{1-6}
$\gamma \gamma_{\rm (L)} $ (uc$|$ct) & $0.53^{+1.37}_{-1.44}$ & $(93,7,0)$& $0.86\pm 0.67$ & $(93.7,6.2,0.2)$& \\
$\gamma \gamma_{\rm (H)} $ (uc$|$ct) &$0.17^{+1.94}_{-1.91}$ & $(67,31,2)$ & $0.92^{+1.1}_{-0.89}$ & $(79.3,19.2,1.4)$ &  \\
$\gamma \gamma_{\rm (L)} $ (uc$|$ec) & $2.51^{+1.66}_{-1.69}$& $(93,7,0)$ &$2.51^{+0.84}_{-0.75}$  &  $(93.2,6.6,0.1)$ & \\
$\gamma \gamma_{\rm (H)} $ (uc$|$ec) & $10.39^{+3.67}_{-3.67}$ & $(65,33,2)$ & $2.69^{+1.31}_{-1.08}$ & $(78.1,20.8,1.1)$& \\
$\gamma \gamma_{\rm (L)} $ (c$|$ct) & $6.08^{+2.59}_{-2.63}$ & $(93,7,0)$ &  $1.37^{+1.02}_{-0.88}$ &$(93.6,6.2,0.2)$&\\
$\gamma \gamma_{\rm (H)} $ (c$|$ct) & $-4.40^{+1.80}_{-1.76}$ & $(67,31,2)$ &$1.99^{+1.50}_{-1.22}$  &$(78.9,19.6,1.5)$ & \\
$\gamma \gamma_{\rm (L)} $ (c$|$ec) & $2.73^{+1.91}_{-2.02}$ & $(93,7,0)$ & $2.21^{+1.13}_{-0.95}$ & $(93.2,6.7,0.1)$&  \\
$\gamma \gamma_{\rm (H)} $ (c$|$ec) & $-1.63^{+2.88}_{-2.88}$ & $(65,33,2)$ & $1.26^{+1.31}_{-1.22}$ & $(77.7,21.2,1.1)$& \cite{ATLASgaga,ATLASgaga1212}\\
$\gamma \gamma $ (c$|$trans.) & $0.35^{+3.56}_{-3.60}$ & $(89,11,0)$ & $2.80^{+1.64}_{-1.55}$ & $(90.7,9.0,0.2)$& \\
$\gamma \gamma $ (dijet) & $2.69^{+1.87}_{-1.84}$ & $(23,77,0)$ & --- & --- & \\
$\gamma \gamma $ (loose high mass $jj$) & --- & --- & $2.76^{+1.73}_{-1.35}$ & $(45,54.9,0.1)$& \\
$\gamma \gamma $ (tight high mass $jj$) & --- & --- & $1.59^{+0.84}_{-0.62}$ & $(23.8,76.2,0)$& \\
$\gamma \gamma $ (low mass $jj$) & --- & --- & $0.33^{+1.68}_{-1.46}$ & $(48.1,49.9,1.9)$& \\
$\gamma \gamma $ ($E_{\rm T}^{\rm miss}$ significance) & --- & --- & $2.98^{+2.70}_{-2.15}$ & $(4.1,83.8,12.1)$& \\
$\gamma \gamma $ (lepton tag) & --- & --- & $2.69^{+1.95}_{-1.66}$ & $(2.2,79.2,18.6)$& \\
\cline{2-4}
\hline
\end{tabular}
\caption{\small ATLAS data used in fits.  Best fits on signal strength modifier $\mu$ with efficiencies $\zeta$ (when provided) for gluon (G), vector (V), and top (T) initiated production.  A best fit marked by $*$ indicates that the fit comes from a likelihood we have constructed directly from event yields; $**$ indicates likelihoods reconstructed from exclusion data (cf. Appendix~\ref{sec:Reconstruction} for details on reconstruction).  In the $\gamma \gamma$ channels, `uc' (`c') corresponds to mostly unconverted (converted) photons, `ct' indicates both photons registered centrally while `ec' indicates one or more of the photons registered in the endcap, and the subscripts (H, L) denote high and low $p_T$.}
\label{tab:ATLAS}
\end{table}
\begin{table}[hhh]
\footnotesize
\centering
\renewcommand{\arraystretch}{1.1}
\begin{tabular}{| l | c | c | c | c | c |}
\hline
Channel & $\hat \mu$ (7 TeV) & $\zeta_i^{\rm (G,V,T)}$ (\%) &  $\hat \mu$ (8 TeV) & $\zeta_i^{\rm (G,V,T)}$(\%) & Refs. \\
\hline\hline 
$b \bar b$ & comb. w/8 & --- & $1.30^{+0.68}_{-0.59}$ & $(0,100,0)$& \cite{CMSbb} \\
$b \bar b$ ($ttH$)& ${-0.81^{+2.05}_{-1.75} }$ & $(0,30,70)$ & ---  & --- & \cite{CMSbbT}\\
\cline{1-6}
$\tau \tau$ ($0/1j$)& comb. w/8 & --- & $0.74^{+0.49}_{-0.52}$ & inclusive &  \\
$\tau \tau$ (VBF) & comb. w/8 & ---& $1.38^{+0.61}_{-0.57}$ & $(0,100,0)$ & \cite{CMStautau} \\
$\tau \tau$ (VH) & comb. w/8 & --- & $0.76^{+1.48}_{-1.43}$ & $(0,100,0)$& \\
\cline{1-6}
$WW\, (0/1 j)$ & comb. w/8 & --- & $0.76\pm 0.21$ & inclusive & \\
$WW\, (2 j)$ & comb. w/8 & ---  & $-0.05^{+0.73}_{-0.56}$ & $(17,83,0)$ & \cite{CMSww} \\
$WW$ (VH)  & comb. w/8 & --- &$-0.31^{+2.24}_{-1.96}$ & $(0,100,0)$& \\
\cline{1-6}
$ZZ$ (untagged) & comb. w/8 & --- &  $0.84^{+0.32}_{-0.26}$ & $(95,5,0)$ & \cite{CMSzz} \\
$ZZ$ (dijet tag) & --- & --- &  $1.22^{+0.84}_{-0.57}$ & $(80,20,0)$ & \\
\cline{1-6}
$\gamma \gamma $ (untagged 0)  & $3.78^{+2.01}_{-1.62}$ & $(61.4, 35.5,3.1)$& $2.12^{+0.92}_{-0.78}$& $(72.9,24.6,2.6)$ & \\
$\gamma \gamma $ (untagged 1) & $0.15^{+0.99}_{-0.92}$ & $(87.6,11.8,0.5)$ & $-0.03^{+0.71}_{-0.64}$ & $(83.5,15.5,1.0)$ &  \\
$\gamma \gamma $ (untagged 2) & $-0.05 \pm 1.21$& $(91.3,8.3,0.3)$ & $0.22^{+0.46}_{-0.42}$ & $(91.7,7.9,0.4)$ &  \\
$\gamma \gamma $ (untagged 3) & $1.38^{+1.66}_{-1.55}$& $(91.3,8.5,0.2)$ & $-0.81^{+0.85}_{-0.42}$ & $(92.5,7.2,0.2)$ &\\
$\gamma \gamma $ (dijet) & $4.13^{+2.33}_{-1.76}$& $(26.8,73.1,0.0)$ & --- & ---  &  \cite{CMSgaga} \\
$\gamma \gamma $ (dijet loose) & --- & --- & $0.75^{+1.06}_{-0.99}$ & $(46.8,52.8,0.5)$ & \\
$\gamma \gamma $ (dijet tight) & --- & --- & $0.22^{+0.71}_{-0.57}$ & $(20.7,79.2,0.1)$ & \\
$\gamma \gamma $ (MET) & --- & --- & $1.84^{+2.65}_{-2.26}$ & $(0.0,79.3,20.8)$ & \\
$\gamma \gamma $ (Electron) & --- & --- & $-0.70^{+2.75}_{-1.94}$ &  $(1.1,79.3,19.7)$ & \\
$\gamma \gamma $ (Muon) & --- & --- & $0.36^{+1.84}_{-1.38}$ &  $(21.1,67.0,11.8)$ & \\
\hline
\end{tabular}
\caption{\small CMS data used in fits.  In the diphoton channels, the notation adopted is that of \cite{CMSgaga}.  Official values for efficiencies are used when quoted, otherwise approximations are made according to a channel's primary topologies.}
\label{tab:CMS}
\end{table}
\begin{table}[hhh]
\footnotesize
\centering
\renewcommand{\arraystretch}{1.1}
\begin{tabular}{| c | l | c | c | c |}
\hline
Exp. & Channel & $\hat \mu$ (2 TeV)  & $\zeta^{\rm (G,V,T)}$ (\%) & Ref. \\
\hline \hline
CDF & $b \bar b$ & $1.72^{+0.92}_{-0.87}$ & $(0,100,0)$ & \cite{CDFD0,TevUpdate} \\
& $\tau \tau$ & $0.00^{+8.44}_{-0.00}$ & $(50,50,0)$  & \\
& $WW$ & $0.00^{+1.78}_{-0.00}$ & inclusive & \\
& $\gamma \gamma $ &  $7.81^{+4.61}_{-4.42}$ & inclusive &\\ 
\hline
D\O & $b \bar b$ & $1.23^{+1.24}_{-1.17}$ & $(0,100,0)$ &  \cite{CDFD0,TevUpdate}  \\
& $\tau \tau$ & $3.94^{+4.11}_{-4.38}$ & $(50,50,0)$ &  \\
& $WW$ & $1.90^{+1.63}_{-1.52}$ & inclusive & \\
& $\gamma \gamma $ &  $4.20^{+4.60}_{-4.20}$ & inclusive &\\ 
\hline
\end{tabular}
\caption{\small CDF/D\O  \ data used in fits.  Efficiencies for $\tau \tau$ channels are approximated from \cite{TevTau}.}
\label{tab:Tevatron}
\end{table}

We note also that in certain cases where results are not presented separately, we've used LHC data from the 7 and 8 TeV runs in its combined form.  Recall that rescaling factors for Higgs production will typically be of the form (neglecting $ttH$ and assuming $c_t \gtrsim c_{b,\tau}$ to dominate gluon fusion)
\beq
R_X = a^2  \times  \zeta^{\rm (V)}_X +  (c_t + c_{gg})^2 \times \zeta^{\rm (G)}_X 
\eeq
where we take $V$ to denote both VBF and VH, G to denote gluon fusion.  Failing to account for the differences in cross-section from energy $E_1$ to energy $E_2$ thus introduces an error in the relative weight of each respective term:
\beq
\epsilon^{(V)}= 1-\left(\frac{\sigma_{\rm V}(E_2)}{\sigma_{\rm V}(E_1)} \times \frac{\sigma_{\rm G}(E_1)}{\sigma_{\rm G}(E_2)}\right); \qquad
\epsilon^{(G)}= 1-\left(\frac{\sigma_{\rm G}(E_2)}{\sigma_{\rm G}(E_1)} \times \frac{\sigma_{\rm V}(E_1)}{\sigma_{\rm V}(E_2)}\right).
\eeq
Between 7 and 8 TeV, these are of order 1\%.  Thus in cases where efficiencies are known to be approximately the same for each run,  the error introduced by using a single likelihood from the combined data is negligible.

\subsection{Electroweak Precision Measurements}
In places where electroweak precision data have been included in global fits, we rely on current results from the Gfitter collaboration \cite{gfitter} which have been recently updated in light of the measured Higgs value.

\begin{figure}[hbt]
\begin{center}
\includegraphics[width=7.1cm]{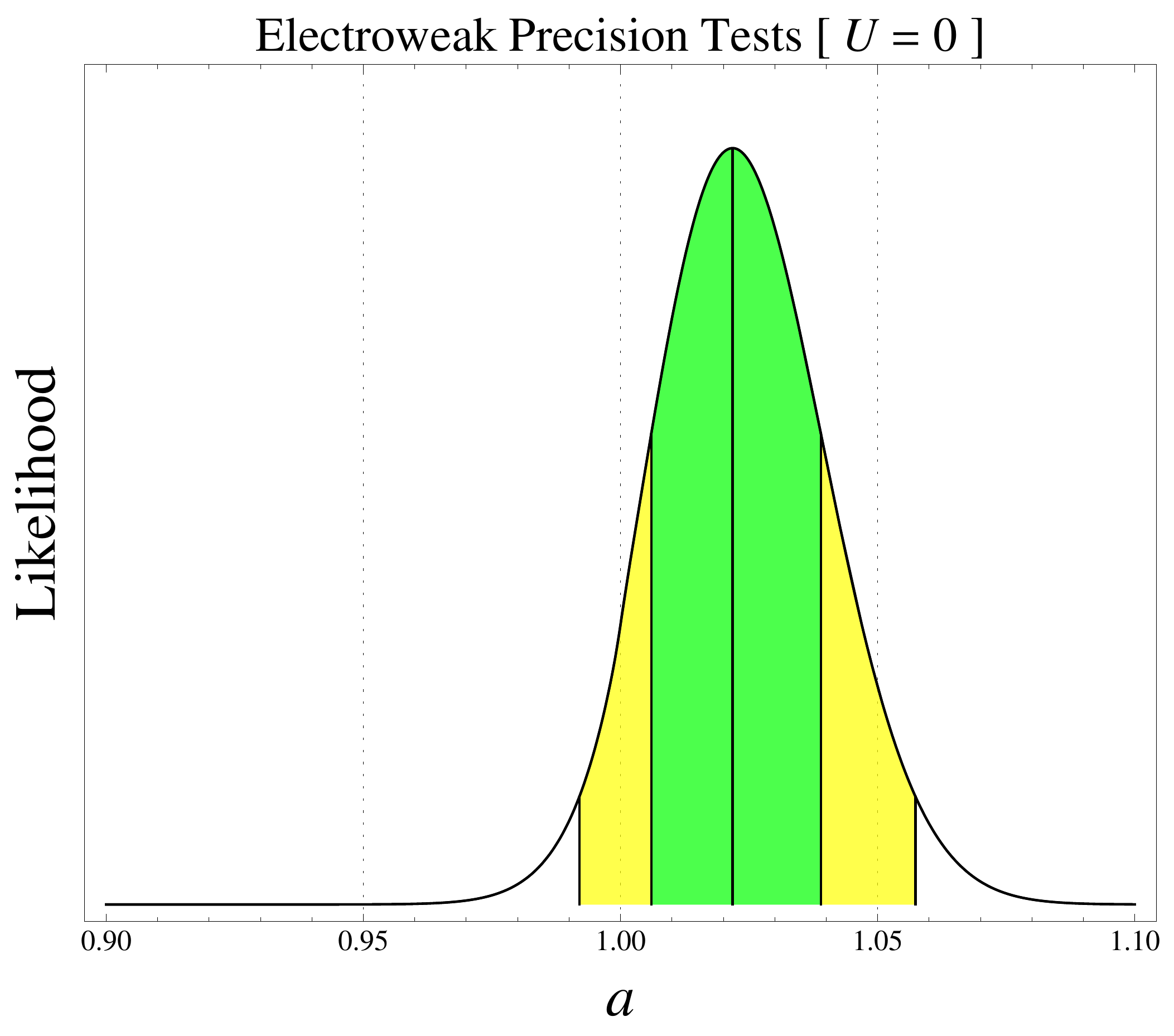} \quad
\includegraphics[width=7.1cm]{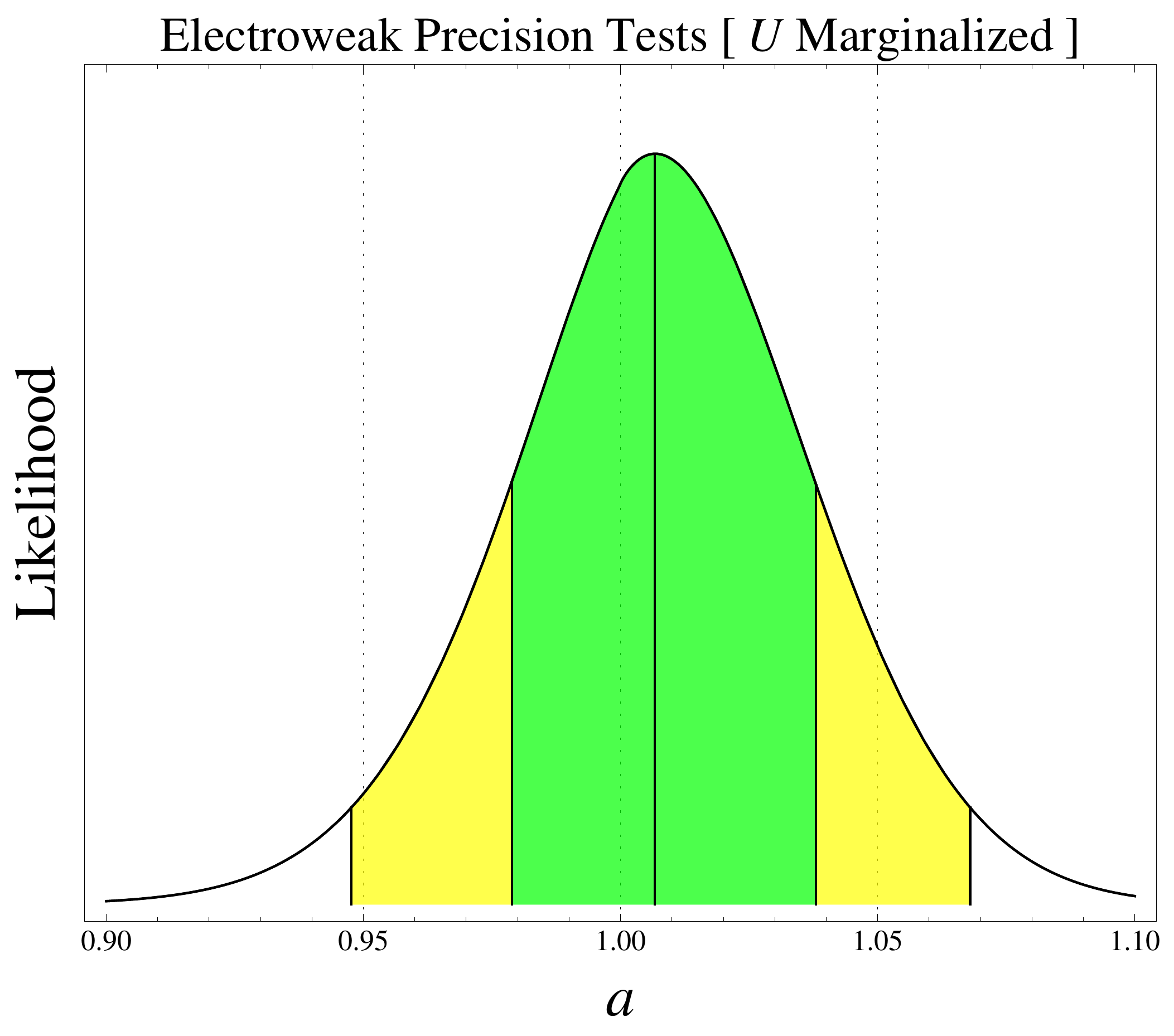}
\caption{\small Likelihood as a function of the Higgs coupling to weak gauge bosons, constructed from precision measurements using central values quoted by Gfitter.  $S$ and $T$ parameters are mapped to functions of the vector coupling $a$ assuming negligible threshold corrections.  The {\bf left panel} shows the likelihood with $U=0$; the {\bf right panel} shows the result of marginalizing over $U$.}
\label{fig:EWPT}
\end{center}
\end{figure}

The central values and uncertainties  for the oblique parameters $S,T,U$ are as follows:
\beq
S&=& 0.03\pm 0.1\nonumber \\
T&=&0.05\pm0.12 \\
U&=&0.03\pm0.1 \nonumber
\eeq
and the following correlation coefficients are specified:
\beq\label{eq:STUcorrelation}
\baa{c}\\ S\\T\\ U\eaa 
\baa{c c c}
\ S&T &U\\
\ 1&0.891&-0.540\\
\ 0.891&1&-0.803\\
\ -0.540&-0.803&1
\eaa
\eeq
From here we can construct  the likelihood:
\beq
P (S,T,U) \propto \exp \left[ -\frac{1}{2} \begin{pmatrix} \frac{S- 0.03}{0.1} \cr \frac{T-0.05}{0.12} \cr \frac{U-0.03}{0.1} \end{pmatrix}^T
\times C^{-1} \times 
\begin{pmatrix} \frac{S- 0.03}{0.1} \cr \frac{T-0.05}{0.12} \cr \frac{U-0.03}{0.1} \end{pmatrix} \right],
\eeq
where $C$ is the correlation matrix of Eq.~(\ref{eq:STUcorrelation}).  In practice, as in Fig.~\ref{fig:ac}, we can  marginalize over $U$ or fix it to a  particular value to recover a reduced likelihood as a function of $S$ and $T$ alone.  As described in Sec.~\ref{sec:EWPT} this likelihood can then be expressed in terms of only the vector coupling, $a$,  provided threshold corrections are assumed to be small.  For reference, we show the resultant likelihood with its 68\% and 95\% CL intervals in Fig.~\ref{fig:EWPT}.

\subsection{A Note on Correlations}
We comment here on the effect of correlated systematic uncertainties in interpreting data from Higgs searches.  At present correlations among sub-channels within a given search mode are not provided by the collaborations.  When combining channels as has been done throughout this work, it is therefore tacitly assumed that all sub-channels are uncorrelated.  This is clearly an approximation that will need to be improved with future data.

The absence of information regarding correlations in fits of Higgs couplings is however not a significant limitation with the current level of statistical uncertainty; satisfactory fits at this point are  obtained  constructing likelihoods for each exclusive channel and combining each assuming zero correlation.  To do otherwise is to use data that has already been combined by the collaborations, such that correlations are properly encoded.  In this case, however, one would not know which efficiencies to use for weighting each coupling in the production rescaling factor.  There is thus a necessary tradeoff between accounting for correlations and accounting for efficiencies.  

\begin{figure}[htb]
\begin{center}
\includegraphics[width=8cm]{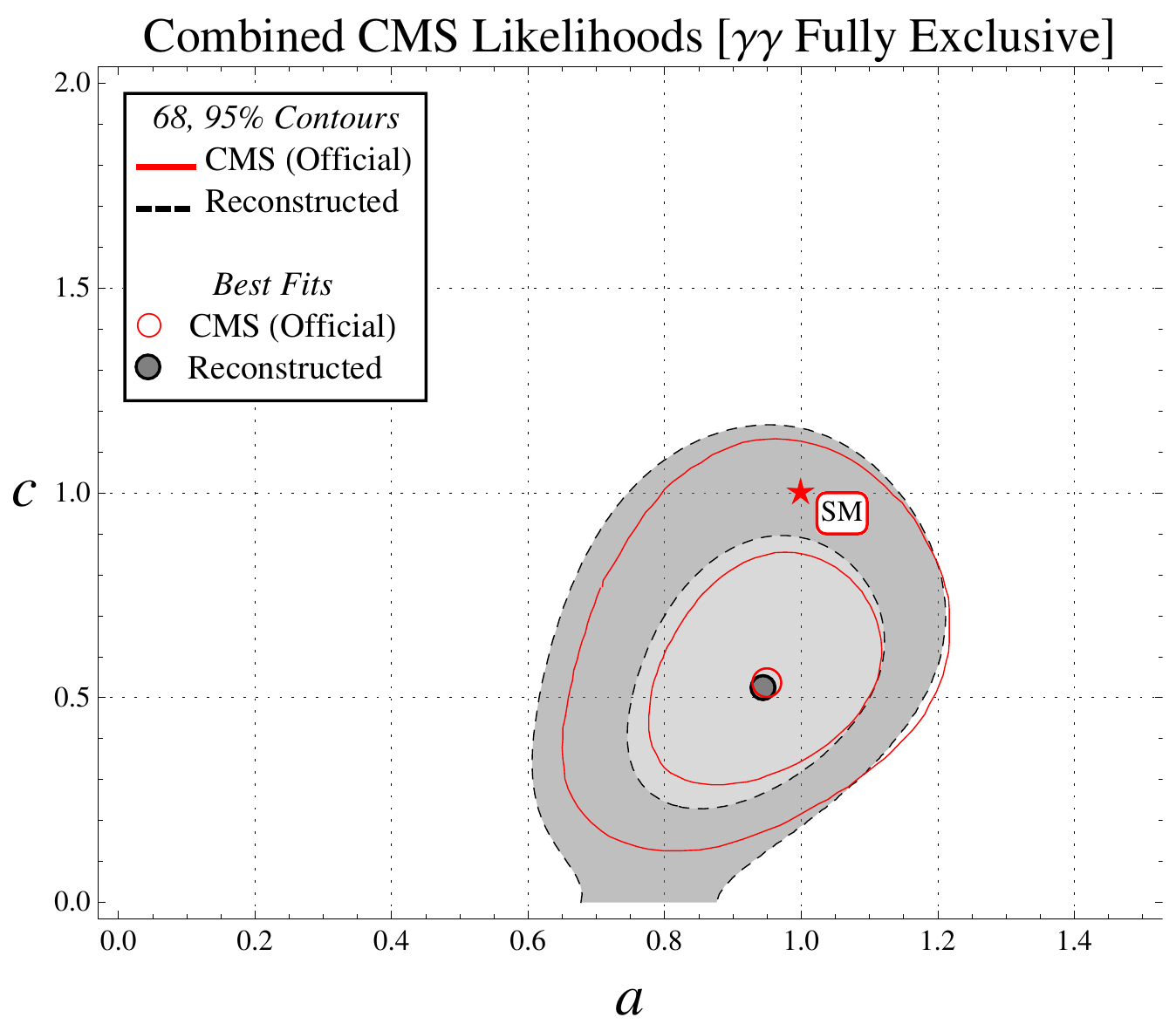}
\caption{\small Higgs couplings to vectors and fermions combining all search modes and using fully exclusive $\gamma \gamma$ channels assuming zero correlation, compared to official results provided from CMS data sets with $\approx 5$/fb at 7 TeV and $\approx 5$/fb at 8 TeV \cite{CMSDiscoveryNote}.  The reproduced best fit point is found to align with the reported best fit at the percent level.}
\label{fig:CMSValidation}
\end{center}
\end{figure}

In Fig.~\ref{fig:CMSValidation} we test the validity of neglecting correlations in favor of proper efficiencies by comparing our reproduced  best fit in $(a,c)$ space to that derived by CMS, using a fully exclusive breakdown for the $\gamma \gamma$ final state.  We use the only official results yet released where all channels entering the fit have been specified and made separately available \cite{CMSDiscoveryNote}.  We have  compared this to the alternative of using pre-combined data for the untagged $\gamma \gamma$ channel under the  assumption of simple inclusive production.  We find an error on the best fit point of less than 1\% proceeding with the exclusive treatment, compared to an error or order 10\% obtained using a semi-inclusive (pre-combined) treatment.   We thus find it preferable to rely on exclusive breakdowns and their quoted efficiencies whenever possible.

\section{CCWZ Formalism for a Composite Higgs}
\label{sec:CCWZ} \setcounter{equation}{0} 
Here we review  the basic structures that we use as ingredients in composite Higgs model building.  Further discussion of the formalism's construction and details can be found in the original literature \cite{CCWZ} as well as in review articles on effective field theory; see for instance \cite{EFT}.

We assume a global symmetry $G$ broken to a subgroup $H$, with transformations of $G$ ($H$)  denoted by $U$ ($V$).   The global group is generated by matrices that we label with capital indices: $T^A$.  When needed, the generators of the unbroken and broken subgroups respectively are specified with unhatted and hatted lowercase indices: $T^a$ and $T^{\hat a}$.  The pions characterizing the coset space are collected in the following matrix transforming as indicated:
\beq
\xi(x) = \exp\, \left(2 i  T^{\hat a} \pi^{\hat a}(x) /f \right) \mapsto U \xi(x) V^\dagger.
\eeq
Gauging global symmetries can be included through a (partially) covariant derivative:
\beq
D_\mu \xi(x) = \partial_\mu \xi(x)  -i g A_\mu^A T^A \xi(x).
\eeq
This form is not quite the covariant object we'd like, since we want to write models in a way that explicitly preserves $H$.  For this we introduce the Cartan form $C_\mu$ and its projections, $E_\mu$ ($d_\mu$), along unbroken (broken) directions:
\beq
C_\mu = i \xi^\dagger(x) D_\mu \xi(x) = E_\mu^a T^a + d_\mu^{\hat a} T^{\hat a}.
\eeq
Explicitly,
\beq
E_\mu^a ={\rm tr} (C_\mu T^a);  \quad d_\mu^{\hat a} =   {\rm tr} (C_\mu T^{\hat a}) .
\eeq
The projections have nice covariant transformation laws as a gauge field and adjoint, respectively, of $H$:
\beq
E_\mu &\equiv& E_\mu^a T^a \mapsto V(E_\mu + i \partial _\mu ) V^\dagger, \\
d_\mu &\equiv& d_\mu^{\hat a} T^{\hat a} \mapsto V d_\mu V^\dagger.
\eeq
 For  reference we note that expanding in pion fields with the conventions adopted thus far, one finds for the $SO(5)/SO(4)$ coset
\beq
d_\mu^{\hat a} &=&A_\mu^{\hat a}(x) - \frac{\sqrt 2}{f} \left[ D_\mu \pi(x) \right]^{\hat a} + \mathcal O(\pi^3)\\
E_\mu^a&=& A_\mu^a(x) + \frac{i}{f^2} \big[ \pi(x) \overleftrightarrow{D_\mu} \pi(x)\big]^a + \mathcal O(\pi^4),
\eeq
where $A_\mu^{\hat a}(x) \equiv  A_\mu^A(x) \, {\rm tr} \, (T^A T^{\hat a})$, etc.

With $E_\mu$ transforming as a gauge field of $H$, it can be used to construct a covariant derivative $\nabla_\mu$ for the other representations of $H$, and the gauged directions' field strength tensor $E_{\mu \nu}$:
\beq
\nabla_\mu &=& \partial_\mu - i E_\mu, \\
E_{\mu \nu} &=& i \left[ \nabla_\mu ,\nabla_\nu \right] 
= \partial_\mu E_\nu -\partial_\nu  E_\mu - i \left[ E_\mu, E_\nu\right]   \nonumber
\eeq
The use of the $H$ covariant derivative allows all Higgs interactions with matter content $\mathcal Q$ of a composite sector to be compactly encoded in kinetic terms, $\Delta \lag = \mathcal Q^\dagger (i \slashed \nabla -M) \mathcal Q$.

\section{Standard Formulas for Loop-Mediated Higgs Couplings}
\label{sec:Loops} \setcounter{equation}{0}
The Higgs production cross sections are given by the following rescaling of the SM quantities (denoting $V = W,Z$):

\begin{equation}
\begin{split}
\sigma(gg\to h) &= \frac{\l|c_tF_{1/2}\l(\frac{4 m_t^2}{m_h^2}\r)+c_bF_{1/2}\l(\frac{4m_b^2}{m_h^2}\r) \r|^2}{\l|F_{1/2}\l(\frac{4 m_t^2}{m_h^2}\r)+F_{1/2}\l(\frac{4m_b^2}{m_h^2}\r) \r|^2} \, \sigma(gg\to h)_{\rm SM} \\[0.15cm]
\sigma(qq\to qq h) &= a^2 \, \sigma(qq\to qq h)_{\rm SM} \\[0.15cm]
\sigma(q\bar q\to Vh) &= a^2 \, \sigma(q\bar q\to Vh)_{\rm SM} \\[0.15cm]
\sigma(gg, q\bar q\to t\bar t h) &= c_t^2 \,\sigma(gg, q\bar q\to t\bar t h)_{\rm SM} .
\end{split}
\end{equation}
Loop contributions from heavy particles (spin one, $1/2$, zero, respectively), are encoded with the functions
\begin{equation}
\begin{split}
F_1(\tau)=&\, 2+3\tau+3\tau(2-\tau)f(\tau)\\
F_{1/2}(\tau)=&\, -2\tau\l[ 1+(1-\tau)f(\tau)\r]  \\
F_0(\tau)=&\, \tau\l[1-\tau f(\tau)\r]\\
\end{split}
\end{equation}

The Higgs branching ratios are likewise determined by rescaling the SM Higgs partial widths. The formulas for these are (denoting by $f$ any of the
SM quarks and leptons):
\begin{equation}
\begin{split}
\Gamma(h\to VV) & = a^2 \, \Gamma(h\to VV)_{\rm SM} \\[0.15cm]
\Gamma(h\to f\bar f) & = c_f^2 \, \Gamma(h\to  f\bar f)_{\rm SM} \\[0.15cm]
\Gamma(h\to gg) & =  \frac{\l|c_tF_{1/2}\l(\frac{4 m_t^2}{m_h^2}\r)+c_bF_{1/2}\l(\frac{4m_b^2}{m_h^2}\r) \r|^2}{\l|F_{1/2}\l(\frac{4 m_t^2}{m_h^2}\r)+F_{1/2}\l(\frac{4m_b^2}{m_h^2}\r) \r|^2} \ \Gamma(h\to  gg)_{\rm SM} \\[0.15cm]
\Gamma(h\to \gamma\gamma) & = \frac{\l| c_t I^t_\gamma(m_h)+c_b I^b_{\gamma}(m_h)+  c_\tau I_\gamma^\tau(m_h)+a J_\gamma(m_h) \r|^2}{\l | I^t_\gamma(m_h)+ I^b_{\gamma}(m_h)+   I_\gamma^\tau(m_h)+ J_\gamma(m_h) \r|^2} 
\, \Gamma(h\to \gamma\gamma)_{\rm SM} \\[0.15cm]
\Gamma(h\to Z\gamma) & = \frac{\left|c_t I^t_{Z\gamma}(m_h)+c_b I^b_{Z\gamma}(m_h)+c_\tau I^\tau_{Z\gamma}(m_h) + a\, J_{Z\gamma}(m_h)\right|^2}{\left|   I^t_{Z\gamma}(M_H)+ I^b_{Z\gamma}(m_h)+ I^\tau_{Z\gamma}(m_h) + \, J_{Z\gamma}(m_h)  \right|^2} \, \Gamma(h\to Z\gamma) _{\rm SM}.
\end{split}
\end{equation}
Thus $\Gamma_{\rm tot}(h)$ is the sum of the above partial widths and ${\rm BR}(h \to X) = \Gamma(h\to X)/\Gamma_{\rm tot}(h)$.
The functions $I$ and $J$ at one-loop level are given by
\begin{align}
I_{\gamma}^t(m_h)=&\, \frac{4}{3}F_{1/2}(4 m_t^2/m_h^2)\l(1-\frac{\alpha_s}{\pi}\r),~~
I_\gamma^b(m_h)=\frac{1}{3}F_{1/2}(4 m_b^2/m_h^2)\l(1-\frac{\alpha_s}{\pi}\r)\nonumber\\
I_{\gamma}^\tau(m_h)=&\, \frac{1}{2}F_{1/2}(4 m_\tau^2/m_h^2),~~
J_\gamma=F_{1}(4 m_W^2/m_h^2)\nonumber\\
I^t_{Z\gamma} (m_h) =&\,  - \frac{ 4 \left(\frac{1}{2} -  \frac{4}{3} \sin^2 \theta_W\right)}{\sin \theta_W \cos \theta_W}
\left[ I_1  \left( \frac{4 m_t^2}{m_h^2}, \frac{4 m_t^2}{m_Z^2} \right) -  I_2  \left( \frac{4 m_t^2}{m_h^2}, \frac{4 m_t^2}{m_Z^2} \right)  \right]\l(1-\frac{\alpha_s}{\pi}\r), \\[0.4cm]
I^b_{Z\gamma} (m_h) =&\,   \frac{ 2 \left(-\frac{1}{2} +  \frac{2}{3} \sin^2 \theta_W\right)}{\sin \theta_W \cos \theta_W}
\left[ I_1  \left( \frac{4 m_b^2}{m_h^2}, \frac{4 m_b^2}{m_Z^2} \right) -  I_2  \left( \frac{4 m_b^2}{m_h^2}, \frac{4 m_b^2}{m_Z^2} \right)  \right]\l(1-\frac{\alpha_s}{\pi}\r), \\[0.4cm]
I^\tau_{Z\gamma} (m_h) =&\,   \frac{ 2 \left(-\frac{1}{2} +  2 \sin^2 \theta_W\right)}{\sin \theta_W \cos \theta_W}
\left[ I_1  \left( \frac{4 m_\tau^2}{m_h^2}, \frac{4 m_\tau^2}{m_Z^2} \right) -  I_2  \left( \frac{4 m_\tau^2}{m_h^2}, \frac{4 m_\tau^2}{m_Z^2} \right)  \right], \\[0.4cm]
\begin{split}
J_{Z\gamma}(m_h) =&\, -\frac{\cos \theta_W}{\sin\theta_W}  \times \left\{  
\left(12 - 4 \tan^2 \theta_W\right) \times I_2 \left(\frac{4 m_W^2}{m_h^2}, \frac{4 m_W^2}{m_Z^2} \right)\right. \\[0.1cm]
 & + \left.  \left[ \left( 1 + \frac{2 m_h^2}{4 m_W^2} \right)  \tan^2 \theta_W  - \left(5+ \frac{2m_h^2}{4 m_W^2} \right) \right]   
    \times I_1 \left( \frac{4 m_W^2}{m_h^2}, \frac{4 m_W^2}{m_Z^2} \right) \right\}\, , 
\end{split}
\end{align}
where
\begin{align}
I_1(a,b) &= \frac{ab}{2(a-b)} + \frac{a^2 b^2}{2(a-b)^2} \left[f(a) -f(b)\right] + \frac{a^2 b}{(a-b)^2} \left[ g(a) -g(b) \right], \\[0.1cm]
I_2(a,b) &= -\frac{ab}{2 (a-b)} \left[ f(a) - f(b) \right] ,
\end{align}
with
\begin{equation}
f(x) = \begin{cases} 
\left[\sin^{-1} \left(1/\sqrt x \right) \right]^2 &{\rm for }  \ x \geq 1 \\[0.1cm]
-\frac{1}{4} \left[  \log \left( \frac{1+\sqrt{1-x}}{1-\sqrt{1-x}} \right) -i \pi \right]^2 &{\rm for }  \ x < 1,
\end{cases} 
\end{equation}
and
\begin{equation}
g(x) = \begin{cases} 
\sqrt{x-1} \sin^{-1} \left( 1/ \sqrt x  \right) &{\rm for }  \ x \geq 1 \\[0.1cm]
\frac{1}{2} \sqrt{1-x} \left[  \log \left( \frac{1+\sqrt{1-x}}{1-\sqrt{1-x}} \right) -i \pi \right]^2 &{\rm for }  \ x < 1.
\end{cases} 
\end{equation}
For discussion of these results, see for instance \cite{HiggsHunters}.

\end{appendix}


\vspace{1cm}






\begin{thebibliography}{9}



\bibitem{ATLASdiscovery} 
  G.~Aad {\it et al.}  [ATLAS Collaboration],
  Phys.\ Lett.\ B {\bf 716}, 1 (2012)
  [arXiv:1207.7214 [hep-ex]].
  
  
  
\bibitem{CMSdiscovery}
  S.~Chatrchyan {\it et al.}  [CMS Collaboration],
  Phys.\ Lett.\ B {\bf 716}, 30 (2012)
  [arXiv:1207.7235 [hep-ex]].
  
  
 \bibitem{Higgs}
    
    P.~W.~Higgs,
  Phys.\ Rev.\ Lett.\  {\bf 13}, 508 (1964); 
  G.~S.~Guralnik, C.~R.~Hagen and T.~W.~B.~Kibble,
  Phys.\ Rev.\ Lett.\  {\bf 13}, 585 (1964);
  F.~Englert and R.~Brout,
  Phys.\ Rev.\ Lett.\  {\bf 13}, 321 (1964);
  P.~W.~Higgs,
  Phys.\ Rev.\  {\bf 145}, 1156 (1966);
  
  
  
  
  
  \bibitem{SM}
  S.~L.~Glashow,
  Nucl.\ Phys.\  {\bf 22}, 579 (1961);
  S.~Weinberg,
  Phys.\ Rev.\ Lett.\  {\bf 19}, 1264 (1967);
  A. Salam, 
  N. Svartholm ed., Almquvist and Wiksell, Stockholm (1968).




\bibitem{lagrangian}
  R.~Contino, C.~Grojean, M.~Moretti, F.~Piccinini and R.~Rattazzi,
  JHEP {\bf 1005} (2010) 089
  [arXiv:1002.1011 [hep-ph]].





\bibitem{LEPprecision}
LEP Collaboration,
  hep-ex/0312023; 
    J.~Alcaraz [ALEPH and CDF and D0 and DELPHI and L3 and OPAL and SLD Collaboration],
  arXiv:0911.2604 [hep-ex].
  
  
  \bibitem{PT}
  M.~E.~Peskin and T.~Takeuchi,
  Phys.\ Rev.\ D {\bf 46}, 381 (1992).
  
  
  
  
  \bibitem{plhc}
  A.~Azatov, R.~Contino and J.~Galloway,
  arXiv:1206.3171 [hep-ph].





  
  \bibitem{fits}
   D.~Carmi, A.~Falkowski, E.~Kuflik and T.~Volansky,
  JHEP {\bf 1207}, 136 (2012)
  [arXiv:1202.3144 [hep-ph]];
  J.~R.~Espinosa, C.~Grojean, M.~Muhlleitner and M.~Trott,
  JHEP {\bf 1205}, 097 (2012)
  [arXiv:1202.3697 [hep-ph]];
  P.~P.~Giardino, K.~Kannike, M.~Raidal and A.~Strumia,
  JHEP {\bf 1206}, 117 (2012)
  [arXiv:1203.4254 [hep-ph]].
  J.~Ellis and T.~You,
  JHEP {\bf 1206}, 140 (2012)
  [arXiv:1204.0464 [hep-ph]];
  P.~P.~Giardino, K.~Kannike, M.~Raidal and A.~Strumia,
  arXiv:1207.1347 [hep-ph].
  J.~Ellis and T.~You,
  JHEP {\bf 1209}, 123 (2012)
  [arXiv:1207.1693 [hep-ph]];
    J.~R.~Espinosa, C.~Grojean, M.~Muhlleitner and M.~Trott,
  arXiv:1207.1717 [hep-ph];
  D.~Carmi, A.~Falkowski, E.~Kuflik, T.~Volansky and J.~Zupan,
  JHEP {\bf 1210}, 196 (2012)
  [arXiv:1207.1718 [hep-ph]].


\bibitem{sfitter}
  M.~Klute, R.~Lafaye, T.~Plehn, M.~Rauch and D.~Zerwas,
  Phys.\ Rev.\ Lett.\  {\bf 109}, 101801 (2012)
  [arXiv:1205.2699 [hep-ph]];
    T.~Plehn and M.~Rauch,
  Europhys.\ Lett.\  {\bf 100}, 11002 (2012)
  [arXiv:1207.6108 [hep-ph]].


\bibitem{fits2}
   A.~Azatov, R.~Contino and J.~Galloway,
  JHEP {\bf 1204}, 127 (2012)
  [arXiv:1202.3415 [hep-ph]].
  
  
  
 
 \bibitem{gfitter}
  M.~Baak, M.~Goebel, J.~Haller, A.~Hoecker, D.~Kennedy, R.~Kogler, K.~Moenig and M.~Schott {\it et al.},
  Eur.\ Phys.\ J.\ C {\bf 72}, 2205 (2012)
  [arXiv:1209.2716 [hep-ph]].
 
  
\bibitem{RychkovEWPT}
   A.~Orgogozo and S.~Rychkov,
  arXiv:1211.5543 [hep-ph].
  
  
  
  \bibitem{ContinoTASI}
    R.~Contino,
  arXiv:1005.4269 [hep-ph].

\bibitem{cohengeorgi}
  A.~G.~Cohen and H.~Georgi,
  Nucl.\ Phys.\ B {\bf 314}, 7 (1989).


\bibitem{RS}
  L.~Randall and R.~Sundrum,
  Phys.\ Rev.\ Lett.\  {\bf 83}, 3370 (1999)
  [hep-ph/9905221].
  
  
  
  \bibitem{CCWZ}
    S.~R.~Coleman, J.~Wess and B.~Zumino,
  Phys.\ Rev.\  {\bf 177}, 2239 (1969);
    C.~G.~Callan, Jr., S.~R.~Coleman, J.~Wess and B.~Zumino,
  Phys.\ Rev.\  {\bf 177}, 2247 (1969).
  
  

  
  
  
  
  \bibitem{MCTC}
    J.~Galloway, J.~A.~Evans, M.~A.~Luty and R.~A.~Tacchi,
  JHEP {\bf 1010}, 086 (2010)
  [arXiv:1001.1361 [hep-ph]].
  
  
    \bibitem{agtr1}
    A.~Falkowski, S.~Rychkov and A.~Urbano,
  JHEP {\bf 1204}, 073 (2012)
  [arXiv:1202.1532 [hep-ph]].
  
  
  
  
    \bibitem{SILH}
    G.~F.~Giudice, C.~Grojean, A.~Pomarol and R.~Rattazzi,
  JHEP {\bf 0706}, 045 (2007)
  [hep-ph/0703164].
  
    
    
 
    
    
    \bibitem{MCHM}
    K.~Agashe, R.~Contino and A.~Pomarol,
  Nucl.\ Phys.\ B {\bf 719}, 165 (2005)
  [hep-ph/0412089];
    K.~Agashe and R.~Contino,
  Nucl.\ Phys.\ B {\bf 742}, 59 (2006)
  [hep-ph/0510164].
  
  
  
  
  \bibitem{GenCH}
    A.~Pomarol and F.~Riva,
  JHEP {\bf 1208}, 135 (2012)
  [arXiv:1205.6434 [hep-ph]];
    M.~Montull and F.~Riva,
  JHEP {\bf 1211}, 018 (2012)
  [arXiv:1207.1716 [hep-ph]].
  
  
    
      \bibitem{ctsign}
    S.~Biswas, E.~Gabrielli and B.~Mele,
  arXiv:1211.0499 [hep-ph];
    M.~Farina, C.~Grojean, F.~Maltoni, E.~Salvioni and A.~Thamm,
  arXiv:1211.3736 [hep-ph].
  
  
  
  
  
  
  
  
  
  \bibitem{ETC}
    S.~Dimopoulos and L.~Susskind,
  Nucl.\ Phys.\ B {\bf 155}, 237 (1979);
    E.~Eichten and K.~D.~Lane,
  Phys.\ Lett.\ B {\bf 90}, 125 (1980).
  
  
  \bibitem{BTC}
    S.~Samuel,
  Nucl.\ Phys.\ B {\bf 347}, 625 (1990);
    A.~Kagan and S.~Samuel,
  Phys.\ Lett.\ B {\bf 252}, 605 (1990).
  
  
  
  
  
  
  \bibitem{twosite}
    R.~Contino, T.~Kramer, M.~Son and R.~Sundrum,
  JHEP {\bf 0705}, 074 (2007)
  [hep-ph/0612180].
  
  
  
  
  
  \bibitem{compfermions}
    R.~Rattazzi and A.~Zaffaroni,
  JHEP {\bf 0104}, 021 (2001)
  [hep-th/0012248].
  
  
  

  
  

\bibitem{LEPfermions}
  P.~Abreu {\it et al.}  [DELPHI Collaboration],
  Eur.\ Phys.\ J.\ C {\bf 8}, 41 (1999)
  [hep-ex/9811005, hep-ex/9811005].


  \bibitem{topcolor}
  C.~T.~Hill,
  Phys.\ Lett.\ B {\bf 345}, 483 (1995)
  [hep-ph/9411426];
      J.~A.~Evans, J.~Galloway, M.~A.~Luty and R.~A.~Tacchi,
  JHEP {\bf 1104}, 003 (2011)
  [arXiv:1012.4808 [hep-ph]];
    H.~Fukushima, R.~Kitano and M.~Yamaguchi,
  JHEP {\bf 1101}, 111 (2011)
  [arXiv:1012.5394 [hep-ph]].





 

  



  
  

  
  

  
  
  
  \bibitem{CompVectors}
   K.~Lane and A.~Martin,
  Phys.\ Rev.\ D {\bf 80}, 115001 (2009)
  [arXiv:0907.3737 [hep-ph]];
  R.~Contino, D.~Marzocca, D.~Pappadopulo and R.~Rattazzi,
  JHEP {\bf 1110}, 081 (2011)
  [arXiv:1109.1570 [hep-ph]];  
  B.~Bellazzini, C.~Csaki, J.~Hubisz, J.~Serra and J.~Terning,
  JHEP {\bf 1211}, 003 (2012)
  [arXiv:1205.4032 [hep-ph]].
  
  
  \bibitem{Higgslowenergy}
   J.~R.~Ellis, M.~K.~Gaillard and D.~V.~Nanopoulos,
  Nucl.\ Phys.\ B {\bf 106}, 292 (1976);
    M.~A.~Shifman, A.~I.~Vainshtein, M.~B.~Voloshin and V.~I.~Zakharov,
  Sov.\ J.\ Nucl.\ Phys.\  {\bf 30}, 711 (1979)
  [Yad.\ Fiz.\  {\bf 30}, 1368 (1979)].
  
  
   \bibitem{GrojeanLET} 
    M.~Gillioz, R.~Grober, C.~Grojean, M.~Muhlleitner and E.~Salvioni,
  JHEP {\bf 1210}, 004 (2012)
  [arXiv:1206.7120 [hep-ph]].
  
  \bibitem{LowVichi}
    I.~Low, R.~Rattazzi and A.~Vichi,
  JHEP {\bf 1004}, 126 (2010)
  [arXiv:0907.5413 [hep-ph]];
    I.~Low and A.~Vichi,
  Phys.\ Rev.\ D {\bf 84}, 045019 (2011)
  [arXiv:1010.2753 [hep-ph]].
  
  
  
  
  
  \bibitem{AGHGG}
    A.~Azatov and J.~Galloway,
  Phys.\ Rev.\ D {\bf 85}, 055013 (2012)
  [arXiv:1110.5646 [hep-ph]].
  
  
  \bibitem{ZGamma}
  A.~Azatov, R.~Contino, J.~Galloway, A.~di Iura; in preparation.
  
  
  \bibitem{contino-pomarol}
    K.~Agashe, R.~Contino, L.~Da Rold and A.~Pomarol,
  Phys.\ Lett.\ B {\bf 641}, 62 (2006)
  [hep-ph/0605341].
  
  
 \bibitem{ZgammaExp}
   G.~Aad {\it et al.}  [ATLAS Collaboration],
  ATLAS-CONF-2013-009;
 S.~Chatrchyan {\it et al.}  [CMS Collaboration],
CMS PAS HIG-13-006.
  
  
  
  
  \bibitem{ZGammaSig}
    J.~S.~Gainer, W.~-Y.~Keung, I.~Low and P.~Schwaller,
  Phys.\ Rev.\ D {\bf 86}, 033010 (2012)
  [arXiv:1112.1405 [hep-ph]].
  
  
  
  

  
    
  \bibitem{MSSMHiggsMass}
    H.~E.~Haber and R.~Hempfling,
  Phys.\ Rev.\ Lett.\  {\bf 66}, 1815 (1991);
    A.~Brignole,
  Phys.\ Lett.\ B {\bf 281}, 284 (1992).
  
  
  \bibitem{GHdecoupling}
    J.~F.~Gunion and H.~E.~Haber,
  Phys.\ Rev.\ D {\bf 67}, 075019 (2003)
  [hep-ph/0207010].
  
  
  
  
      \bibitem{HeavyHiggs}
  L.~Maiani, A.~D.~Polosa and V.~Riquer,
  New J.\ Phys.\  {\bf 14}, 073029 (2012)
  [arXiv:1202.5998 [hep-ph]];
  Phys.\ Lett.\ B {\bf 718}, 465 (2012)
  [arXiv:1209.4816 [hep-ph]];
    N.~Craig and S.~Thomas,
  JHEP {\bf 1211}, 083 (2012)
  [arXiv:1207.4835 [hep-ph]];
    W.~Altmannshofer, S.~Gori and G.~D.~Kribs,
  arXiv:1210.2465 [hep-ph].
  
  
  
  \bibitem{ACCG}
    A.~Azatov, S.~Chang, N.~Craig and J.~Galloway,
  Phys.\ Rev.\ D {\bf 86}, 075033 (2012)
  [arXiv:1206.1058 [hep-ph]].
  
  
    

  
  
  
  \bibitem{DawsonTASI}
    S.~Dawson,
  hep-ph/9712464.
  
  
  
  \bibitem{BlumPredicts}
    K.~Blum, R.~T.~D'Agnolo and J.~Fan,
  arXiv:1206.5303 [hep-ph].
  


  
  
   

  
  
  
  
  \bibitem{MSSMQuartics}
    M.~S.~Carena, J.~R.~Espinosa, M.~Quiros and C.~E.~M.~Wagner,
  Phys.\ Lett.\ B {\bf 355}, 209 (1995)
  [hep-ph/9504316].
  
  
  


  
  
  
  
  
  
  \bibitem{NMSSM}
  M.~Maniatis,
  Int.\ J.\ Mod.\ Phys.\ A {\bf 25}, 3505 (2010)
  [arXiv:0906.0777 [hep-ph]];
    U.~Ellwanger, C.~Hugonie and A.~M.~Teixeira,
  Phys.\ Rept.\  {\bf 496}, 1 (2010)
  [arXiv:0910.1785 [hep-ph]].
  
  
  
  \bibitem{NMSSMmass}
  G.~L.~Kane, C.~F.~Kolda and J.~D.~Wells,
  Phys.\ Rev.\ Lett.\  {\bf 70}, 2686 (1993)
  [hep-ph/9210242];
    J.~R.~Espinosa and M.~Quiros,
  Phys.\ Lett.\ B {\bf 302}, 51 (1993)
  [hep-ph/9212305].
  
  
  
  
  
  
  \bibitem{NMSSMrecent}
    L.~J.~Hall, D.~Pinner and J.~T.~Ruderman,
  JHEP {\bf 1204}, 131 (2012)
  [arXiv:1112.2703 [hep-ph]];
    J.~-J.~Cao, Z.~-X.~Heng, J.~M.~Yang, Y.~-M.~Zhang and J.~-Y.~Zhu,
  JHEP {\bf 1203}, 086 (2012)
  [arXiv:1202.5821 [hep-ph]].
  
  
  
  \bibitem{NDDT}
    P.~Batra, A.~Delgado, D.~E.~Kaplan and T.~M.~P.~Tait,
  JHEP {\bf 0402}, 043 (2004)
  [hep-ph/0309149].
  
  
  
  
  
  
  \bibitem{SCTC}
    A.~Azatov, J.~Galloway and M.~A.~Luty,
  Phys.\ Rev.\ Lett.\  {\bf 108}, 041802 (2012)
  [arXiv:1106.3346 [hep-ph]];
    A.~Azatov, J.~Galloway and M.~A.~Luty,
  Phys.\ Rev.\ D {\bf 85}, 015018 (2012)
  [arXiv:1106.4815 [hep-ph]].
  
  
  \bibitem{GP}
    T.~Gherghetta and A.~Pomarol,
  JHEP {\bf 1112}, 069 (2011)
  [arXiv:1107.4697 [hep-ph]].
  
  
  \bibitem{DSSM}
    J.~J.~Heckman, P.~Kumar, C.~Vafa and B.~Wecht,
  JHEP {\bf 1201}, 156 (2012)
  [arXiv:1108.3849 [hep-ph]].
  
  
  
  \bibitem{FatMagnetic}
    N.~Craig, D.~Stolarski and J.~Thaler,
  JHEP {\bf 1111}, 145 (2011)
  [arXiv:1106.2164 [hep-ph]].
  
  
  \bibitem{SUSYpc}
    C.~Csaki, Y.~Shirman and J.~Terning,
  Phys.\ Rev.\ D {\bf 84}, 095011 (2011)
  [arXiv:1106.3074 [hep-ph]];
    C.~Csaki, L.~Randall and J.~Terning,
  Phys.\ Rev.\ D {\bf 86}, 075009 (2012)
  [arXiv:1201.1293 [hep-ph]];
    R.~Kitano, M.~A.~Luty and Y.~Nakai,
  JHEP {\bf 1208}, 111 (2012)
  [arXiv:1206.4053 [hep-ph]].
  
  
  
  \bibitem{stealth}
    J.~Fan, M.~Reece and J.~T.~Ruderman,
  JHEP {\bf 1111}, 012 (2011)
  [arXiv:1105.5135 [hep-ph]];
    J.~Fan, M.~Reece and J.~T.~Ruderman,
  JHEP {\bf 1207}, 196 (2012)
  [arXiv:1201.4875 [hep-ph]].
  
  
  
  
   \bibitem{staus}
   M.~Carena, S.~Gori, N.~R.~Shah and C.~E.~M.~Wagner,
  JHEP {\bf 1203}, 014 (2012)
  [arXiv:1112.3336 [hep-ph]].
  
  
  
  
  
  \bibitem{Newgaga}
    U.~Ellwanger,
  JHEP {\bf 1203}, 044 (2012)
  [arXiv:1112.3548 [hep-ph]];
    M.~Kadastik, K.~Kannike, A.~Racioppi and M.~Raidal,
  JHEP {\bf 1205}, 061 (2012)
  [arXiv:1112.3647 [hep-ph]];
   J.~F.~Gunion, Y.~Jiang and S.~Kraml,
  Phys.\ Lett.\ B {\bf 710}, 454 (2012)
  [arXiv:1201.0982 [hep-ph]];
    M.~Carena, I.~Low and C.~E.~M.~Wagner,
  JHEP {\bf 1208}, 060 (2012)
  [arXiv:1206.1082 [hep-ph]];
  N.~Bonne and G.~Moreau,
  Phys.\ Lett.\ B {\bf 717}, 409 (2012)
  [arXiv:1206.3360 [hep-ph]];
      L.~G.~Almeida, E.~Bertuzzo, P.~A.~N.~Machado and R.~Z.~Funchal,
  JHEP {\bf 1211}, 085 (2012)
  [arXiv:1207.5254 [hep-ph]];
   K.~Schmidt-Hoberg and F.~Staub,
  JHEP {\bf 1210}, 195 (2012)
  [arXiv:1208.1683 [hep-ph]]. 
  
  
  
  
   \bibitem{GluonFusion}
   B.~A.~Dobrescu, G.~D.~Kribs and A.~Martin,
  Phys.\ Rev.\ D {\bf 85}, 074031 (2012)
  [arXiv:1112.2208 [hep-ph]];
    K.~Kumar, R.~Vega-Morales and F.~Yu,
  arXiv:1205.4244 [hep-ph].
  
  
  
  
    
  \bibitem{ReeceSign}
    M.~Reece,
  arXiv:1208.1765 [hep-ph].
  
  
  
  \bibitem{Moreau}
    G.~Moreau,
  arXiv:1210.3977 [hep-ph].
 
 

  

  
 
 
 
 \bibitem{staupheno}
   M.~Carena, S.~Gori, I.~Low, N.~R.~Shah and C.~E.~M.~Wagner,
  arXiv:1211.6136 [hep-ph].
 
 
 
 \bibitem{stauTheory}
    N.~Arkani-Hamed, K.~Blum, R.~T.~D'Agnolo and J.~Fan,
  arXiv:1207.4482 [hep-ph];
    M.~Carena, S.~Gori, I.~Low, N.~R.~Shah and C.~E.~M.~Wagner,
  arXiv:1211.6136 [hep-ph].
  
  
\bibitem{impostor}
W. D.~Goldberger, B.~Grinstein and W.~Skiba,
 Phys.\ Rev.\ Lett.\  {\bf 100}, 111802 (2008)
 [arXiv:0708.1463 [hep-ph]];
J.~Fan, W. D.~Goldberger, A.~Ross and W.~Skiba,
 Phys.\ Rev.\ D {\bf 79}, 035017 (2009)
 [arXiv:0803.2040 [hep-ph]].
    I.~Low, J.~Lykken and G.~Shaughnessy,
  Phys.\ Rev.\ D {\bf 86}, 093012 (2012)
  [arXiv:1207.1093 [hep-ph]];
    D.~Elander and M.~Piai,
  arXiv:1208.0546 [hep-ph];
  Z.~Chacko, R.~Franceschini and R.~K.~Mishra,
  arXiv:1209.3259 [hep-ph];
  B.~Bellazzini, C.~Csaki, J.~Hubisz, J.~Serra and J.~Terning,
  arXiv:1209.3299 [hep-ph].
  
  
  
  \bibitem{invisible}
    Y.~Bai, P.~Draper and J.~Shelton,
  JHEP {\bf 1207}, 192 (2012)
  [arXiv:1112.4496 [hep-ph]];
     J.~R.~Espinosa, M.~Muhlleitner, C.~Grojean and M.~Trott,
  JHEP {\bf 1209}, 126 (2012)
  [arXiv:1205.6790 [hep-ph]];
    C.~Englert, M.~Spannowsky and C.~Wymant,
  Phys.\ Lett.\ B {\bf 718}, 538 (2012)
  [arXiv:1209.0494 [hep-ph]].
  
  
  \bibitem{friends}
    D.~Bertolini and M.~McCullough,
  arXiv:1207.4209 [hep-ph].
  
  
 

  

  


\bibitem{ATLASbb} 
  [ATLAS Collaboration],
  ATLAS-CONF-2012-161.


\bibitem{ATLASbbT}
  [ATLAS Collaboration],
  ATLAS-CONF-2012-135.
  
  
\bibitem{ATLAStautau} 
  G.~Aad {\it et al.}  [ATLAS Collaboration],
  JHEP {\bf 1209}, 070 (2012)
  [arXiv:1206.5971 [hep-ex]]; 
  ATLAS-CONF-2012-160.



\bibitem{ATLASww} 
  G.~Aad {\it et al.}  [ATLAS Collaboration],
  Phys.\ Lett.\ B {\bf 716}, 62 (2012)
  [arXiv:1206.0756 [hep-ex]]; ATLAS-CONF-2013-030.


\bibitem{ATLASzz} 
  G.~Aad {\it et al.}  [ATLAS Collaboration],
  Phys.\ Lett.\ B {\bf 710}, 383 (2012)
  [arXiv:1202.1415 [hep-ex]];
ATLAS-CONF-2013-013.






\bibitem{ATLASgaga} 
  G.~Aad {\it et al.}  [ATLAS Collaboration],
  Phys.\ Rev.\ Lett.\  {\bf 108}, 111803 (2012)
  [arXiv:1202.1414 [hep-ex]].
  
  
 \bibitem{ATLASgaga1212}
 ATLAS Collaboration,   ATLAS-CONF-2013-012.
  
  
  \bibitem{CMSbb}
    S.~Chatrchyan {\it et al.}  [CMS Collaboration],
  Phys.\ Lett.\ B {\bf 710}, 284 (2012)
  [arXiv:1202.4195 [hep-ex]].
  
  
  
    \bibitem{CMSbbT}
CMS Collaboration,  CMS-PAS-HIG-12-025.



\bibitem{CMStautau}
CMS Collaboration,  CMS-PAS-HIG-12-043; CMS-PAS-HIG-13-004.


  \bibitem{CMSww}
CMS Collaboration,  CMS-HIG-12-042;  CMS-HIG-13-003; CMS-HIG-13-009.
  
  
  \bibitem{CMSzz}
  CMS Collaboration,
  CMS-HIG-12-023;
    CMS-HIG-13-002. 
  
  
  
  
  
\bibitem{CMSgaga}
  CMS Collaboration,
  CMS-PAS-HIG-12-015; CMS-PAS-HIG-13-001.
  
  
  


\bibitem{CDFD0}
  C.~a.~D.~C.~a.~t.~T.~N.~P.~a.~H.~W.~Group [Tevatron New Physics Higgs Working Group and CDF and D0 Collaborations],
  arXiv:1207.0449 [hep-ex].
  
  
\bibitem{TevUpdate}  
See talks from L.~Zivkovic and W.-M.~Yao at ``Rencontres de Moriond", March 2013: {\tt https://indico.in2p3.fr/conferenceOtherViews.py?view=standard\&confId=7411}  
  
    
\bibitem{TevTau}
  V.~M.~Abazov {\it et al.}  [D0 Collaboration],
  arXiv:1211.6993 [hep-ex].




    
  \bibitem{CMSDiscoveryNote} 
 CMS Collaboration,
  CMS-PAS-HIG-12-020.  
  
  
  
    \bibitem{EFT}
  A.~V.~Manohar,
  [hep-ph/9606222];
    G.~Ecker,
  Prog.\ Part.\ Nucl.\ Phys.\  {\bf 35}, 1 (1995)
  [hep-ph/9501357].
  
  
  
  \bibitem{HiggsHunters}
  J.~F.~Gunion, H.~E.~Haber, G.~L.~Kane and S.~Dawson,
  Front.\ Phys.\  {\bf 80}, 1 (2000).
  




\end{thebibliography}
\end{document}